\pdfoutput=1

\documentclass[twocolumn]{aastex63}

\received{2021 July 14}
\revised{2021 August 31}
\accepted{2021 September 02}

\shorttitle{NPOI Be-Star Survey}
\shortauthors{Hutter et al.}

\begin{document}

\title{Surveying the Bright Stars by Optical Interferometry III: \\
A Magnitude-Limited Multiplicity Survey of Classical Be-Stars}

\correspondingauthor{D. J. Hutter}
\email{hutte1dj@cmich.edu}

\author[0000-0002-3580-9251]{D. J. Hutter}
\affiliation{Central Michigan University \\
Department of Physics \\
Mount Pleasant, MI 48859, USA}

\author[0000-0001-7339-4870]{C. Tycner}
\affiliation{Central Michigan University \\
Department of Physics \\
Mount Pleasant, MI 48859, USA}

\author[0000-0002-9402-2870]{R. T. Zavala}
\affiliation{U.S. Naval Observatory, Flagstaff Station \\
10391 West Naval Observatory Road \\
Flagstaff, AZ 86005-8521, USA}

\author[0000-0002-6985-3955]{J. A. Benson}
\affiliation{U.S. Naval Observatory, Flagstaff Station \\
10391 West Naval Observatory Road \\
Flagstaff, AZ 86005-8521, USA}
\affiliation{retired}

\author[0000-0002-4308-0763]{C. A. Hummel}
\affiliation{European Southern Observatory \\
Karl-Schwarzschild-Str. 2 \\
Garching, 85748, Germany}

\author{H. Zirm}
\affiliation{private address}

\begin{abstract}

We present the results of a multiplicity survey for a magnitude-limited sample of 31 classical Be stars conducted with the Navy Precision Optical Interferometer and the Mark III Stellar Interferometer.  The interferometric observations were used to detect companions in ten previously known binary systems.  For two of these sources (66 Oph and $\beta$ Cep) new orbital solutions were obtained, while for a third source ($\upsilon$ Sgr) our observations provide the first direct, visual detection of the hot sdO companion to the Be primary star.  Combining our interferometric observations with an extensive literature search, we conclude that an additional four sources (o Cas, 15 Mon, $\beta$ Lyr, and $\beta$ Cep) also contain wider binary components that are physical companions to the narrow binaries, thus forming hierarchical multiple systems.  Among the sources not previously confirmed as spectroscopic or visual binaries, BK Cam was resolved on a number of nights within a close physical proximity of another star with relative motion possibly suggesting a physical binary.  Combining our interferometric observations with an extensive literature search, we provide a detailed listing of companions known around each star in the sample, and discuss the multiplicity frequency in the sample.  We also discuss the prospects for future multiplicity studies of classical Be stars by long baseline optical interferometry.

\end{abstract}


\section{Introduction} \label{intro}

A Be star, defined in the broadest sense, is ``a non-supergiant B star whose spectrum has, or had at some time, one or more Balmer lines in emission'' \citep{jsj81,gwc87}.  Be star emission features are observed to be transient, but when present, are highly variable on all time scales exceeding a few minutes \citep{rcm13}. A narrower definition, applicable to ``classical'' Be stars is that they are ``very rapidly rotating main sequence B stars, which, through a still unknown, but increasingly constrained process, form an outwardly diffusing gaseous, dust-free Keplerian disk'' \citep{rcm13}. This latter definition distinguishes the classical Be stars from other emission line objects such as Herbig Ae/Be and B[e] stars, and Oe and A -- F shell stars, and in particular intends to exclude systems in which a secondary component $\it currently$ fills its Roche lobe and is actively accreting onto the Be primary \citep{rcm13}.

Classical Be stars are far from uncommon, constituting about 15\% of all B stars \citep{jaj90}, and are among the most rapidly rotating classes of non-degenerate stars in terms of $v \sin i$, and possibly in terms of fractional critical rotation \citep{rcm13}.  Indeed, the measured $v \sin i$ values may underestimate the rotational velocity of some stars because gravity darkening causes the most rapidly rotating part of the star to become increasingly less luminous at rotational velocities approaching critical \citep{toh04}.  Interferometric results seem to confirm this effect \citep{auf06,auf07}.

The disks that are the source of the emission features in classical Be stars are highly variable in structure, sometimes disappearing altogether for extended periods of time \citep{gbb07}.  This structural variability suggests that gas injection into the disks \citep[``decretion,''][]{rcm13} cannot be entirely due to the rapid rotation of the Be star \citep{toh04}, but nonetheless it greatly assists other processes such as magnetic fields, stellar pulsations, and stellar winds in lifting material from the stellar surface into the orbiting disk \citep{sow05}.

Therefore, it appears that rapid stellar rotation is a necessary, but likely not sufficient, component in the emergence of the Be star phenomenon.  However, the question of why the central stars of Be systems rotate so rapidly does not have a generally agreed answer.  While there is evidence that many Be stars were ``spun up'' by earlier mass transfer from a binary companion that later evolved into a neutron star, white dwarf, or a hot subdwarf \citep{pcw91, drg00}, studies of B and Be stars by spectroscopy \citep{aac84}, speckle interferometry \citep{mas97,sgb20}, and adaptive optics \citep[AO;]{oap10} do not support any significant difference in binary frequency.  In this regard, the work of \citet{oap10} is particularly noteworthy in that it probed binaries at component magnitude differences of up to 10~mag, down to angular separations of 100 mas, corresponding to 20~AU at typical target distances.  However, binaries at even closer separations are not precluded.  For example, studies of classical Be stars in the radio that show spectral energy distribution (SED) turndowns, thought to be indicative of disk truncation due to close binary companions~\citep{kcr19}, point to a population of binary companions at separations of only a few tens to perhaps several hundred radii of the primary Be star (equivalent to $< 10$~AU).  At distances typical of the brighter Be stars, these separations are equivalent to angular separations ($\la$ 30~mas) that are generally inaccessible to either speckle interferometry or AO.  Likewise, the rotationally broadened spectral lines of classical Be stars may make it difficult to detect the low-amplitude radial velocity (RV) shifts ($\la 10$~km~s$^{-1}$) caused by such low-mass companions \citep{kcr19}. Therefore, long baseline optical interferometry (LBOI) appears ideally suited to the detection of very close companions to Be star primaries.  

LBOI has been extremely successful in the direct resolution of Be stars disks, with many examples provided by interferometers such as the Mark III Stellar Interferometer \citep[][$\zeta$ Tau]{qbm94} and the Navy Precision Optical Interferometer \citep[NPOI;][including o Aqr, $\gamma$ Cas, $\beta$ CMi, $\chi$ Oph, 48 Per, $\phi$ Per, $\eta$ Tau, and $\zeta$ Tau]{thm03,tha04,tlh05,tgz06,tjs08,stj15,jsg17}.  These studies utilized measurements obtained simultaneously in many spectral channels covering a wide spectral range where the source is unresolved, or weakly resolved (the continuum emission from the photosphere of the central star) to calibrate the data in one or two spectral channels containing a strong signal in the H$\alpha$ line originating in the optically thick circumstellar envelope that is clearly resolved in nearer sources.  Modeling of these data has quantified the extent, orientation, and brightness profile of these disks, and in some cases directly constrained non-LTE models of disk temperature and density distribution \citep{jsg17}.  Currently archived NPOI data exist for perhaps another half-dozen sources where the circumstellar emmision is sufficiently resolved for such studies, including binary systems where it may be possible to study the mutual inclination between the Be star disk and a companion star's orbit.  

While LBOI observations have generally not been used to examine the general question of the multiplicity of Be stars, having concentrated on only a few exceptional binary systems \citep[e.g.,][]{taz11}, there exist extensive archives of past observations that can be utilized for the purpose of such a study. For example, the data archives of the Mark III and NPOI contain observations of Be stars on hundreds of nights spanning a period of over 30 years (including the observations used in the circumstellar disk studies cited above).

Here we report on such a multiplicity study, based on archival observations of 31 stars from a magnitude-limited sample of classical Be stars.  We first summarize the capabilities of the NPOI and Mark III for such a survey in \S~\ref{npoi}.  Based on these capabilities, we next discuss the selection of the targets for the survey in \S~\ref{targsel}.  The observing procedures and data reduction are described in \S~\ref{obs}.  In \S~\ref{modeling}, we discuss the systematic examination of the calibrated visibility data resulting from the NPOI observations of the program sources for statistically significant evidence of stellar companions and the detailed modeling of the separation, position angle, and magnitude difference of the detected binaries.  The results for each of the program sources are then presented in \S~\ref{bindet}, where for two sources (66 Oph and $\beta$ Cep) there were sufficient number of new observations to justify fitting for updated orbital parameters, while for a third source ($\upsilon$ Sgr) our observations provide the first direct, visual resolution of the binary.  For the close binary in the BK Cam system resolved by the NPOI (BK Cam Aa,Ab), the relative motion and close physical proximity of the components indicate a possible physical binary with a period of many decades.

To place our results in the larger context of previous multiplicity studies, we also include in \S~\ref{bindet} the results of an extensive search of the relevant literature.  This, in turn, allows a discussion of the multiplicity statistics of our sample (\S~\ref{multstat}), as well as a comparison of the sensitivity and completeness of our survey with other surveys made at complementary ranges of component angular separation and magnitude difference (\S~\ref{senscomp}).  Lastly, the future role of LBOI in multiplicity studies of Be stars is discussed (\S~\ref{future}).

\section{The NPOI and Mark III Interferometers} \label{npoi}

The NPOI \citep{arm98} located on Anderson Mesa, AZ, is a joint project of the U. S. Naval Observatory and the Naval Research Laboratory in collaboration with Lowell Observatory.  A detailed description of the relevant features of the instrument can be found in the first paper of this series \citep[][hereafter Paper I, and references therein]{hzt16}, and therefore, here we will only note changes in the system configuration and reemphasize a few key features.  

\begin{figure}
\epsscale{1.1}
\plotone{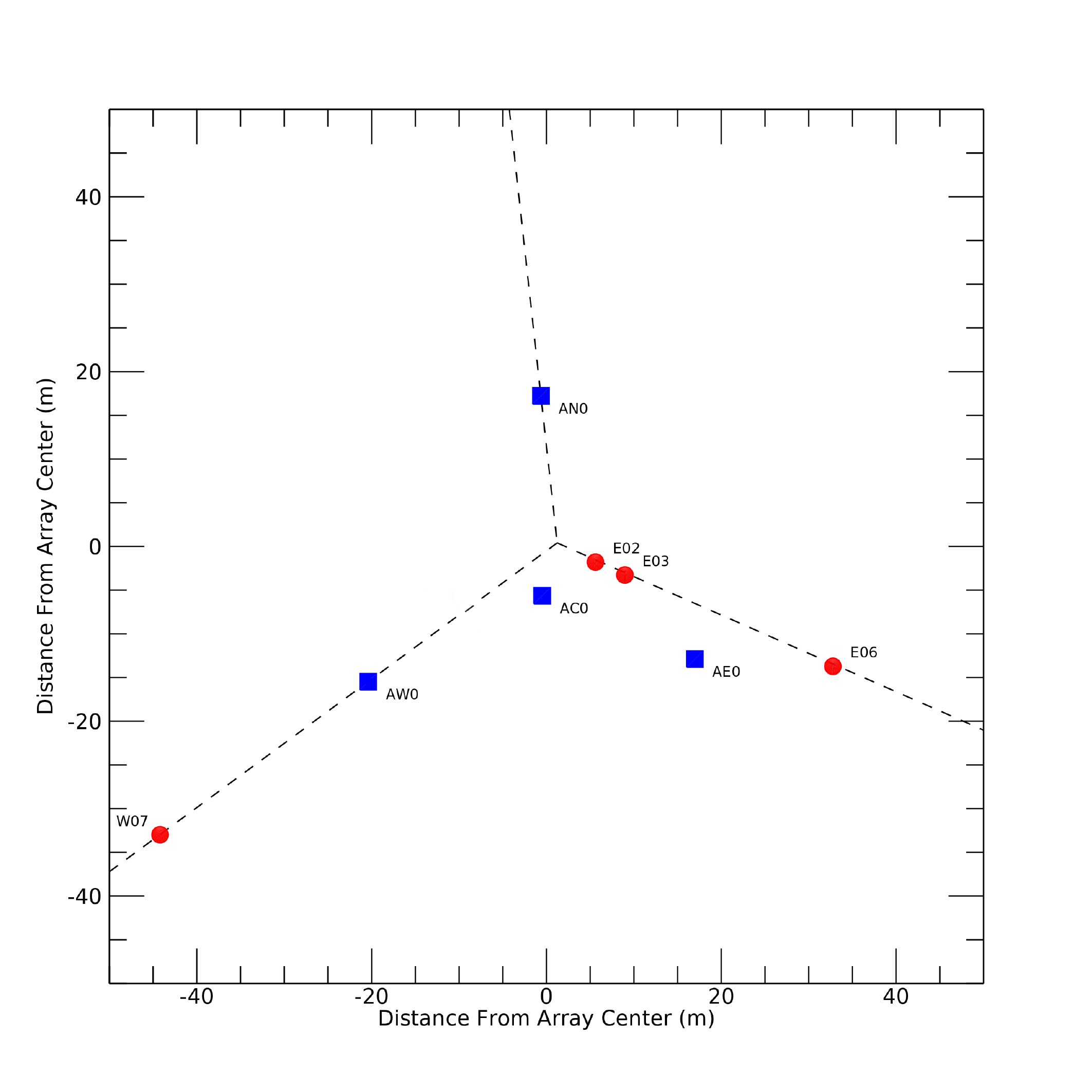}
\caption{The locations and designations of the NPOI siderostat stations used, in various combinations, for the 
observations reported in this paper.  The locations of the astrometric (filled squares) and imaging (filled circles) 
stations are plotted with respect to the nominal array center.}
\label{Be-star-stations}
\end{figure}

The four stations of the astrometric array (AC, AE, AN, and AW), plus four additional imaging siderostats at stations E02, E03, E06, and W07 (Figure~\ref{Be-star-stations}), were used for the observations reported here.  The resulting baselines range from 7.2~m (AC--E02) to 79.4~m (E06--W07).  A list of all the baselines used in this study appears in Table~\ref{baselines}, along with their lengths~($B$), orientations, and nominal angular resolutions at 550~nm \citep[$\lambda/(2B)$;][]{trb00}.

%% array stations

%% Table 1 -- Baseline Parameters

%%\floattable
%%\startlongtable
\begin{deluxetable}{crrr}
\tablecaption{Baseline Parameters \label{baselines}}
\tablewidth{0pt}
\tablehead{
\colhead{Baseline}  &   \colhead{Length}    &   \colhead{Orientation}   &   \colhead{Resolution}    \\
\colhead{}          &   \colhead{(m)}       &   \colhead{(deg)}         & \colhead{(mas)}           \\
\colhead{(1)}       &   \colhead{(2)}       &   \colhead{(3)}           & \colhead{(4)}             \\
}
\startdata
%% BL      Length       PA      Res.
AC-AE	&	18.9	&	113	&	3.0	\\
AC-AN	&	22.9	&	  0	&	2.5	\\
AC-AW	&	22.2	&	244	&	2.6	\\
AC-E02	&	 7.2	&	 58	&	7.9	\\
AC-E03	&    9.8	&	 76	&	5.8	\\
AC-E06	&	34.3	&	104	&	1.7	\\
AC-W07	&	51.6	&	238	&	1.1	\\
AE-AN	&	34.9	&	330	&	1.6	\\
AE-AW	&	37.5	&	266	&	1.5	\\
AE-E02	&	15.9	&	314	&	3.6	\\
AE-E06	&	15.9	&	 93	&	3.6	\\
AE-W07	&	64.4	&	252	&	0.9	\\
AN-AW	&	38.2	&	211	&	1.5	\\
AN-E02	&	20.0	&	162	&	2.8	\\
AN-E06	&	45.6	&	133	&	1.2	\\
AN-W07	&	66.5	&	221	&	0.9	\\
AW-E02	&	29.4	&	 62	&	1.9	\\
AW-E03	&	31.9	&	 68	&	1.8	\\
AW-E06	&	53.3	&	 88	&	1.1	\\
AW-W07	&	29.6	&	234	&	1.9	\\
E06-W07	&	79.4	&	256	&	0.7	\\
\enddata
\tablecomments{Col. (1): baseline name, composed of siderostat station names, in 
alphabetical order.  Col. (2): approximate baseline length (meters) in the horizontal (east -- 
north) plane.  Col. (3): position angle (degrees) of second station with respect to the first 
station, measured from north through east.  Col. (4): nominal angular resolution (mas) at  
550 nm.}
\end{deluxetable}

The unvignetted aperture is the same for all siderostats (35 cm), but is stopped down to a 12.5~cm diameter by the feed-system optics.  All stations are equipped with wave-front tip-tilt correctors.  The feed-system that directs light from each siderostat to the beam-combining lab is evacuated and contains remotely actuated mirrors that allow the configuration of the light path from each siderostat to a corresponding delay line (up to 6).  The NPOI has six vacuum delay lines that can add up to 35~m of continuous additional optical path length to each beam (these are referred to as fast delay lines; FDLs).  These FDLs are used to track the atmospheric and sidereal motion of the fringes.

The beam combiner used for all the observations reported here is a pupil-plane, free-space, bulk-optics system that can coherently combine a maximum of six input beams.  The three beam combiner outputs each contain light from up to four stations~\citep[see the schematic diagram in Figure~2 of][and Figure~1 of Paper I]{tgz06}.  The combined beams, after spatial filtering using pinholes, pass through prism spectrometers and are then collimated onto lenslet arrays and detected by photon-counting avalanche photodiodes (APDs).  The spectral range covered is from 550 to 860~nm in 16 channels equally-spaced in wave number and centered near 700~nm, approximating the original Johnson \textit{R} band \citep{jon66, hin11}.  As the beam combiner does not use single-mode optical fibers in either the beam combination process or for spatial filtering, it does not suffer from the various effects that effectively limit the {$\it interferometric$} field of view~(FOV) for combiners employing single-mode fibers (typically resulting in FOV $\leq$ 50 -- 100~mas).  However, the finite bandwidth of the beam combiner's 16 spectral channels reduces visibility contrast (via bandwidth smearing), which can be estimated and usefully corrected for angular separations out to $\sim$ 860~mas (Paper I).  Additionally, the 50~$\mu$m diameter pinholes used as spatial filters in the spectrometers serve to restrict the {$\it photometric$} field of view to a radius of $\approx$ 1.5 arcseconds \citep{moz05}, effectively blocking incoherent flux from wider binary/multiple system components, such as the known, wide components discussed in (\S~\ref{bindet})\footnote{In the unlikley event of there being tertiary or higher order components in the range of angular separation 860~mas $\le$ $\rho$ $\le$ 1.5~arcsec, these could contribute incoherent flux.  Such flux could lower the contrast of the fringes for a detected binary pair ($\rho$ $\le$ 860~mas) and thus affect the measurement of magnitude difference of the closer binary.}.

Each FDL imposes on its beam a triangle-wave modulation in the optical path difference.  The resulting delay modulation on a baseline is the difference between two FDL modulations.  The modulation sweeps the fringe pattern of a corresponding baseline in time causing an intensity that varies sinusoidally with time.  Changing the amplitude of the modulations changes the frequencies of the fringes.  Since the three output beams of our hybrid six-beam combiner contain contributions from up to four input beams, there are a maximum of six baselines present on each output beam.  There are many stroke amplitude choices for input beams that will produce the signal corresponding to different baselines (i.e., beam pairs) at separate frequencies.  Custom electronics bin the detected photons in synchrony with the delay modulation and compute real-time fringe tracking error signals.  The real and imaginary components of the Fourier transforms along the bin direction (at up to eight frequencies) provide the science data, while the real and imaginary components of the Fourier transform along the wavelength direction at each of the eight frequencies provide the group delay, which in turn is used to calculate the fringe tracking error signals.  

As discussed in Paper I, the NPOI has the capability to operate in a Fourier transform spectrometer (FTS) mode since 2005 that allows for very accurate measurement of the central wavelengths of all spectrometer channels at the level of $\approx 0.6$~nm RMS (i.e., $\pm$ 0.1$\%$ at 656~nm).  The systematic errors in the wavelength scale for observations obtained before 2005 are likely to be of the order $\pm$ 0.5$\%$. With the ability to measure precisely the central wavelength of each of the 16 spectral channels, some observations utilized in this study were optimized for the H$\alpha$ emission line at 656.28~nm (e.g., the Be star circumstellar disk studies cited in \S~\ref{intro}) by translating the lenslet array such that the H$\alpha$ line was centered in a 15~nm wide spectral channel, with the remaining spectral channels at both shorter and longer wavelengths. Even for observations that were not specifically optimized for H$\alpha$ emission work, and where the H$\alpha$ line was not necessarily centered in a spectral channel, it was still possible to clearly isolate the interferometric channels that contained the H$\alpha$ emission from the circumstellar disks associated with Be stars~\citep{thm03}.  To avoid any contribution from Be star disks with strong H$\alpha$ emission, and sources where the circumstellar disks have been clearly resolved in the H$\alpha$, we have excluded the H$\alpha$ containing channel from our binary-star analysis where we observed that the H$\alpha$ squared visibility amplitude ($V^2$) values were more than $\sim$15\% lower than the surrounding continuum channels. It should be noted, this is the same approach that has been utilized when the H$\alpha$ channel was excluded from analysis when the binary signature was modeled in NPOI observations of the Be star $\delta$~Sco (\S~\ref{delsco}), which has significant H$\alpha$ emission from its circumstellar disk~\citep{taz11}.  Thus, in summary, the visibility data in the spectral channel containing the H$\alpha$ emission, which are of paramount importance in Be star disk studies such as those cited in \S~\ref{intro}, are here, in the case of a stellar multiplicity study, {\it deliberately excluded}, the surrounding continuum channels containing the data relevant to the detection of a binary companion (\S~\ref{modeling}).

Lastly, a small number of additional observations (8\% of the total obtained; see also \S~\ref{obslog} and \S~\ref{betcep}) were obtained at the Mark III Stellar Interferometer \citep{sch88,haq93}, the NPOI's immediate predecessor, which operated between 1986 and 1992.  The Mark III was a phase-tracking interferometer capable of observations on a single baseline (one siderostat pair) at a time, but with a selection of baselines spanning 3 -- 31 m in length.  Fringe tracking was performed in a broadband channel (650 -- 900 nm) with science observations obtained simultaneously in three narrow-band (22 -- 25 nm) channels centered at 800, 550, and 500 nm (450 nm in 1989).

\section{Program Star Selection} \label{targsel}

For the purposes of our multiplicity survey of Be stars, the target list (``Program Stars,'' see Table~\ref{targetlist}) was limited to a small subset of the 2330 stars currently listed in the {\it Be Star Spectra (BeSS) Database} \citep{nei11}.  Five selection criteria were used to produce a modest list of sources that could be practically observed given the capabilities of the NPOI, and for which a significant number of observations exist in the NPOI data archive:

1) First, the list was culled to include only ``classical'' Be stars, defined by \citet{nei11} as non-supergiant B stars that have displayed Balmer line emission on at least once occasion (and including some late O and early A stars showing such emission).  Thus, all Herbig Ae/Be stars and B[e] supergiants were excluded.

2) Second, the list was next culled to include only stars with an apparent magnitude $m_V$ $\leq$ 5.0.  While the NPOI has under exceptionally good observing conditions successfully obtained interferometric observations of stellar targets as faint as $m_{\rm V} \sim 6.0$, consistently good quality data has generally only been obtained over multiple nights for stars with $m_{\rm V} \leq 5.0$.

3) Third, we further restricted the sample to include only stars with declinations $\delta$ $\geq$ -23$\arcdeg$, the accessible range of observation on the most commonly used baselines of the NPOI (e.g., Paper I, Figure 2).

4) Fourth, the penultimate cull of the sample was made to include only stars for which the archive of NPOI observations (through 2019 October 1 UT, inclusive) indicated at least one night of four or more coherent observations on one or more baselines.  We judged this to be the absolute minimum amount of data needed for meaningful binary detection and modeling (\S~\ref{modeling}).

5) Lastly, six stellar targets were excluded from analysis either because they had originally been observed as calibrators for sources on other observing programs, and there being no other stars observed that could be used as substitute calibrators (within angular distance $\lesssim 30 \arcdeg$) on any night, or the data were of such poor quality that reasonable calibration and modeling was not possible (\S~\ref{obs}, \S~\ref{modeling}).

The application of all these criteria resulted in a final {$\it magnitude-limited$} sample list of 31 program stars (Table~\ref{targetlist}).  While we would have preferred a volume-limited sample, such was not feasible given the low spatial density of Be stars and the $m_V$ $\leq$ 5.0 limit of NPOI observations.  For example, had we selected classical Be stars from BeSS using the same declination limit, but had imposed a distance limit at the {$\it median$} HIPPARCOS\footnote{For purpose of comparison, we also include the GAIA EDR3 or DR2 parallax in Table~\ref{targetlist}, where available.  These and the corresponding distances are consistent, on average, with the HIPPARCOS values at the 2$\sigma$ level, and in no case would the substitution of the GAIA values change any of our conclusions in \S~\ref{bindet}.} distance of our Table~\ref{targetlist} sample (151 pc), the result would have been a similar number of stars (32), but only 22 (69\%) of these would have $m_V$ $\leq$ 5.0 and thus be observable.  Had we further limited our sample to the same 75 pc limit of a proposed volume-limited sample of all B-type stars \citep{rdr14} in hopes of a higher percentage of observable stars, only five Be stars would remain, with only four having $m_V$ $\leq$ 5.0.

%% Table 2 -- Program Stars

\begin{longrotatetable}
\begin{deluxetable}{lcrccrcccccc}
\tablecaption{Program Stars \label{targetlist}}
\tablewidth{0pt}
\tablehead{
\\
\colhead{} & \colhead{} & \colhead{} & \colhead{Other} & \colhead{$m_V$} & \colhead{$\bv$} & \colhead{GAIA EDR3} & \colhead{Hipparcos} &
\colhead{Hipparcos} & \colhead{$M_V$} & \colhead{Spectral} & \colhead{Multiplicity}\\
\colhead{HIP} & \colhead{HR} & \colhead{FK5} & \colhead{Name} & \colhead{(mag)} & \colhead{(mag)} & \colhead{Parallax (mas)} & \colhead{Parallax (mas)} &
\colhead{Distance (pc)} & \colhead{(mag)} & \colhead{Type} & \colhead{}\\
\colhead{(1)} & \colhead{(2)} & \colhead{(3)} & \colhead{(4)} & \colhead{(5)} & \colhead{(6)} & \colhead{(7)} & \colhead{(8)} &
\colhead{(9)} & \colhead{(10)} & \colhead{(11)} & \colhead{(12)}\\
}
\startdata
%% HIP       HR         FK5           Name               V            B-V        GAIA EDR3 plx              Hipp plx                  dist                       Mv         Sp Type.  # Comp
  3504	&	 193	&	  25	&	o Cas			&	4.50	&	-0.060	& 4.70$\pm$0.43$^{\rm a}$   &   4.64$\pm$0.38	&	215.5$^{+19.2}_{-16.3}$	&	$-2.17$	&	B5IIIe		&	4	\\
  4427	&	 264	&	  32	&	$\gamma$ Cas	&	2.39	&	-0.100	&	 \nodata               &   5.94$\pm$0.12	&	168.4$^{+3.5}_{-3.3}$	&	$-3.74$	&	B0IVpe		&	8	\\
  8068	&	 496	&	  57	&	$\phi$ Per		&	4.06	&	-0.040	&	 5.40$\pm$0.20          &   4.54$\pm$0.20	&	220.3$^{+10.2}_{-9.3}$	&	$-2.65$	&	B2Vpe		&	2	\\
  8886	&	 542	&	  63	&	$\epsilon$ Cas	&	3.37	&	-0.150	&	 7.00$\pm$0.25          &   7.92$\pm$0.43	&	126.3$^{+7.2}_{-6.5}$	&	$-2.14$	&	B3IIIe		&	1	\\
 15520	&	 985	&	\nodata	&	BK Cam			&	4.85	&	-0.150	&	 4.06$\pm$0.32          &   4.28$\pm$0.48	&	233.6$^{+29.5}_{-23.6}$	&	$-1.99$	&	B2.5Vne		&	2	\\
 16826	&	1087	&	\nodata	&	$\psi$ Per		&	4.23	&	-0.060	&	 5.96$\pm$0.13          &   5.59$\pm$0.22	&	178.9$^{+7.3}_{-6.8}$	&	$-2.03$	&	B5Ve		&	1	\\
 17499	&	1142	&	 136	&	17 Tau			&	3.70	&	-0.110	&	 8.35$\pm$0.44          &   8.06$\pm$0.25	&	124.1$^{+4.0}_{-3.7}$	&	$-1.77$	&	B6IIIe		&	1	\\	
 17608	&	1156	&	\nodata	&	23 Tau			&	4.18	&	-0.060	&	 7.07$\pm$0.29          &   8.58$\pm$0.37	&	116.6$^{+5.3}_{-4.8}$	&	$-1.15$	&	B6IVe		&	1	\\
 17702	&	1165	&	 139	&	$\eta$ Tau		&	2.87	&	-0.090	&	 \nodata               &   8.09$\pm$0.42	&	123.6$^{+6.8}_{-6.1}$	&	$-2.59$	&	B7IIIe		&	1	\\
 19343	&	1273	&	 152	&	48 Per			&	4.03	&	-0.030	&	 6.38$\pm$0.29          &   6.84$\pm$0.16	&	146.2$^{+3.5}_{-3.3}$	&	$-1.79$	&	B3Ve		&	1	\\
 26451	&	1910	&	 211	&	$\zeta$ Tau		&	3.03	&	-0.190	&	 \nodata               &   7.33$\pm$0.82	&	136.4$^{+17.2}_{-13.7}$	&	$-2.64$	&	B2IVe		&	2	\\
 26594	&	1934	&	2423	&	$\omega$ Ori	&	4.59	&	-0.110	&	 1.98$\pm$0.15          &   2.36$\pm$0.29	&	423.7$^{+59.4}_{-46.4}$	&	$-3.55$	&	B2IIIe		&	1	\\
 31978	&	2456	&	\nodata	&	15 Mon			&	4.64	&	-0.240	&	 \nodata               &   3.55$\pm$0.50	&	281.7$^{+46.2}_{-34.8}$	&	$-2.61$	&	B1Ve		&	4	\\
 36188	&	2845	&	 285	&	$\beta$ CMi		&	2.89	&	-0.090	&	 \nodata               &   20.17$\pm$0.20	&	49.58$^{+0.50}_{-0.49}$	&	$-0.59$	&	B8Ve		&	1	\\
 55084	&	4368	&	1292	&	$\phi$ Leo		&	4.45	&	 0.204	&	17.28$\pm$0.19          &   17.71$\pm$0.25	&	56.47$^{+0.81}_{-0.79}$	&	$ 0.69$	&	A7IVne		&	1	\\
 61281	&	4787	&	 472	&	$\kappa$ Dra	&	3.89	&	-0.140	&	 7.00$\pm$0.30          &   6.65$\pm$0.34	&	150.4$^{+8.1}_{-7.3}$	&	$-2.00$	&	B6IIIpe		&	2	\\
 76041	&	5774	&	\nodata	&	53 Boo			&	4.98	&	 0.096	&	 7.39$\pm$0.18          &   8.40$\pm$0.77	&	119.1$^{+12.0}_{-10.0}$	&	$-0.40$	&	A5Ve		&	2	\\
 78207	&	5941	&	1417	&	48 Lib			&	4.87	&	-0.100	&	 7.15$\pm$0.12          &   6.97$\pm$0.24	&	143.5$^{+5.1}_{-4.8}$	&	$-0.91$	&	B8IIe		&	1	\\
 78401	&	5953	&	 594	&	$\delta$ Sco	&	2.32	&	-0.120	&	 \nodata               &   6.64$\pm$0.89	&	150.6$^{+23.3}_{-17.8}$	&	$-3.57$	&	B0.2IVe		&	2	\\
 80569	&	6118	&	3298	&	$\chi$ Oph		&	4.43	&	 0.280	&	 6.54$\pm$0.20          &   6.21$\pm$0.23	&	161.0$^{+6.2}_{-5.8}$	&	$-1.60$	&	B2Vne		&	1	\\
 81377	&	6175	&	 622	&	$\zeta$ Oph		&	2.56	&	 0.020	&	 7.41$\pm$0.66          &   8.91$\pm$0.20	&	112.2$^{+2.6}_{-2.5}$	&	$-2.69$	&	O9Ve		&	1	\\
 88149	&	6712	&	\nodata	&	66 Oph			&	4.60	&	-0.020	&	 4.90$\pm$0.37          &   5.01$\pm$0.26	&	199.6$^{+10.9}_{-9.8}$	&	$-1.90$	&	B2Ve		&	2	\\
 88794	&	6779	&	 681	&	o Her			&	3.83	&	-0.025	&	 9.35$\pm$0.15          &   9.65$\pm$0.16	&	103.6$^{+1.7}_{-1.7}$	&	$-1.25$	&	B9.5Ve		&	1	\\
 92420	&	7106	&	 705	&	$\beta$ Lyr		&	3.42	&	 0.000	&	 3.60$\pm$0.18          &   3.39$\pm$0.17	&	295.0$^{+15.6}_{-14.1}$	&	$-3.93$	&	B7Ve		&	4	\\
 95176	&	7342	&	 727	&	$\upsilon$ Sgr	&	4.61	&	 0.100	&	 2.09$\pm$0.17          &   1.83$\pm$0.23	&	546.5$^{+78.6}_{-61.0}$	&	$-4.08$	&	Bpshe		&	2	\\
103632	&	8047	&	1551	&	59 Cyg			&	4.75	&	-0.040	&	 1.47$\pm$0.37          &   2.30$\pm$0.42	&	434.8$^{+97.1}_{-67.1}$	&	$-3.44$	&	B1.5Vnne	&	3	\\
105138	&	8146	&	1559	&	$\upsilon$ Cyg	&	4.42	&	-0.100	&	 5.01$\pm$0.19          &   5.08$\pm$0.55	&	196.9$^{+23.9}_{-19.2}$	&	$-2.05$	&	B2Vne		&	1	\\
106032	&	8238	&	 809	&	$\beta$ Cep		&	3.23	&	-0.220	&	 \nodata               &    4.76$\pm$0.30	&	210.1$^{+14.1}_{-12.5}$	&	$-3.38$	&	B2IIIev		&	3	\\
108874	&	8402	&	3765	&	o Aqr			&	4.69	&	-0.050	&	 7.00$\pm$0.14          &   7.49$\pm$0.23	&	133.5$^{+4.2}_{-4.0}$	&	$-0.94$	&	B7IVe		&	1	\\
110386	&	8520	&	 843	&	31 Peg			&	4.99	&	-0.100	&	 \nodata               &   2.01$\pm$0.28	&	497.5$^{+80.5}_{-60.8}$	&	$-3.49$	&	B2IVe		&	1	\\ 
113889	&	8773	&	1602	&	$\beta$ Psc		&	4.52	&	-0.120	&	 8.05$\pm$0.13          &   7.99$\pm$0.22	&	125.2$^{+3.5}_{-3.4}$	&	$-0.97$	&	B6Ve		&	1	\\ 
\enddata
\tablecomments{Cols. (1), (2), (3), (5), and (6): HIPPARCOS number \citep{esa97}, HR number \citep{hof91}, FK5 number \citep{frk88}, apparent $m_V$ magnitude, and $\bv$ color index from the $\mathit{SIMBAD}$ database \citep{wen00}, and references therein.  Col. (4): other name (Bayer or Flamsteed designation) from the $\mathit{SIMBAD}$ database \citep{wen00} or the BeSS database 
\citep{nei11}.  Col. (7): GAIA EDR3 parallaxes ($^{\rm a}$ -- except for HR 193 where the reported EDR3 parallax was erroneous and was replaced with DR2 value). Not all targets had 
parallaxes available in the GAIA archive (https://gea.esac.esa.int/archive/). Col. (8): parallax from \citet{vL07}.  Col. (9): distance, derived from Col. (8).  Col. (10): absolute $M_V$ 
magnitude, derived from Cols. (5) and (9).  Col. (11): spectral type from the BeSS database \citep{nei11}.  Col. (12): multiplicity; estimated number of physical components (\S~\ref{bindet}).}
\end{deluxetable}
\end{longrotatetable}

\section{Observations and Data Reduction} \label{obs}

\subsection{Observation Sequence and Data Reduction} \label{seqred}

The observation sequence of program and calibrator stars was identical to that discussed in Paper I.  Basic data on each of the calibrator stars are listed in Table~\ref{calibs}.  However, the process used to estimate the adopted limb darkened angular diameters ($\theta_{LD}$) for most of the calibrator stars (Table~\ref{calibs}, Column~8) from the V magnitude and V - K color index was modified to include a correction for interstellar extinction $A_V$ by a process described in \citet{sch16}.  The value of $A_V$ was subtracted from the V magnitude of the calibrator ($m_V$, Table~\ref{calibs}, Column~5) and the V - K color index was recomputed assuming that the correction in K was negligible.  The angular diameters for four of the calibrator stars (denoted with ``b'' in Table~\ref{calibs}) were estimated from the infrared flux method \citep{blk91}.  

%% Table 3 -- Calibrator Stars
\floattable
\tabletypesize{\small}
\startlongtable
\begin{deluxetable}{lrrrrrll}
\tablecaption{Calibrator Stars \label{calibs}}
\tablewidth{0pt}
\tablehead{
\colhead{} & \colhead{} & \colhead{} & \colhead{} & \colhead{} & \colhead{} & \colhead{} & \colhead{Adopted} \\
\colhead{} & \colhead{} & \colhead{} & \colhead{Other} & \colhead{$m_V$} & \colhead{${\it V - K}$} & \colhead{Spectral} &
\colhead{$\theta_{LD}$} \\
\colhead{HIP} & \colhead{HR} & \colhead{FK5} & \colhead{Name} & \colhead{(mag)} & \colhead{(mag)} & \colhead{Type} &
\colhead{(mas)} \\
\colhead{(1)} & \colhead{(2)} & \colhead{(3)} & \colhead{(4)} & \colhead{(5)} & \colhead{(6)} & \colhead{(7)} & \colhead{(8)} \\
}
\startdata
%% HIP       HR          FK5            Name                 V           V-K        Sp Type	      Adop. Diam(LD)
  1067  &     39    &	   7	&	$\gamma$ Peg	    &	2.83	&	-0.940	&	B2IV		&	0.498		\\
  2920	&    153	&	  17	&	$\zeta$ Cas		    &	3.66	&	-0.587	&	B2IV		&	0.424		\\
  8886	&    542	&	  63	&	$\epsilon$ Cas	    &	3.38	&	-0.583	&	B3III		&	0.486$^a$	\\
 15110	&    972	&	1089	&	$\zeta$ Ari		    &	4.89	&	 0.065	&	A1V			&	0.361		\\
 17776	&	1172	&	\nodata	&	HD 23753		    &	5.45	&	-0.110	&	B8V			&	0.248		\\
 17851	&	1180	&	\nodata	&	28 Tau			    &	5.09	&	 0.153	&	B8Vpe		&	0.350		\\
 17959	&	1148	&	 138	&	$\gamma$ Cam	    &	4.63	&	 0.272	&	A2IVn		&	0.475		\\
 18532	&	1220	&	 147	&	$\epsilon$ Per	    &	2.89	&	-0.823	&	B0.5V+A2V	&	0.519		\\
 20635	&	1387	&	\nodata	&	$\kappa$$^1$ Tau    &	4.22	&	 0.143	&	A7IV-V		&	0.531		\\
 23497	&	1620	&	 184	&	$\iota$ Tau		    &	4.64	&	 0.395	&	A7V			&	0.513		\\
 25336	&	1790	&	 201	&	$\gamma$ Ori		&	1.64	&	-0.735	&	B2III		&	0.958		\\
 31216	&	2385	&	1174	&	13 Mon			    &	4.50	&	 0.199	&	A0Ib		&	0.460		\\
 33018	&	2540	&	 261	&	$\theta$ Gem		&	3.60	&	 0.437	&	A3III		&	0.797$^b$	\\
 35350	&	2763	&	 277	&	$\lambda$ Gem	    &	3.58	&	 0.040	&	A3V			&	0.672		\\
 55434	&	4386	&	 427	&	$\sigma$ Leo		&	4.05	&	-0.089	&	B9.5Vs		&	0.491		\\
 59774	&	4660	&	 456	&	$\delta$ UMa		&	3.31	&	 0.206	&	A3V			&	0.843		\\
 65477	&	5062	&	\nodata	&	80 UMa			    &	4.01	&	 0.865	&	A5V			&	0.674		\\ 
 72220	&	5511	&	 547	&	109 Vir			    &	3.72	&	 0.074	&	A0V			&	0.640		\\
 77622	&	5892	&	 588	&	$\epsilon$ Ser		&	3.71	&	 0.285	&	A2Vm		&	0.735		\\
 78554	&	5972	&	3268	&	$\pi$ Ser			&	4.83	&	 0.211	&	A3V			&	0.417		\\
 79375	&	6031	&	3280	&	$\psi$ Sco			&	4.94	&	 0.308	&	A3IV		&	0.417		\\
 80170	&	6095	&	 609	&	$\gamma$ Her		&	3.75	&	 0.806	&	A9III		&	0.944$^b$	\\
 81377	&	6175	&	 622	&	$\zeta$ Oph		    &	2.56	&	-0.124	&	O9.5Vn		&	0.966$^a$	\\
 84379	&	6410	&	 641	&	$\delta$ Her		&	3.14	&	 0.332	&	A3IV		&	0.987		\\
 87108	&	6629	&	 668	&	$\gamma$ Oph		&	3.75	&	 0.128	&	A0Vnp		&	0.650		\\
 88771	&	6771	&	 680	&	72 Oph			    &	3.73	&	 0.318	&	A4IVs		&	0.732$^b$	\\
 93194	&	7178	&	 713	&	$\gamma$ Lyr		&	3.24	&	 0.118	&	B9III		&	0.817		\\
 93747	&	7235	&	 716	&	$\zeta$ Aql		    &	2.99	&	 0.114	&	A0Vn		&	0.919		\\
 95168	&	7340	&	\nodata	&	$\rho$$^1$ Sgr		&	3.93	&	 0.521	&	F0IV-V		&	0.772		\\
 95853	&	7420	&	 733	&	$\iota$$^2$ Cyg	    &	3.79	&	 0.192	&	A5Vn		&	0.668		\\
102724  &   7977    &	\nodata	&	55 Cyg			    &	4.84	&	 0.894	&	B3Ia		&	0.638		\\
105102	&	8143	&	1558	&	$\sigma$ Cyg		&	4.23	&	 0.547	&	B9Iab		&	0.599		\\
109410	&	8454	&	 835	&	$\pi$$^2$ Peg		&	4.29	&	 1.173	&	F5III		&	0.945$^b$	\\
109427	&	8450	&	 834	&	$\theta$ Peg		&	3.53	&	 0.153	&	A2Vp		&	0.733		\\
110609	&	8541	&	\nodata	&	4 Lac			    &	4.57	&	 0.287	&	B9Iab		&	0.462		\\
111169	&	8585	&	 848	&	$\alpha$ Lac		&	3.77	&	-0.081	&	A1V			&	0.568		\\
111497	&	8597	&	 850	&	$\eta$ Aqr			&	4.02	&	-0.216	&	B9IV-Vn		&	0.462		\\
112029	&	8634	&	 855	&	$\zeta$ Peg		    &	3.40	&	-0.166	&	B8V			&	0.635		\\
\enddata
\tablecomments{Cols. (1) through (4) per Table~\ref{targetlist}.  Cols. (5) and (7): $m_V$ magnitude (uncorrected 
for extinction) and spectral type from $\mathit{The\ Bright\ Star\ Catalogue,}$ 5th Rev. Ed. \citep{hof91}.  
Col. (6): ${\it V - K}$ color index calculated from Col. (5) value and ${\it K}$ magnitude from the 
$\mathit{2MASS}$ catalog \citep{skr06}.  Col. (8): Adopted limb darkened angular diameter (\S~\ref{seqred}).}
\tablenotetext{a}{also program star; see Table~\ref{targetlist}}
\tablenotetext{b}{\citet{blk91}}
\end{deluxetable}

Two calibrator stars are also program stars ($\epsilon$ Cas; \S~\ref{epscas} and $\zeta$ Oph; \S~\ref{zetoph}, denoted with ``a'' in Table~\ref{calibs}).  Working with archival data, we occasionally encountered instances where some of our program stars were designated, or had to be used, as calibrators for observations of other stars in our sample (on nights of otherwise plentiful, good quality data), there being no alternative calibrators observed within reasonable angular distance (in this case, within 25$\arcdeg$ -- 60$\arcdeg$).  Such was the case on four of the five nights of observations of 48 Lib, for all four nights of observations of $\chi$~Oph, and on one night of 66 Oph observations (Table~\ref{progstarsobs}), where $\zeta$ Oph (FK5 622) was the only available calibrator, and on one night for $\beta$ Cep where $\epsilon$ Cas (FK5 63) was one of two calibrators used.  However, both $\epsilon$~Cas and $\zeta$~Oph were also observed independently, as program stars, on other nights using other calibrators, and in both cases were found to be unresolved (see \S~\ref{modeling} for further discussion).  Based on this analysis, we concluded that $\epsilon$~Cas and $\zeta$~Oph can be both used as calibrators in observations of other program stars. In all other respects, the calibration of the NPOI and Mark III visibility data was performed in a manner identical to that reported in Paper I and references therein.

\subsection{Observation Logs} \label{obslog}

Observations of the program stars, with the exception of $\delta$ Sco (\S~\ref{delsco}), which were reported by \citet{taz11} and \citet{cmt12}, were obtained at the NPOI and Mark III over 150 nights from 1989 September 10 through 2018  December 12 UT, and the observation logs of the program stars are listed in Table~\ref{progstarsobs} (in the same order as in Table~\ref{targetlist}).  In total, 1857 coherent, multi-baseline observations were obtained for 30 of the 31 program stars, an average of 12 per night, an excellent cadence for an optical interferometer.  Program stars were, on average, each observed on six nights, with all the stars but three (15 Mon, 53 Boo, and 59 Cyg) having at least three nights of observations.  Table~\ref{progstarsobs} lists the UT dates of the observations, the number of multi-baseline observations on each night, the baselines used, the total number of $V^2$ and (when available) closure phase measurements on each night, and the calibrator star observed.  For the program stars, a total of 68,238 $V^2$ measurements and 13,709 closure phase measurements were obtained over all spectral channels and spectrometer outputs.

%% Table 4 -- Observation Log

%%\floattable
%%\tabletypesize{\small}
%%\startlongtable
\begin{longrotatetable}
\begin{deluxetable*}{llcclccl}
\tablecaption{Program Stars - Observation Log \label{progstarsobs}}
%%\tablewidth{800pt}
%%\tabletypesize{scriptsize}
\tablehead{
\colhead{Object} & \colhead{Other} & \colhead{UT Date} & \colhead{No. of} & \colhead{Baselines} & \colhead{No. of} & \colhead{No. of} & \colhead{Calibrator} \\
\colhead{FK5/HR} & \colhead{Name} & \colhead{} & \colhead{Observations} & \colhead{} & \colhead{Visibilities} & \colhead{Closure} & \colhead{FK5/HR} \\
\colhead{} & \colhead{} & \colhead{} & \colhead{} & \colhead{} & \colhead{} & \colhead{Phases} & \colhead{} \\
\colhead{(1)} & \colhead{(2)} & \colhead{(3)} & \colhead{(4)} & \colhead{(5)} & \colhead{(6)} & \colhead{ (7)} & \colhead{(8)} \\
}
\startdata
%%FK5/HR         Name                Obs Date     #obs       Baselines                                  # V^2       # Cl. Ph.       Calib. 
FK5 25	    &	o Cas		    &	2018 Sep 27	&	25	&	AC-AE, AC-AW							&	 654    &   \nodata     &   FK5 17      \\
		    &				    &	2018 Sep 28	&	23	&	AC-AE, AC-AW							&	 652	&	\nodata     &   FK5 17	    \\
		    &				    &	2018 Sep 29	&	25	&	AC-AE, AC-AW							&	 720	&	\nodata     &   FK5 17	    \\
		    &				    &	2018 Nov 15	&	 9	&	AC-AE, AC-AW							&	 258	&	\nodata     &   FK5 17	    \\
		    &				    &	2018 Nov 16	&	15	&	AC-AE, AC-AW							&	 419	&	\nodata     &   FK5 17	    \\
		    &				    &	2018 Nov 21	&	14	&	AC-AE, AC-AW							&	 391	&	\nodata     &   FK5 17	    \\
		    &				    &	2018 Nov 27	&	16	&	AC-AE, AC-AW							&	 438	&	\nodata     &   FK5 17	    \\
		    &				    &	2018 Dec 04	&	10	&	AC-AE, AC-AW							&	 283	&	\nodata     &   FK5 17	    \\
		    &				    &	2018 Dec 12	&	15	&	AC-AE, AC-AW							&	 394	&	\nodata     &   FK5 17	    \\
		    &				    &				&		&										    &			&               &               \\
FK5 32	    &	$\gamma$ Cas	&	2004 Dec 02	&	 9	&	AE-AW, AE-W07					    	&	 261	&	\nodata     &   FK5 17	    \\
		    &				    &	2004 Dec 03	&	10	&	AE-AW, AE-W07						    &	 285	&	\nodata     &   FK5 17	    \\
		    &				    &	2004 Dec 11	&	11	&	AE-AW, AE-W07						    &	 319	&	\nodata     &   FK5 17	    \\
		    &				    &	2004 Dec 19	&	 8	&	AE-AW, AE-W07						    &	 232	&	\nodata     &   FK5 17	    \\
		    &				    &	2004 Dec 23	&	12	&	AC-AE, AE-AW							&	 359	&	\nodata     &   FK5 17	    \\ 		
		    &				    &				&		&					    					&			&               &               \\
FK5 57	    &	$\phi$ Per		&	2004 Nov 03	&	 4	&	AE-AW, AE-W07		    				&	  86	&	\nodata     &   FK5 17	    \\
		    &				    &	2004 Nov 04	&	 9	&	AE-AW, AE-W07			    			&	 260	&	\nodata     &   FK5 17	    \\
		    &				    &	2004 Nov 05	&	 8	&	AE-AW, AE-W07				    		&	 215	&	\nodata     &   FK5 17	    \\
		    &				    &	2004 Dec 01	&	 8	&	AE-AW, AE-W07						    &	 232	&	\nodata     &   FK5 17	    \\
		    &				    &	2004 Dec 02	&	14	&	AE-AW, AE-W07					    	&	 403	&	\nodata     &   FK5 17	    \\
		    &				    &	2004 Dec 23	&	10	&	AC-AE, AE-AW							&	 271	&	\nodata     &   FK5 17	    \\
		    &				    &				&		&								    		&			&               &               \\
FK5 63	    &	$\epsilon$ Cas	&	2004 Nov 04	&	 6	&	AE-AW, AE-W07			    			&	 181	&	\nodata     &   FK5 17	    \\
		    &				    &	2004 Nov 05	&	 7	&	AE-AW, AE-W07		    				&	 203	&	\nodata     &   FK5 17	    \\
		    &				    &	2004 Dec 02	&	 9	&	AE-AW, AE-W07					    	&	 276	&	\nodata     &   FK5 17	    \\
		    &				    &	2004 Dec 03	&	10	&	AE-AW, AE-W07				    		&	 303	&	\nodata     &   FK5 17	    \\
		    &				    &	2004 Dec 11	&	11	&	AE-AW, AE-W07			    			&	 340	&	\nodata     &   FK5 17	    \\
		    &				    &	2004 Dec 19	&	 9	&	AE-AW, AE-W07					    	&	 269	&	\nodata     &   FK5 17	    \\
		    &				    &	2004 Dec 23	&	12	&	AE-AW, AE-W07				    		&	 384	&	\nodata     &   FK5 17	    \\
		    &				    &				&		&	    									&			&               &               \\
HR 985	    &	BK Cam		    &	2006 Nov 08	&	 6	&	AC-AE, AC-AW							&	 185	&	\nodata     &   FK5 147	    \\
		    &				    &	2006 Nov 11	&	 3	&	AC-AE, AC-AW							&	  89	&	\nodata     &   FK5 147	    \\
		    &				    &	2009 Nov 25	&	12	&	AC-AE, AC-AW							&	 369	&	\nodata     &   FK5 138	    \\
		    &				    &	2009 Dec 01	&	18	&	AC-AE, AC-AW							&	 558	&	\nodata     &   FK5 138	    \\
		    &				    &	2009 Dec 03	&	13	&	AC-AE, AC-AW							&	 385	&	\nodata     &   FK5 138	    \\
		    &				    &	2010 Oct 09	&	15	&	AC-AE, AC-AW							&	 446	&	\nodata     &   FK5 138	    \\
		    &				    &	2010 Oct 12	&	13	&	AC-AE, AC-AW							&	 375	&	\nodata     &   FK5 138	    \\
		    &				    &				&		&   										&			&               &               \\
HR 1087	    &	$\psi$ Per		&	2006 Nov 17	&	10	&	AC-AE, AC-AW, AC-W07, AE-W07, AW-W07	&	 510	&	    215     &   FK5 147     \\
		    &				    &	2017 Oct 27	&	25	&	AC-AE, AC-AW, AC-E06, AW-E06			&	1371	&	    336     &   FK5 147	    \\
		    &				    &	2017 Nov 10	&	34	&	AC-AE, AC-AW, AC-E06, AW-E06			&	1802	&	    442     &   FK5 147	    \\		
		    &				    &	2017 Nov 22	&	36	&	AC-AE, AC-AW, AC-E06, AW-E06			&	1826	&	    440     &   FK5 147	    \\		
		    &				    &	2017 Nov 23	&	34	&	AC-AE, AC-AW, AC-E06, AW-E06			&	1708	&	    341     &   FK5 147	    \\		
		    &				    &				&		&										    &			&               &               \\
FK5 136	    &	17 Tau		    &	2012 Dec 06	&	 4	&	AC-AW								    &	  59	&	\nodata     &   FK5 1089	\\
		    &				    &	2012 Dec 08	&	 3	&	AC-AE, AC-AW							&	  85	&	\nodata     &   FK5 1089	\\
		    &				    &	2012 Dec 09	&	 4	&	AC-AE, AC-AW							&	  97	&	\nodata     &   FK5 1089	\\
		    &				    &	2012 Dec 10	&	 2	&	AC-AW								    &	  30	&	\nodata     &   FK5 1089	\\
		    &				    &				&		&										    &		    &               &               \\
HR 1156	    &	23 Tau		    &	2013 Mar 13	&	10	&	AC-AE, AC-AW							&	 245	&	\nodata     &   HR 1387	    \\
		    &				    &	2013 Mar 14	&	 8	&	AC-AE, AC-AW							&	 213	&	\nodata     &   HR 1387	    \\
		    &				    &	2013 Mar 18	&	 3	&	AC-AE, AC-AW							&	  70	&	\nodata     &   HR 1387	    \\
		    &				    &	2013 Mar 20	&	 6	&	AC-AE, AC-AW							&	 170	&	\nodata     &   HR 1387	    \\
		    &				    &	2013 Mar 22	&	 4	&	AC-AE, AC-AW							&	  94	&	\nodata     &   HR 1387	    \\
		    &				    &				&		&										    &			&               &               \\
FK5 139	    &	$\eta$ Tau		&	2002 Dec 14	&	 3	&	AC-AW, AE-AW, AE-AN, AN-AW			    &	 138	&	     15     &   HR 1180	    \\
		    &				    &	2012 Oct 09	&	17	&	AC-AE, AC-AW							&	 510	&	\nodata     &   FK5 147	    \\
		    &				    &	2013 Feb 05	&	14	&	AC-AE, AC-AW							&	 333	&	\nodata     &   HR 1172	    \\
		    &				    &	2013 Feb 17	&	11	&	AC-AE, AC-AW							&	 319	&	\nodata     &   FK5 147	    \\
		    &				    &	2013 Feb 18	&	15	&	AC-AE								    &	 225	&	\nodata     &   FK5 147	    \\
		    &				    &				&		&										    &		    &               &               \\		
FK5 152	    &	48 Per		    &	2006 Nov 08	&	 7	&	AC-AE, AC-AW							&	 183	&	\nodata     &   FK5 147	    \\
		    &				    &	2006 Nov 09	&	 8	&	AC-AE, AC-AW, AC-W07, AE-W07, AW-W07	&	 379	&	    131     &   FK5 147	    \\
		    &				    &	2006 Nov 16	&	 7	&	AC-AE, AC-AW, AC-W07, AE-W07, AW-W07	&	 306	&	    107     &   FK5 147	    \\
		    &				    &	2006 Nov 17	&	12	&	AC-AE, AC-AW, AC-W07, AE-W07, AW-W07	&	 565	&	    235     &   FK5 147	    \\
		    &				    &	2006 Nov 20	&	 7	&	AC-AE, AC-AW, AC-W07, AE-W07, AW-W07	&	 456	&	    141     &   FK5 147	    \\
		    &				    &				&		&										    &			&               &               \\
FK5 211	    &	$\zeta$ Tau	    &	2006 Dec 05	&	16	&	AE-AW, AE-E06, AW-E06, AW-W07, E06-W07	&	 834	&	    473     &   FK5 184	    \\
		    &				    &	2006 Dec 06	&	24	&	AE-AW, AW-W07						    &	 609	&	\nodata     &   FK5 184	    \\
		    &				    &	2006 Dec 07	&	19	&	AE-AW, AE-E06, AW-E06, AW-W07, E06-W07	&	 825	&	    368     &   FK5 184	    \\
		    &				    &	2006 Dec 08	&	23	&	AE-AW, AW-W07						    &	 643	&	\nodata     &   FK5 184	    \\
		    &				    &	2007 Feb 06	&	31	&	AC-AE, AC-AW, AC-W07, AE-W07, AW-W07	&	2042	&	   1281     &   FK5 201     \\
		    &				    &				&		&										    &			&               &               \\
HR 1934	    &	$\omega$ Ori	&	2012 Jan 07	&	 5	&	AC-AE, AE-E06							&	 148	&	\nodata     &   FK5 1174	\\
		    &				    &	2017 Nov 22	&	 6	&	AC-AE, AC-AW, AC-E06, AW-E06 			&	 300	&	     54     &   FK5 201	    \\
		    &				    &	2017 Nov 23	&	 4	&	AC-AE, AC-AW, AC-E06, AW-E06			&	 196	&	     35     &   FK5 201	    \\
		    &				    &	2017 Dec 02	&	 1	&	AC-AE, AC-AW, AC-E06, AW-E06			&	  38	&	\nodata     &   FK5 201	    \\
		    &				    &	2017 Dec 03	&	 3	&	AC-AE, AC-AW, AC-E06, AW-E06			&	 169	&	     40     &   FK5 201	    \\
		    &				    &				&		&										    &			&               &			    \\
HR 2456	    &	15 Mon		    &	2002 Dec 13	&	 1	&	AE-AW, AN-AW, AW-E02					&	  48	&	\nodata     &   FK5 261	    \\
		    &				    &	2002 Dec 14	&	 1	&	AE-AN, AE-E02, AN-E02, AW-E02			&	  76	&	\nodata     &   FK5 261	    \\
		    &				    &				&		&										    &			&               &			    \\
FK5 285	    &	$\beta$ CMi	    &	2015 Mar 28	&	22	&	AC-AE, AC-AW, AC-E03, AW-E03			&	1232	&       286     &   FK5 277	    \\
		    &				    &	2015 Mar 30	&	15	&	AC-AE, AC-AW, AC-E03, AW-E03			&	 854	&	    195     &   FK5 277	    \\
		    &				    &	2015 Mar 31	&	13	&	AC-AE, AC-AW, AC-E03, AW-E03			&	 715	&	    169     &   FK5 277	    \\
		    &				    &	2015 Apr 04	&	 8	&	AC-AE, AC-AW							&	 226	&	\nodata     &   FK5 277	    \\
		    &				    &	2015 Apr 05	&	17	&	AC-AE, AC-AW							&	 474	&	\nodata     &   FK5 277	    \\
		    &				    &				&		&										    &			&               &               \\
FK5 1292	&	$\phi$ Leo		&	2013 Jan 17	&	11	&	AC-AE, AC-AW							&	 317	&	\nodata     &   FK5 427	    \\
		    &				    &	2013 Jan 19	&	12	&	AC-AE, AC-AW							&	 344	&	\nodata     &   FK5 427	    \\
		    &				    &	2013 Jan 20	&	12	&	AC-AE, AC-AW							&	 329	&	\nodata     &   FK5 427	    \\
		    &				    &	2013 Jan 21	&	11	&	AC-AE, AC-AW							&	 268	&	\nodata     &   FK5 427	    \\
		    &				    &	2018 Mar 20	&	14	&	AC-AE, AC-AW							&	 372	&	\nodata     &   FK5 427	    \\
		    &				    &				&		&										    &			&               &               \\
FK5 472	    &	$\kappa$ Dra	&	2002 Jun 19	&	 3	&	AC-AW, AC-E02, AE-AN, AE-AW, AE-E02,	&	 205	&	    101     &   HR 5062	    \\
		    &				    &				&		&	AN-AW, AN-E02, AW-E02					&			&               &               \\
		    &				    &	2006 Mar 16	&	19	&	AC-AE, AC-AN, AC-AW, AE-AN				&	1139	&	    266     &   FK5 456	    \\
		    &				    &	2011 Jan 11	&	 8	&	AC-AE, AC-E06						    &	 218	&	\nodata     &   FK5 456	    \\
		    &				    &	2011 Jan 14	&	 3	&	AC-AE, AC-E06		    				&	  80	&	\nodata     &   FK5 456	    \\
		    &				    &	2011 Jan 15	&	 6	&	AC-AE, AC-E06			    			&	 174	&	\nodata     &   FK5 456	    \\
		    &				    &				&		&								    		&			&               &               \\
HR 5774	    &	53 Boo		    &	2013 Mar 12	&	11	&	AC-AE, AC-AW, AC-E06, AW-E06			&	 599	&       109     &   FKV 3268	\\
		    &				    &				&		&									    	&			&               &               \\
FK5 1417	&	48 Lib		    &	2007 Jun 30	&	 8	&	AE-W07, E06-W07						    &	 229	&	\nodata     &   FK5 3280	\\
		    &				    &	2016 May 28	&	14	&	AC-AE, AC-AW							&	 377	&	\nodata     &   FK5 622	    \\
		    &				    &	2016 May 29	&	12	&	AC-AE, AC-AW							&	 313	&	\nodata     &   FK5 622	    \\
		    &				    &	2016 May 30	&	15	&	AC-AE, AC-AW							&	 420	&	\nodata     &   FK5 622	    \\
		    &				    &	2016 Jun 04	&	12	&	AC-AE, AC-AW							&	 308	&	\nodata     &   FK5 622	    \\
		    &				    &				&		&									    	&			&               &               \\
FK5 594     &   $\delta$ Sco$^a$    &	\nodata &   \nodata &   \nodata                             &   \nodata &   \nodata     &   \nodata     \\
		    &				    &				&		&									    	&			&               &               \\
HR 6118	    &	$\chi$ Oph	    &	2006 Jun 11	&	 5	&	AC-AE, AC-AW, AC-W07, AE-W07, AW-W07	&	 427	&	    167     &   FK5 622	    \\
		    &				    &	2006 Jun 13	&	 4	&	AC-AE, AC-AW, AC-W07, AE-W07, AW-W07	&	 332	&       144     &   FK5 622	    \\
		    &				    &	2006 Jun 17	&	 6	&	AC-AE, AC-AW, AC-W07, AE-W07, AW-W07	&	 480	&	    288     &   FK5 622	    \\		
		    &				    &	2006 Jun 18	&	 7	&	AC-AE, AC-AW, AC-W07, AE-W07, AW-W07	&	 554	&	    313     &   FK5 622	    \\
		    &				    &				&		&										    &			&               &			    \\
FK5 622	    &	$\zeta$ Oph	    &	2007 May 08	&	11	&	AC-AE, AE-E06							&	 329	&	\nodata     &   FK5 588	    \\
		    &				    &	2007 May 09	&	11	&	AC-AE, AC-AN, AE-AN					    &	 402	&	     84     &   FK5 588	    \\
		    &				    &	2007 May 11	&	 9	&	AC-AE, AC-AN, AE-AN, AE-E06, AN-E06		&	 783	&	    378     &   FK5 588	    \\		
		    &				    &	2007 May 15	&	 9	&	AC-AE, AC-AN, AE-AN, AE-E06, AN-E06		&	 613	&	    315     &   FK5 588	    \\
		    &				    &	2011 Apr 17	&	 8	&	AC-AE, AC-AW, AC-E06, AW-E06			&	 469	&	    119     &   FK5 547	    \\
		    &				    &				&		&										    &			&               &               \\
HR 6712	    &	66 Oph		    &	2006 May 18	&	10	&	AC-AE, AC-AN, AE-AN, AE-AW, AN-AW		&	 911	&       318     &   FK5 668	    \\
		    &				    &	2006 May 25	&	 3	&	AC-AE, AC-AW, AC-W07, AE-W07, AW-W07	&	 216	&	    109     &   FK5 668	    \\
		    &				    &	2007 Jun 13	&	 5	&	AN-W07, E06-W07						    &	 139	&	\nodata     &   FK5 668	    \\		
		    &				    &	2007 Jun 16	&	 3	&	AN-W07, E06-W07					    	&	  90	&	\nodata     &   FK5 668	    \\	
		    &				    &	2007 Jun 17	&	10	&	AN-W07, E06-W07						    &	 307	&	\nodata     &   FK5 668	    \\
		    &				    &	2007 Jun 18	&	12	&	AN-W07, E06-W07					    	&	 368	&	\nodata     &   FK5 668	    \\
		    &				    &  	2007 Jun 20	&	 4	&	AN-W07, E06-W07						    &	 106	&	\nodata     &   FK5 668	    \\
		    &				    &	2007 Jun 21	& 	 6	&	AN-W07, E06-W07			    			&	 180	&	\nodata     &   FK5 668	    \\		
		    &				    &	2007 Jun 22	&	 6	&	AN-W07, E06-W07				    		&	 184	&	\nodata     &   FK5 668	    \\
		    &                   &   2007 Jun 24 &    5  &   AN-W07, E06-W07                         &    137    &   \nodata     &   FK5 668     \\
		    &				    &	2007 Jun 26	&	 3	&	AN-W07, E06-W07						    &	  90	&	\nodata     &   FK5 668	    \\		
		    &				    &	2011 May 25	&	15	&	AC-AE						    		&	 221	&	\nodata     &   FK5 622	    \\
		    &				    &				&		&									    	&			&               &               \\
FK5 681	    &	o Her		    &	2004 Jun 23	&	 7	&	AC-AE, AC-AW, AC-W07, AW-W07		    &	 425	&	    192     &   FK5 680	    \\
		    &				    &	2004 Jul 01	&	10	&	AC-AE, AC-AW, AC-W07, AE-W07, AW-W07	&	 833	&	    478     &   FK5 680	    \\
		    &				    &	2004 Jul 05	&	12	&	AC-AE, AC-AW, AC-W07, AE-W07, AW-W07	&	1090	&	    693     &   FK5 680	    \\		
		    &				    &	2004 Jul 06	&	 6	&	AC-AE, AC-AW, AC-W07, AE-W07, AW-W07	&	 491	&	    276     &   FK5 680	    \\
		    &				    &				&		&										    &			&               &               \\
FK5 705	    &	$\beta$ Lyr	    &	2006 Sep 17	&	 7	&	AC-AE, AC-W07, AE-W07					&	 390	&       127     &   FK5 713	    \\
		    &				    &	2006 Sep 18	&	20	&	AC-AE, AC-W07, AE-W07					&	1196	&	    301     &   FK5 713	    \\
		    &				    &	2007 Jun 29	&	 6	&	AE-W07, AE-AN, AN-W07, E06-W07			&	 306	&        60     &   FK5 713	    \\		
		    &				    &	2007 Jul 02	&	13	&	AE-W07, AE-AN, AN-W07, E06-W07			&	 663	&	    126     &   FK5 713	    \\	
		    &				    &	2007 Jul 03	&	 5	&	AE-W07, AE-AN, AN-W07, E06-W07			&	 193	&	     26     &   FK5 713	    \\
		    &				    &	2007 Jul 04	&	13	&	AE-W07, AE-AN, AN-W07, E06-W07			&	 652	&	    140     &   FK5 713	    \\
		    &				    &	2007 Jul 12	&	 6	&	AE-W07, AE-AN, AN-W07, E06-W07			&	 246	&	     30     &   FK5 713	    \\
		    &				    &				&		&										    &			&               &               \\
FK5 727	    &	$\upsilon$ Sgr	&	2011 Jul 01	&	 8	&	AE-AW, AE-E06, AW-E06, AW-W07, E06-W07	&	 432	&       182     &   HR 7340	    \\
		    &				    &	2011 Jul 02	&	 5	&	AE-AW, AE-E06, AW-E06, AW-W07, E06-W07	&	 363	&	    162     &   HR 7340	    \\
		    &				    &	2011 Jul 13	&	 6	&	AE-AW, AE-E06, AW-E06, AW-W07, E06-W07	&	 378	&	    193     &   HR 7340	    \\		
		    &				    &				&		&										    &			&               &			    \\
FK5 1551    &	59 Cyg		    &	2009 Jul 07	&	 6	&	AC-AE, AC-AN, AE-AN, AE-AW, AN-AW		&	 556	&	    272     &   FK5 1558	\\
		    &				    &	2009 Jul 22	&	14	&	AC-AE, AC-AN, AE-AN, AE-AW, AN-AW		&	1259	&	    798     &   FK5 1558	\\
		    &				    &				&		&										    &			&               &			    \\
FK5 1559	&	$\upsilon$ Cyg	&	2013 Sep 21	&	22	&	AC-AE, AE-E06							&	 625	&	\nodata     &   HR 7977	    \\
		    &			    	&	2013 Sep 22	&	21	&	AC-AE, AE-E06							&	 630	&	\nodata     &   HR 7977	    \\
		    &				    &	2013 Sep 29	&	22	&	AC-AE, AE-E06							&	 613	&	\nodata     &   HR 7977	    \\		
		    &				    &	2013 Oct 02	&	24	&	AC-AE, AE-E06							&	 712	&	\nodata     &   HR 7977	    \\
		    &				    &	2013 Oct 03	&	20	&	AC-AE, AE-E06							&	 595	&	\nodata     &   HR 7977	    \\
		    &				    &				&		&						    				&			&               &			    \\
FK5 809	    &	$\beta$ Cep	    &	1989 Sep 10	&	 6	&	NVF-SVC$^b$				    			&	  18	&	\nodata     &   FK5 848	    \\
		    &				    &	1991 Aug 07	&	24	&	NVF-SVC$^c$					    		&	  72	&	\nodata     &   FK5 848	    \\
		    &				    &	1991 Sep 13	&	 7	&	NVD-SVC$^d$						    	&	  21	&	\nodata     &   FK5 848	    \\		
		    &				    &	1991 Sep 14	&	10	&	NVD-SVC$^d$							    &	  30	&	\nodata     &   FK5 848	    \\
		    &				    &	1991 Sep 15	&	19	&	NVD-SVC$^d$		    					&	  57	&	\nodata     &   FK5 848	    \\		
		    &				    &	1991 Sep 30	&	18	&	NAS-SAS$^e$			    				&	  54	&	\nodata     &   FK5 848	    \\
		    &				    &	1991 Dec 06	&	13	&	NVF-SVC$^c$				    			&	  39	&	\nodata     &   FK5 848	    \\		
		    &				    &	1991 Dec 07	&	13	&	NVD-SVC$^d$					    		&	  39	&	\nodata     &   FK5 848	    \\ 
		    &				    &	1992 Jul 06	&	25	&	NAS-SAS$^e$						    	&	  75	&	\nodata     &   FK5 848	    \\
		    &				    &	1992 Aug 06	&	14	&	NVD-SVC$^d$							    &	  42	&	\nodata     &   FK5 848	    \\ 
		    &				    &	1997 Jul 03	&	 2	&	AC-AE, AC-AW, AE-AW$^f$ 				&	 190	&	     62     &   FK5 713,	\\
		    &				    &				&		&							    			&			&               &	FK5 835,	\\
		    &				    &				&		&								    		&			&               &	FK5 848	    \\
		    &				    &	1997 Jul 16	&	 1	&	AC-AE, AC-AW, AE-AW$^f$			    	&	  95	&	     31     &   FK5 641,	\\
		    &				    &				&		&										    &			&               &	FK5 733,	\\
		    &				    &				&		&										    &			&               &	FK5 835,	\\
		    &				    &				&		&										    &			&               &	FK5 848	    \\
		    &				    &	1997 Jul 18	&	 2	&	AC-AE, AC-AW, AE-AW$^f$	    			&	 190	&	     62     &   FK5 733,	\\
		    &				    &				&		&								    		&			&               &	FK5 835,	\\
		    &				    &				&		&									    	&			&               &	FK5 848	    \\		
		    &				    &	1997 Sep 30	&	 1	&	AC-AE, AC-AW, AE-AW$^f$				    &	  95	&	     31     &   FK5 7,	    \\
		    &				    &				&		&									    	&			&               &   FK5 63	    \\
		    &				    &	1998 Jun 30	&	 1	&	AC-AE, AC-AW, AE-AW$^f$				    &	  95	&	     31     &   FK5 716,	\\
		    &				    &				&		&									    	&			&               &   FK5 834,	\\
		    &				    &				&		&										    &			&               &   FK5 848	    \\		
		    &				    &	1998 Oct 03	&	 3	&	AC-AE, AC-AW, AE-AW$^f$				    &	 285	&	     93     &   FK5 641,	\\
		    &				    &				&		&										    &			&               &   FK5 848	    \\
		    &				    &	2001 Jun 20	&	 3	&	AC-AE, AC-AW, AE-AW$^f$				    &	 285	&	     93     &   HR 8541,	\\
		    &				    &				&		&										    &			&               &	FK5 609,	\\
		    &				    &				&		&										    &			&               &	FK5 668,    \\
		    &				    &				&		&										    &			&               &	FK5 848	    \\		
		    &				    &	2002 Jun 18	&	 2	&	AC-AW, AC-E02, AE-AN, AE-AW,			&	 232	&	    123     &   FK5 713	    \\ 
		    &				    &				&		&	AE-E02, AN-AW, AN-E02, AW-E02			&			&               &               \\
		    &				    &	2002 Jun 19	&	 5	&	AC-AW, AC-E02, AE-AN, AE-AW,			&	 320	&	    225     &   FK5 713	    \\ 
		    &				    &				&		&	AE-E02, AN-AW, AN-E02, AW-E02   		&			&               &			    \\
		    &				    &	2002 Jun 20	&	 2	&	AC-AW, AC-E02, AE-AN, AE-AW,			&	 148	&	     88     &   FK5 713	    \\ 
		    &				    &				&		&	AE-E02, AN-AW, AN-E02, AW-E02			&			&               &			    \\
		    &				    &	2002 Jun 22	&	 5	&	AC-AW, AC-E02, AE-AN, AE-AW,			&	 371	&	     96     &   FK5 713	    \\ 
		    &				    &				&		&	AE-E02, AN-AW, AN-E02, AW-E02   		&			&               &			    \\
		    &				    &	2002 Jun 23	&	 5	&	AC-AW, AC-E02, AE-AN, AE-AW,			&	 304	&	     93     &   FK5 713	    \\ 
		    &				    &				&		&	AE-E02, AN-AW, AN-E02, AW-E02   		&			&               &			    \\
		    &				    &	2002 Jun 24	&	 2	&	AC-AW, AC-E02, AE-AN, AE-AW,			&	 254	&	     13     &   FK5 713	    \\ 
		    &				    &				&		&	AE-E02, AN-AW, AN-E02, AW-E02   		&			&               &			    \\
		    &				    &				&		&									    	&			&               &			    \\
HR 8402	    &	o Aqr		    &	2007 Jun 18	&	11	&	AN-W07, E06-W07						    &	 307	&	\nodata     &   FK5 850	    \\
		    &				    &	2012 Oct 14	&	18	&	AC-AW, AE-AW							&	 529	&	\nodata     &   FK5 850	    \\
		    &				    &	2012 Oct 15	&	14	&	AC-AW, AE-AW							&	 363	&	\nodata     &   FK5 850	    \\		
		    &				    &	2012 Oct 16	&	15	&	AC-AW, AE-AW							&	 408	&	\nodata     &   FK5 850	    \\
		    &				    &	2012 Oct 19	&	17	&	AC-AW, AE-AW							&	 444	&	\nodata     &   FK5 850	    \\	
		    &				    &				&		&										    &			&               &			    \\
FK5 843	    &	31 Peg		    &	2009 Jul 07	&	 1	&	AC-AE, AC-AN, AE-AN, AE-AW, AN-AW		&	  87	&	     49     &   FK5 855	    \\
		    &				    &	2009 Jul 09	&	 5	&	AC-AE, AC-AN, AE-AN, AE-AW, AN-AW		&	 274	&	     57     &   FK5 855	    \\
		    &				    &	2009 Jul 10	&	 7	&	AC-AE, AC-AN, AE-AN, AE-AW, AN-AW		&	 410	&	    142     &   FK5 855	    \\		
		    &				    &	2009 Jul 15	&	 6	&	AC-AE, AC-AN, AE-AN, AE-AW, AN-AW		&	 415	&	    170     &   FK5 855	    \\
		    &				    &	2009 Jul 22	&	 6	&	AC-AE, AC-AN, AE-AN, AE-AW, AN-AW		&	 390	&	    179     &   FK5 855	    \\
		    &				    &				&		&										    &			&               &			    \\
FK5 1602	&	$\beta$ Psc	    &	2006 Nov 09	&	21	&	AC-AE, AC-AW							&	 576	&	\nodata     &   FK5 850	    \\
		    &				    &	2006 Nov 10	&	21	&	AC-AE, AC-AW							&	 570	&	\nodata     &   FK5 850	    \\
		    &				    &	2006 Nov 14	&	17	&	AC-AE, AC-AW							&	 480	&	\nodata     &   FK5 850	    \\
		    &				    &	2006 Nov 16	&	23	&	AC-AE, AC-AW							&	 669	&	\nodata     &   FK5 850	    \\
		    &				    &	2006 Nov 17	&	24	&	AC-AE, AC-AW							&	 709	&	\nodata     &   FK5 850	    \\
\enddata
\tablecomments{Cols. (1) and (8): FK5 \citep{frk88} or HR number \citep{hof91}, as per NPOI scheduling and archiving software.  Col. (2): Per 
Table~\ref{targetlist}.  Col. (3): UT date of NPOI observations.  Col. (4): Total number of multi-baseline interferometric observations.  Col. (5): 
Baselines used.  Col. (6): Number of $V^2$ measurements made over available spectral channels, over all recorded baselines (less edits). 
Col. (7): Number of closure phase measurements made over available spectral channels, over all recorded beam combiner outputs (less edits). 
Col. (8)): Calibrator star.}
\tablenotetext{a}{See \citet{taz11,cmt12}.}
\tablenotetext{b}{19.7m north-south Mark III baseline.  Spectral channels at 450.2 nm, 550.0 nm, and 800.0 nm effective wavelengths.}
\tablenotetext{c}{19.7m north-south Mark III baseline.  Spectral channels at 500.0 nm, 550.0 nm, and 800.0 nm effective wavelengths.}
\tablenotetext{d}{15.2m north-south Mark III baseline.  Spectral channels at 500.0 nm, 550.0 nm, and 800.0 nm effective wavelengths.}
\tablenotetext{e}{12.0m north-south Mark III baseline.  Spectral channels at 500.0 nm, 550.0 nm, and 800.0 nm effective wavelengths.}
\tablenotetext{f}{Data collected using earlier three-station beam combiner, producing 32 spectral channel data, covering 450--860 nm, 
on each of three spectrometer outputs (less one defective channel, for a total of 95 $V^2$ measurements per observation).} 
\end{deluxetable*}
\end{longrotatetable}

\section{Data Modeling and Results} \label{modeling}

Following the methodology of Paper I, once the data reduction procedures were completed, we systematically examined all of the calibrated visibility data for our program stars (Table~\ref{progstarsobs}) for the possible sinusoidal signature of a binary companion \citep[][\S~\ref{bindet}]{brn74} using the GRIDFIT program, developed for this survey using IDL\footnote{\url{https://www.l3harrisgeospatial.com/Software-Technology/IDL}}.  The output plots from GRIDFIT (e.g., Paper I, Figure 8) showing the minimum value of the reduced $\chi_{\nu}^2$ for a binary star model over a grid search in component separation ($\rho$) and position angle ($\theta$) at each of a series of component magnitude difference ($\Delta m$) values were examined for evidence of a ``significant'' global $\chi_{\nu}^2$ minimum ($\chi_{\nu}^2$(min)), here defined as a minimum with a corresponding 68\% confidence interval \citep[$\chi_{\nu}^2$(min)+3.53 for a three-parameter $\Delta m$, $\rho$, $\theta$ fit][]{yav76,jvw96} with a half-width of $\pm$1 or less in $\Delta m$.\footnote{While this is a less stringent standard than that used in Paper I (a 99\% confidence interval), experience has shown \citep[][hereafter Paper II, and \S~\ref{bindet}]{htz19} that it generally results in $\rho$, $\theta$, and $\Delta m$ values of sufficient consistency and accuracy to allow subsequent detailed modeling.}  For those systems where a significant binary detection was obtained, we then proceeded to detailed modeling using the OYSTER\footnote{\url{https://www.eso.org/~chummel/oyster/oyster.html}} modeling software, which has been the standard for displaying, editing, averaging, calibrating and modeling interferometry data from the NPOI for many years \citep[e.g.,][]{hbh03}.  OYSTER was used to fit binary models on a night-by-night basis to the visibility and closure phase data as a function of $\it u$ and $\it v$ using a Levenberg-Marquardt non-linear least-squares technique, where the $\rho$, $\theta$ and $\Delta m$ values corresponding to the global $\chi_{\nu}^2$(min) in GRIDFIT were used as starting values.  The results of the OYSTER binary model fits are presented in Table~\ref{programfit}, which lists the fitted angular separations and position angles for each system, on each night, along with the parameters of the fit error ellipse and the fitted component magnitude difference at $\lambda$ = 700~nm ($\Delta m_{700}$).  The uncertainty ellipses correspond to one-seventh of the synthesized beam, which has been shown to give realistic estimates of the astrometric accuracy of NPOI multi-baseline observations \citep{hbh03, hrn13}.  In some cases, where closure phase data is lacking (Table~\ref{progstarsobs}), there is a $\pm$180$\arcdeg$ ambiguity in the fitted position angles.  In some instances, the fitted position angles required adjustment by 180$\arcdeg$ for consistency with those listed in the Fourth Catalog of Interferometric Measurements of Binary Stars \citep {hmw01b} at similar epochs or with the predictions of published orbits (\S~\ref{bindet}).

In an attempt to verify some of the GRIDFIT results by an independent method, the $V^2$ and, when available, closure phase data were also examined using the CANDID algorithm \citep{gal15}\footnote{\url{https://github.com/amerand/CANDID}} for a randomly selected subset of six sources (\S~\ref{bindet}), but over a much smaller angular field (a radius of $\leq$ 57 mas, as compared to $\leq$ 500 mas for GRIDFIT).  While no convincing detections were obtained for any of the 30 nights of data examined, the average 3$\sigma$ detection limit for a companion produced by CANDID ($\Delta m_{700}$ $\approx$ 3.5, the average of the ``Absil'' and ``injection'' methods) is consistent with the results of Paper~I ($\Delta m_{700}$ = 3.0 -- 3.5).

\newpage

%% Table 5 -- Program Stars - Binary Model Fits

\startlongtable
%%\begin{longrotatetable}
\begin{deluxetable}{lcccrrrrrc}
%%\tabletypesize{\scriptsize}
\tablecaption{Binary Program Stars - Binary Model Fits \label{programfit}}
\tablewidth{0pt}
\tablehead{
\colhead{Object} & \colhead{Other} & \colhead{UT Date} & \colhead{MJD} & \colhead{$\rho$} & \colhead{$\theta$} & 
\colhead{$\sigma_{maj}$} & \colhead{$\sigma_{min}$} & \colhead{$\phi$} & \colhead{$\Delta m_{700}$}  \\
\colhead{FK5/HR} & \colhead{Name} & \colhead{} & \colhead{} & \colhead{(mas)} & \colhead{(deg)} & \colhead{(mas)} & 
\colhead{(mas)} & \colhead{(deg)} & \colhead{} \\
\colhead{(1)} & \colhead{(2)} & \colhead{(3)} & \colhead{(4)} & \colhead{(5)} & \colhead{(6)} & \colhead{(7)} & \colhead{(8)} & 
\colhead{(9)} & \colhead{(10)} \\
}
\startdata
FK5 25	    &	o Cas Aa,Ab	    &	2018 Sep 27	&	58388.30	&	  15.81	&	279.32	&	0.34	&	0.20	&	155.8   &   2.91 $\pm$ 0.02 \\
		    &				    &	2018 Sep 28	&	58389.28	&	  15.90	&	278.50	&	0.34	&	0.20	&	155.8	&   2.92 $\pm$ 0.02 \\
		    &				    &	2018 Sep 29	&	58390.31	&	  15.87	&	278.89	&	0.34	&	0.20	&	155.0	&   2.90 $\pm$ 0.01 \\
		    &				    &	2018 Nov 15	&	58437.30	&	  16.89	&	271.20	&	0.38	&	0.21	&	119.5	&   2.79 $\pm$ 0.02 \\
		    &				    &	2018 Nov 16	&	58438.28	&	  16.93	&	271.73	&	0.34	&	0.20	&	136.9	&   2.82 $\pm$ 0.02 \\
		    &				    &	2018 Nov 21	&	58443.28	&	  16.87	&	269.36	&	0.38	&	0.20	&	136.6	&   2.88 $\pm$ 0.03 \\
		    &				    &	2018 Nov 27	&	58449.29	&	  16.95	&	269.34	&	0.36	&	0.20	&	139.8	&   2.90 $\pm$ 0.02 \\
		    &				    &	2018 Dec 04	&	58456.29	&	  16.98	&	270.09	&	0.40	&	0.19	&	151.4	&   2.90 $\pm$ 0.03 \\
		    &				    &	2018 Dec 12	&	58464.28	&	  16.88	&	267.38	&	0.37	&	0.20	&	129.3	&   2.96 $\pm$ 0.03 \\
		    &				    &				&   			&		    &			&		    &		    &           &                   \\
FK5 57	    &	$\phi$ Per      &	2004 Nov 03	&	53312.30	&	   5.28	&	289.29	&	0.22	&	0.07	&	147.3   &   3.09 $\pm$ 0.05 \\
		    &				    &	2004 Nov 04	&	53313.29	&	   4.95	&	288.16	&	0.28	&	0.07	&	146.9   &   3.40 $\pm$ 0.04 \\
		    &				    &	2004 Nov 05	&	53314.30	&	   5.76	&	291.51	&	0.25	&	0.07	&	138.0   &   2.28 $\pm$ 0.04 \\
		    &				    &	2004 Dec 01	&	53340.30	&	   3.85	&	311.08	&	0.38	&	0.06	&	156.5   &   3.07 $\pm$ 0.09 \\
		    &				    &	2004 Dec 02	&	53341.29	&	   3.65	&	312.45	&	0.20	&	0.07	&	142.8   &   3.03 $\pm$ 0.02 \\
		    &				    &	2004 Dec 23	&	53362.29	&	   2.64	&    91.08	&	0.31	&	0.12	&	143.9   &   3.03 $\pm$ 0.02 \\
		    &				    &				&			    &			&			&		    &		    &           &                   \\
HR 985	    &	    BK Cam	    &	2006 Nov 08	&	54047.30	&	 116.59	&	 44.51	&	0.41	&	0.19	&	175.2   &   2.46 $\pm$ 0.02 \\
		    &				    &	2006 Nov 11	&	54050.29	&	 116.42	&	 44.54	&	0.43	&	0.19	&	174.1   &   2.60 $\pm$ 0.05 \\
		    &				    &	2009 Nov 25	&	55160.31	&	 132.05	&	 50.33	&	0.37	&	0.18	&	126.6   &   2.45 $\pm$ 0.02 \\
		    &				    &	2009 Dec 01	&	55166.28	&	 132.26	&	 50.22	&	0.39	&	0.19	&	120.1   &   2.60 $\pm$ 0.02 \\
		    &				    &	2009 Dec 03	&	55168.29	&	 131.91	&	 50.29	&	0.41	&	0.18	&	127.0   &   2.39 $\pm$ 0.02 \\
		    &				    &	2010 Oct 09	&	55478.29	&	 137.34	&	 51.26	&	0.39	&	0.18	&	156.5   &   2.45 $\pm$ 0.01 \\
		    &				    &	2010 Oct 12	&	55481.28	&	 137.34	&	 51.14	&	0.40	&	0.18	&	145.7   &   2.37 $\pm$ 0.02 \\		    
		    &				    &				&			    &			&			&		    &		    &           &                   \\
HR 2456	    &	15 Mon Aa,Ab    &	2002 Dec 13	&	52621.30	&	  72.36	&	239.63	&	0.27	&	0.16	&	100.7   &   1.57 $\pm$ 0.05 \\
		    &				    &	2002 Dec 14	&	52622.28	&	  72.24	&	239.95	&	0.73	&	0.14	&	 48.0   &   1.51 $\pm$ 0.05 \\
		    &				    &				&			    &			&			&		    &		    &           &                   \\
HR 5774	    &	53 Boo AB	    &	2013 Mar 12	&	56363.29	&	  27.70	&	156.22	&	0.35	&	0.09	&	  5.2   &   1.98 $\pm$ 0.02 \\
		    &				    &				&			    &			&			&		    &		    &           &                   \\
HR 6712	    &	66 Oph AB 	    &	2006 May 18	&	53873.28	&	 110.28	&	154.13	&	0.20	&	0.15	&	101.7   &   2.62 $\pm$ 0.01 \\
		    &				    &	2006 May 25	&	53880.28	&	 109.37	&	154.42	&	2.43	&	0.50	&	153.4   &   \nodata         \\
		    &				    &	2007 Jun 13	&	54264.30	&	 121.14	&	157.42	&	0.20	&	0.08	&	152.8	&   2.53 $\pm$ 0.03 \\
		    &				    &	2007 Jun 16	&	54267.29	&	 121.33	&	157.38	&	0.23	&	0.08	&	152.9	&   2.53 $\pm$ 0.05 \\
		    &				    &	2007 Jun 17	&	54268.31	&	 121.44	&	157.41	&	0.19	&	0.08	&	144.4	&   2.58 $\pm$ 0.02 \\
		    &				    &	2007 Jun 18	&	54269.30	&	 121.38	&	157.39	&	0.18	&	0.07	&	143.8	&   2.68 $\pm$ 0.02 \\
		    &				    &	2007 Jun 20	&	54271.30	&	 121.74	&	157.41	&	0.23	&	0.08	&	145.8	&   2.60 $\pm$ 0.04 \\
		    &				    &	2007 Jun 21	&	54272.28	&	 121.63	&	157.47	&	0.21	&	0.07	&	152.9	&   2.61 $\pm$ 0.03 \\
		    &				    &	2007 Jun 22	&	54273.31	&	 121.41	&	157.48	&	0.18	&	0.06	&	150.2	&   2.54 $\pm$ 0.03 \\
		    &				    &	2007 Jun 24	&	54275.27	&	 121.53	&	157.53	&	0.19	&	0.06	&	155.4	&   2.77 $\pm$ 0.04 \\
		    &				    &	2007 Jun 26	&	54277.28	&	 121.45	&	157.51	&	0.26	&	0.08	&	154.8   &   2.53 $\pm$ 0.04 \\
		    &				    &	2011 May 25	&	55706.31	&	 132.41	&	165.76	&	3.63	&	0.18	&	 18.7	&   2.55 $\pm$ 0.02 \\
		    &				    &				&			    &			&			&		    &		    &           &                   \\
FKV 705	    & $\beta$ Lyr Aa1,2 &	2006 Sep 17	&	53995.31	&	   0.85	&	 60.97	&	0.36	&	0.08	&	109.5   &   1.20 $\pm$ 0.09 \\
		    &				    &	2006 Sep 18	&	53996.29	&	   0.57	&	 58.84	&	0.21	&	0.08	&	115.6   &   \nodata         \\
		    &				    &	2007 Jun 29	&	54280.31	&	   0.77	&	 61.75	&	0.16	&	0.06	&	148.5   &   0.68 $\pm$ 0.01 \\
		    &				    &	2007 Jul 02	&	54283.30	&	   0.71	&	233.18	&	0.16	&	0.05	&	149.4   &   1.27 $\pm$ 0.02 \\
		    &				    &	2007 Jul 03	&	54284.28	&	   0.96	&	222.33	&	0.41	&	0.03	&	160.6   &   0.73 $\pm$ 0.02 \\
		    &				    &	2007 Jul 04	&	54285.30	&	   0.96	&	252.46	&	0.16	&	0.05	&	149.8   &   1.21 $\pm$ 0.004    \\
		    &				    &	2007 Jul 12	&	54293.28	&	   0.93	&	 87.75	&	0.19	&	0.05	&	155.1   &   0.87 $\pm$ 0.01 \\
		    &				    &				&			    &			&			&		    &		    &           &                   \\
FK5 727	    &   $\upsilon$ Sgr	&	2011 Jul 01	&	55743.31	&	   1.60	&	352.42	&	1.83	&	0.08	&	  2.8   &   3.72 $\pm$ 0.05 \\
		    &			    	&	2011 Jul 02	&	55744.30	&	   1.36	&	358.56	&	0.73	&	0.07	&	176.9   &   \nodata         \\
		    &				    &	2011 Jul 13	&	55755.31	&	   2.10	&	 29.25	&	1.54	&	0.08	&	  2.4   &   3.45 $\pm$ 0.02 \\		
		    &				    &				&			    &			&			&		    &   		&           &                   \\		
FK5 1551    &	59 Cyg Aa,Ab    &	2009 Jul 07, 22 &	55026.80$^a$    &   156.0   &   358.8   &   1.9 &   0.7 &    30.0   &   3.01 $\pm$ 0.12$^b$ \\
		    &				    &				&			    &			&			&		    &	    	&           &                   \\
FK5 809	    & $\beta$ Cep Aa,Ab &	1989 Sep 10	&	47778.97	&	  60.27	&	 52.04	&	0.54	&	0.15	&	 91.4   &   \nodata$^c$     \\
		    &				    &	1991 Aug 07	&	48475.01	&	  28.67	&	 53.77	&	0.33	&	0.10	&	 81.0   &   \nodata         \\
		    &				    &	1991 Sep 13	&	48512.02	&	  27.72	&	 54.63	&	0.35	&	0.09	&	227.1   &   \nodata         \\
		    &				    &	1991 Sep 14	&	48513.00	&	  27.47	&	 54.82	&	0.22	&	0.08	&	 65.1   &   \nodata         \\
		    &				    &	1991 Sep 15	&	48513.98	&	  27.16	&	 54.40	&	0.22	&	0.10	&	 72.4   &   \nodata         \\
		    &				    &	1991 Sep 30	&	48529.00	&	  26.93	&	 55.21	&	0.27	&	0.11	&	 73.7   &   \nodata         \\
		    &				    &	1991 Dec 06	&	48595.97	&	  23.25	&	 56.12	&	0.41	&	0.11	&	212.5   &   \nodata         \\
		    &				    &	1991 Dec 07	&	48597.00	&	  24.24	&	 55.73	&	0.30	&	0.10	&	228.8   &   \nodata         \\
		    &				    &	1992 Jul 06	&	48809.00	&	  14.09	&	 61.08	&	0.30	&	0.10	&	 95.9   &   \nodata         \\
		    &				    &	1992 Aug 06	&	48839.99	&	  12.64	&	 62.04	&	0.19	&	0.08	&	104.8   &   \nodata         \\
		    &				    &	1997 Jul 03	&	50632.31	&	  54.83	&	227.58	&	0.66	&	0.16	&	  4.5   &   \nodata         \\
		    &				    &	1997 Jul 16	&	50645.29	&	  54.97	&	227.51	&	0.48	&	0.17	&	  4.0   &   \nodata         \\
		    &				    &	1997 Jul 18	&	50647.29	&	  55.02	&	227.41	&	0.45	&	0.17	&	172.6   &   \nodata         \\
		    &				    &	1997 Sep 30	&	50721.31	&	  54.63	&	227.51	&	0.89	&	0.16	&	168.2   &   \nodata         \\
		    &				    &	1998 Jun 30	&	50994.31	&	  46.31	&	228.23	&	0.61	&	0.17	&	  9.0   &   \nodata         \\
		    &				    &	1998 Oct 03	&	51089.28	&	  40.73	&	228.64	&	0.32	&	0.19	&	150.3   &   \nodata         \\
		    &				    &	2001 Jun 20	&	52080.29	&	  39.68	&	 45.26	&	0.71	&	0.16	&	  1.0   &   \nodata         \\
		    &				    &	2002 Jun 18	&	52443.31	&	  66.29	&	 46.18	&	0.36	&	0.19	&	161.1   &   \nodata         \\
		    &				    &	2002 Jun 19	&	52444.29	&	  66.42	&	 46.16	&	0.33	&	0.19	&	  2.9   &   \nodata         \\
		    &				    &	2002 Jun 20	&	52445.27	&	  66.58	&	 46.25	&	0.34	&	0.19	&	 16.4   &   \nodata         \\
		    &				    &	2002 Jun 22	&	52447.28	&	  66.53	&	 46.39	&	0.40	&	0.18	&	177.6   &   \nodata         \\
		    &				    &	2002 Jun 23	&	52448.31	&	  67.29	&	 45.40	&	0.27	&	0.22	&	157.8   &   \nodata         \\
		    &				    &	2002 Jun 24	&	52449.29	&	  66.66	&	 46.57	&	0.34	&	0.18	&	144.6   &   \nodata         \\
\enddata
\tablecomments{Col. (1): FK5 \citep{frk88} or HR number \citep{hof91}, as per NPOI scheduling and archiving software.  Col. (2): per 
Table~\ref{targetlist}.  Col. (3): UT date of observations.  Col. (4): MJD of observations.  Col. (5): Fitted binary separation.  Col. (6): Fitted 
binary position angle.  Col. (7): Semimajor axis of error ellipse.  Col. (8): Semiminor axis of error ellipse.  Col. (9): Position angle of error 
ellipse.  Col. (10): Fitted component magnitude difference at 700~nm.}
\tablenotetext{a}{Mean MJD of observations.}
\tablenotetext{b}{Weighted mean of values on each date: 2.89 $\pm$ 0.01 (2009 Jul 07) and 3.13 $\pm$ 0.01 (2009 Jul 22).}
\tablenotetext{c}{Simultaneous fit for all nights: $\Delta m_{500}$ = 2.07 $\pm$ 0.30, $\Delta m_{550}$ = 2.03 $\pm$ 0.20, and 
$\Delta m_{800}$ = 1.84 $\pm$ 0.10 (\S~\ref{betcep}).}
\end{deluxetable}
%%\end{longrotatetable}

\section{Results for Individual Systems} \label{bindet}

To place the results of our binary star modeling into the larger context of previous multiplicity studies, we also include the results of an extensive search of the relevant literature.  We searched the Washington Double Star Catalog \citep[WDS;][]{mas01}, including the associated notes file (``WDS Notes''), the Second Catalog of Rectilinear Elements (LIN2)\footnote{\url{www.astro.gsu.edu/wds/lin2.html}}, the Sixth Catalog of Orbits of Visual Binary Stars \citep[ORB6;][]{hmw01a}, the Fourth Catalog of Interferometric Measurements of Binary Stars \citep[INT4;][]{hmw01b}, the HIPPARCOS Catalogue Double and Multiple Systems Annex \citep[DMSA;][]{lld97}, The Ninth Catalogue of Spectroscopic Binary Orbits \citep[SB9;][]{pou04}, the Multiple Star Catalog\footnote{\url{http://www.ctio.noao.edu/~atokovin/stars/index.html}} \citep[MSC;][]{tok18a}, and papers 1 through 265 of Griffin \citep[e.g.,][]{grf19} for references to binary companions (although none of the program stars appeared in any of the papers of Griffin).  In evaluating these results we assessed the likely number of physical components in each system (Table~\ref{targetlist}, Column 12).  In this assessment, and unless contradicted by our own astrometry or by more recent results from the bibliographic references for each system in the $\it SIMBAD$ literature database \citep[][searching from 2010 through mid-2020, plus secondary references derived there from]{wen00}, we only included pairs clearly identified in the WDS as physical, and pairs listed in the ORB6, DMSA, SB9, or MSC catalogs.  Lastly, pairs listed in the LIN2 catalog were assumed to be nonphysical, while those listed as having common proper motions (a CPM pair) were assumed to be physical.

The combined results of our modeling and literature search efforts indicate that the program stars can be placed into one of three groups.  The first group is made up of 10 sources (\S~\ref{knownbin-det}), each containing a previously confirmed spectroscopic or visual binary at separations $\lesssim$ 0$\farcs$8 that were also detected in our NPOI data using GRIDFIT and subsequently modeled using OYSTER.  For two of the sources (66 Oph; \S~\ref{66oph} and $\beta$ Cep; \S~\ref{betcep}) the number and timespan of our new relative position fits justified attempting new orbit solutions.  Four of the 10 sources (o Cas; \S~\ref{omicas}, 15 Mon; \S~\ref{15mon}, $\beta$ Lyr; \S~\ref{betlyr}, and $\beta$ Cep; \S~\ref{betcep}) also contain wider binary components that our literature search indicated are physical companions to the narrow binaries, thus forming hierarchical multiple systems. The second group consists of three sources (\S~\ref{knownbin-nondet}), each containing a previously confirmed spectroscopic binary, that were {\it not} detected in our NPOI data using GRIDFIT.  The likely reasons for these non-detections are discussed in detail for each source, but it appears most likely they were not detected due to a very large $\Delta m$ difference between the components rather than lack of angular resolution associated with the NPOI observations.  One of the sources in the second group ($\gamma$~Cas; \S~\ref{gamcas}) also contains wide binary components that our literature search indicates are physical companions to the narrow binary, thus forming an extended hierarchical multiple system. Lastly, the third group with 18 sources (\S~\ref{nondet}) contains targets that were not previously confirmed to be either spectroscopic or visual binaries at separations $\lesssim$ 0$\farcs$8.  For 17 sources from third group, binary signatures were not detected in our NPOI data using GRIDFIT, nor were these sources indicated as having wider, physical binary companions based on our literature search.  The remaining source, BK Cam (\S~\ref{bkcam}), was resolved to be a binary system on a number of nights in the NPOI data using GRIDFIT, and modeled in detail using OYSTER, but the resulting position fits indicate only linear relative motion between the stars over the relatively short timespan of our observations.  However, the close physical proximity of the stars suggests a physical binary.

\subsection{Known Binaries - Detected by NPOI} \label{knownbin-det}

\subsubsection{o Cas} \label{omicas}

o Cas (HIP 3504, HR 193, FK5 25, HD 4180, WDS 00447+4817A): This star does not appear in the LIN2 or MSC.  The WDS lists pair AB ($\theta$ = 301$\fdg$60, $\rho$ = 34$\farcs$380 in 2011, with $\Delta m$ = 6.66) without comment regarding it being a physical pair, and pair Aa,Ab ($\theta$ = 269$\fdg$10, $\rho$ = 17 mas in 2007, with $\Delta m$ = 2.90) as having an orbit, which is listed in the ORB6 \citep[$P$ = 2.835 yr, $a$ = 17 mas, $i$ = 113$\fdg$4, $e$ = 0.019;][]{dac19}.  The WDS Notes indicate that the HIPPARCOS astrometric solution for Aa,Ab in the DMSA has similar $P$, but $a$ = 7.15 mas and $e$ = 0.1325, based on some adopted elements from the orbit of \citet{aal78}.  This binary appears as system 43 in the SB9, which identifies it as a SB1 and lists the orbital solution of \citet[][$P$ = 1033~d, $e$ = 0.11]{aal78}, and also notes it appears to share a common proper motion with the B component.  \citet{wgp18}, searching archival $\it IUE$ (International Ultraviolet Explorer) spectra for hot sdO companions to Be stars, found no indication of such in this system.

\citet{khh10} used NPOI observations from 2005, 2006, and 2007 to spatially resolve the Aa,Ab pair for the first time at $\Delta m_{700}$ = 2.9 $\pm$ 0.1.  Utilizing these observations, along with new spectroscopic data, they determined a combined interferometric and spectroscopic orbit ($P$ = 2.824~yr,  $a$ = 17.0~mas, $i$ = 115$\fdg$0, $e$ = 0), the parallax of the system, and a primary mass of $M_{p}$ = 6.2 $M_{\sun}$.  Combining additional UBV photometry (from a time interval when the spectra lacked emission features) with the $\Delta m_{700}$ and the new parallax, they determined an absolute visual magnitude for the primary of $M_{V}$ = $-2.55$, consistent with the spectral classification of B5III (B5IIIe; BeSS, Table~\ref{targetlist}).  They further noted the large $\Delta m_{700}$ is inconsistent with the nearly equal masses of the Aa,Ab components determined, and suggest that the secondary might in turn be a submilliarcsecond binary of equal mass, early A dwarfs each at $M_{V}$ $\approx$ 1.1.  Subsequently, \citet{tgs13} noted that \citet{edg07} ``observed spectral features corresponding to two similar late-B or early early-A stars that showed Doppler shifts on a timescale of approximately 4 days.''  For $M_{V}$ $\approx$ 1.1, the tables of \citet{sck82} indicate a spectral type of  A1V -- A2V with a mass $\sim$ 2.7$M_{\sun}$.  A mass sum of 5.4$M_{\sun}$ and $P = 4$~d would then predict a separation of 0.09 AU, or 0.3 mas at the parallax of \citet{khh10} (or 0.4 mas at the GAIA parallax listed in Table~\ref{targetlist}), considerably below the angular resolution of the current NPOI observations.

%% omi Cas astrometry fits

\begin{figure}
\epsscale{1.0}
\plotone{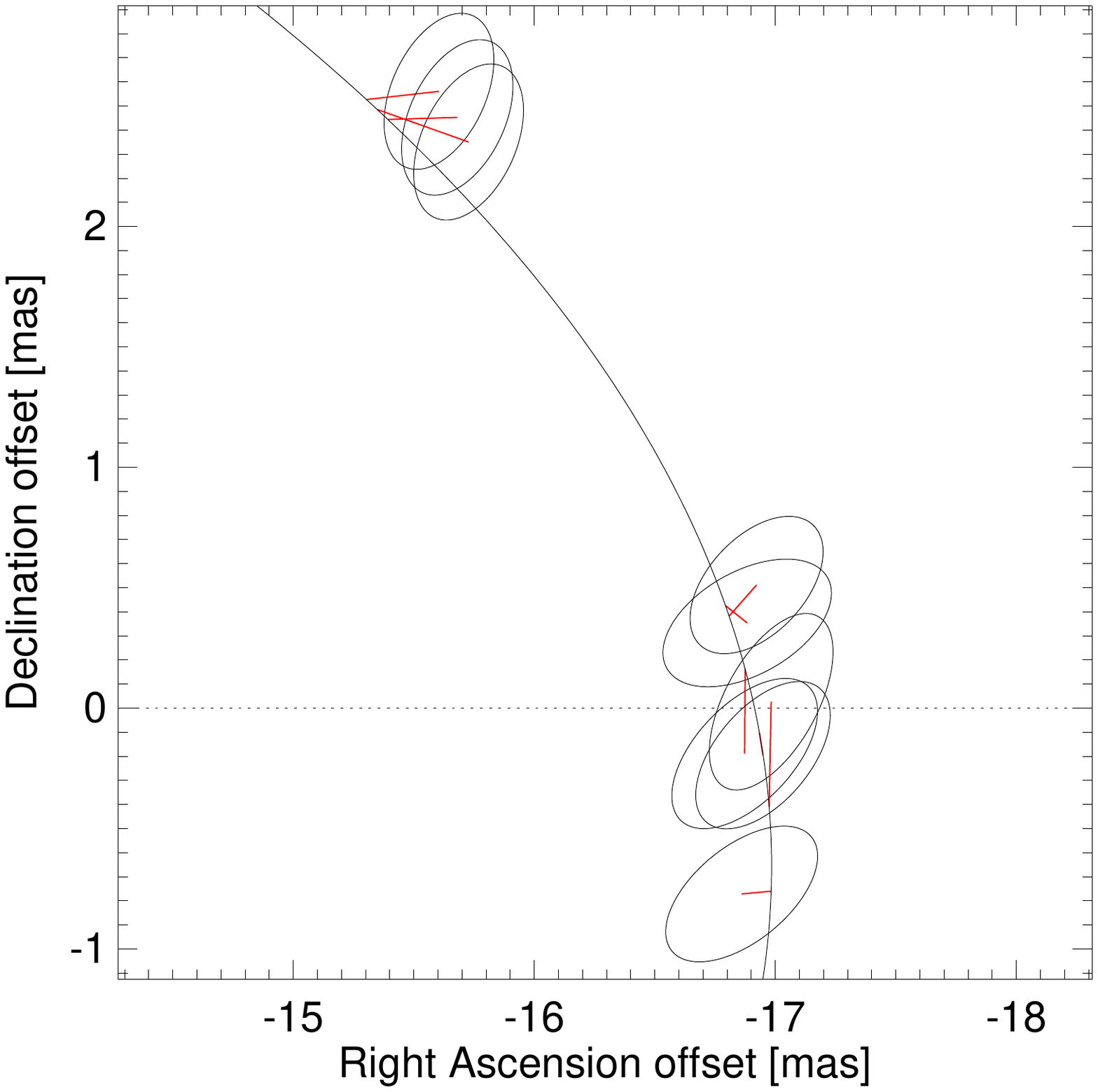}
\caption{Orbit of o Cas Aa,Ab: Plot of the fitted relative positions at each of the nine epochs of NPOI observations (Table~\ref{programfit}) with error ellipses plotted to scale, along with a section of the apparent orbit of \citet{khh10}.  The red line segments indicate the offset of each measurement with respect to the position predicted by the orbit of \citet{khh10} at the epoch of observation.}
\label{omi-Cas_plot}
\end{figure}

The INT4 lists six resolved observations of the Aa,Ab pair by the NPOI in 2005 -- 2007 \citep{khh10}.  Here, we first examine an additional nine nights of NPOI observations from 2018 (Table~\ref{progstarsobs}) using GRIDFIT.  While the GRIDFIT results indicate significant detections (\S~\ref{modeling}) on only four of the nights, all the nightly minima follow a consistent trend for the relative positions of the components.  Unlike the earlier NPOI observations of \citet{khh10}, the observations of 2018 do not include closure quantities, and we thus adjusted the position angles of the GRIDFIT results by 180$\arcdeg$ for consistency with the predictions of the \citet{khh10} orbit.  We then proceeded to detailed modeling of the data from each of the nine nights with OYSTER using fixed stellar angular diameters of $\theta_{P}$ = 0.5 mas (from the NPOI planning software) and $\theta_{S}$ = 0.5 mas (assumed), the nightly $\rho$ and $\theta$ values from GRIDFIT, and $\Delta m_{700}$ = 2.9 \citep{khh10} as inputs.  The results of the nightly fits are presented in Table~\ref{programfit} and Figure~\ref{omi-Cas_plot}.  The O-C values of the final $\rho$, $\theta$ fits with respect to the orbit of \citet{khh10} show good agreement (0.10 -- 0.36 mas), and they and the earlier NPOI results \citep{khh10} show much better agreement with the \citep{khh10} orbit than with the later orbit of \citet{dac19}.  The weighted mean of the component magnitude difference fits for the nine nights was $\Delta m_{700}$ = 2.89 $\pm$ 0.02, in excellent agreement with the result of \citet{khh10}.

Based on the available evidence, we conclude o Cas is, at a minimum, a hierarchical triple system consisting of components Aa, Ab, and B, or more likely four components: Aa, Ab1, Ab2, and B.

\subsubsection{$\phi$ Per} \label{phiper}

$\phi$ Per (HIP 8068, HR 496, FK5 57, HD 10516, WDS 01437+5041): This star does not appear in the LIN2, DMSA, or MSC.  The spectrum of a hot sdO subdwarf companion was detected in $\it IUE$ data by \citet{tbg95}, and the WDS lists it as a binary with an orbit ($\theta$ = 226$\fdg$6, $\rho$ = 1 mas in 2013, with $\Delta m$ = 3.74).  The WDS Notes list it as a Be+sdO binary with a combined interferometric/spectroscopic orbital solution \citep{mmm15}, the orbital parameters also appearing in the ORB6 ($P$ = 126.6982~d, $a$ = 5.89~mas, $e$ = 0.0).  The SB9 lists it as system 88 and a SB2, including the orbital solution of  \citet{poe81}, with $P$ = 126.696~d and $e$ = 0.02.  The INT4 lists this star as unresolved by speckle interferometry in the visible on 31 dates (1975 -- 1996) implying a limit of $\rho$ $<$ 23 mas, but resolved by LBOI on 10 dates (2011 -- 2013) at 1.6 $\mu$m \citep{mmm15}.

%% phi Per astrometry fits

\begin{figure}
\epsscale{1.0}
\plotone{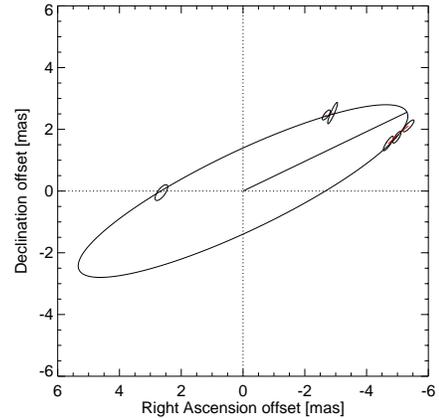}
\caption{Orbit of $\phi$ Per: Plot of the fitted relative positions at each of the six epochs of NPOI observations (Table~\ref{programfit}) with error ellipses plotted to scale, along with the apparent orbit of \citet{mmm15}.  The line segment extending from the origin to the orbit marks periastron, while the red line segments indicate the offset of each measurement with respect to the position predicted by the orbit of \citet{mmm15} at the epoch of observation.}
\label{phi-Per_plot}
\end{figure}

We examined 12 nights of archival NPOI data (2004) using GRIDFIT, obtaining significant detections (\S~\ref{modeling}) on six nights (Table~\ref{progstarsobs}).  We then proceeded to detailed modeling of the data from each of the six nights with OYSTER using fixed stellar angular diameters of $\theta_{P}$ = 0.3 mas and $\theta_{S}$ = 0.056 mas \citep{mmm15}, the orbit of \citet{mmm15}, and the nightly $\rho$ and $\theta$ values and the average $\Delta m_{700}$ obtained from GRIDFIT as inputs.  The results of the nightly fits are presented in Table~\ref{programfit} and Figure~\ref{phi-Per_plot}.  The O-C values of the final $\rho$, $\theta$ fits with respect to the orbit of \citet{mmm15} show good agreement (0.06 -- 0.43 mas).  The weighted mean of the component magnitude difference fits for the six nights was $\Delta m_{700}$ = 3.00 $\pm$ 0.11.  We can check whether this result is reasonable using known parameters of the stellar components (and ignoring the effects of the extended disk emission).  From the Be (primary) star mass of $M_{p}$ = 9.6 $\pm$ 0.3 $M_{\sun}$ and luminosity log($L_{p}$/$L_{\sun}$) = 4.16 $\pm$ 0.10 \citep{mmm15}, Table 1 of \citet{toh04} indicates a spectral type of $\sim$ B1.5, consistent with that of B1.5 V:e-shell \citep{slt82} quoted by \citet{mmm15}. A B1.5 spectral type, along with the large values for the stellar rotation velocity and inclination (assumed equal to the orbital inclination) also reported by \citet[][$v/v_{\rm crit}$ = 0.93 and $i$ $\approx$ 78$\arcdeg$, respectively]{mmm15} allows the use of Figure 3 of \citet{toh04} to predict an absolute visual magnitude $M_{V}$ $\sim -2.9 \pm 0.3$.  For the sdO secondary star, the values of the luminosity log($L_{s}$/$L_{\sun}$) = 3.80 $\pm$ 0.13 and effective temperature $T_{\rm eff}$ = 53000 $\pm$ 3000K reported by \citet{mmm15} can be used \citep[e.g.,][]{sck82} to compute an absolute bolometric magnitude $M_{\rm bol}$(secondary) = $-4.86$ and a bolometric correction B.C. $\approx$ $-4.8$, respectively. Thus we obtain $M_{V}$ (secondary) $\approx -0.1$ and $\Delta m_{V}$ $\approx$ 2.8 $\pm$ 0.3 for the binary, consistent with our measurement and the visible flux ratio reported by \citet[][7.5 $\pm$ 1.7\%, or $\Delta m$ = 2.8 $\pm$ 0.3]{mmm15}. 

From the available evidence, we conclude $\phi$ Per consists of two physical stellar components.

\subsubsection{15 Mon} \label{15mon}

15 Mon (HIP 31978, HR 2456, HD 47839, WDS 06410+0954Aa,Ab): A luminous O star and part of the NGC\,2264 \citep{cngc} Christmas Tree cluster adjacent to the Cone Nebula star forming region. 15 Mon is an irregular variable star, S Mon \citep{gcvs}, and is beautifully seen in a wide field of view image taken with the ESO 2.2~m telescope at 
La Silla\footnote{\url{https://www.eso.org/public/news/eso0848/}}. \citet{sto85} convincingly argue for 15 Mon as the ionizing source of the Cone Nebula using observations spanning the infrared, millimeter and radio bands. 15 Mon does not appear in the LIN2.  The WDS lists 27 wide pairs (AB -- NK; $\rho$ $>$ 3$\arcsec$ and $\Delta m$ $\geq$ 0.3) without comment as to any of them being physical, and many of these components are likely independent stars in the NGC\,2264 cluster.  The DMSA lists pair AB ($\theta$ = 212$\fdg$8, $\rho$ = 2$\farcs$910 in 1991.25) with a type-F solution [``fixed double or multiple system (identical proper motions and parallaxes)''].  The WDS additionally lists a close pair Aa,Ab ($\theta$ = 268$\fdg$7, $\rho$ = 135 mas in 2018) noted as having an orbit, and the ORB6 lists the orbital elements of \citet[][P = 74.28 yr, a = 95.6 mas, e = 0.716]{cvn10} for this pair.  The SB9 lists system 1725 \citep[][which the WDS identifies with Aa,Ab]{gmh93} as a SB1 with $P = 9247$~d (25.3~yr) and $e = 0.67$.  The MSC lists pairs AB and AB,C as CPM pairs, and identifies Aa,Ab as the SB1 and visual binary, along with yet a third orbital solution for Aa,Ab \citep[][with $P = 190.5$~yr, $a = 170$~mas, $e = 0.851$]{tok18b}.  \citet{jma19} notes there have been $\it five$ visual orbits for Aa,Ab published \citep{gmh93,gmb97,cvn09,cvn10,tok18b} agreeing on a high eccentricity (0.67 -- 0.85) and a periastron passage close to 1996, but with widely different periods (23.6 -- 190.5~yr), semimajor axes (33.9 -- 170~mas), and other parameters.  \citet{jma19} goes on to produce a new orbital solution for Aa,Ab ($P = 108.0$~yr, $a = 112.5$~mas, $i = 47\arcdeg$, $e = 0.770$).  \citet{mbs18} used ``lucky spectroscopy'' to assign separate spectral types to Aa,Ab [O7 V((f))z var] and B (B2: Vn), but were not able to separate Aa and Ab.  Subsequently, \citet{mab20} used the Space Telescope Imaging Spectrograph (STIS) of HST to separate the spectra, classifying Aa as O7 V((f))z and Ab as B1: Vn.  \citet{mab20} also provide the Galactic O-Star Spectroscopic Survey \citep[GOSSS,][]{msw11} spectral classification for component C, A3V, with the spectrum shown in their Figure 3. 

%% 15 Mon astrometry fits

\begin{figure*}[ht]
\epsscale{0.45}
\plotone{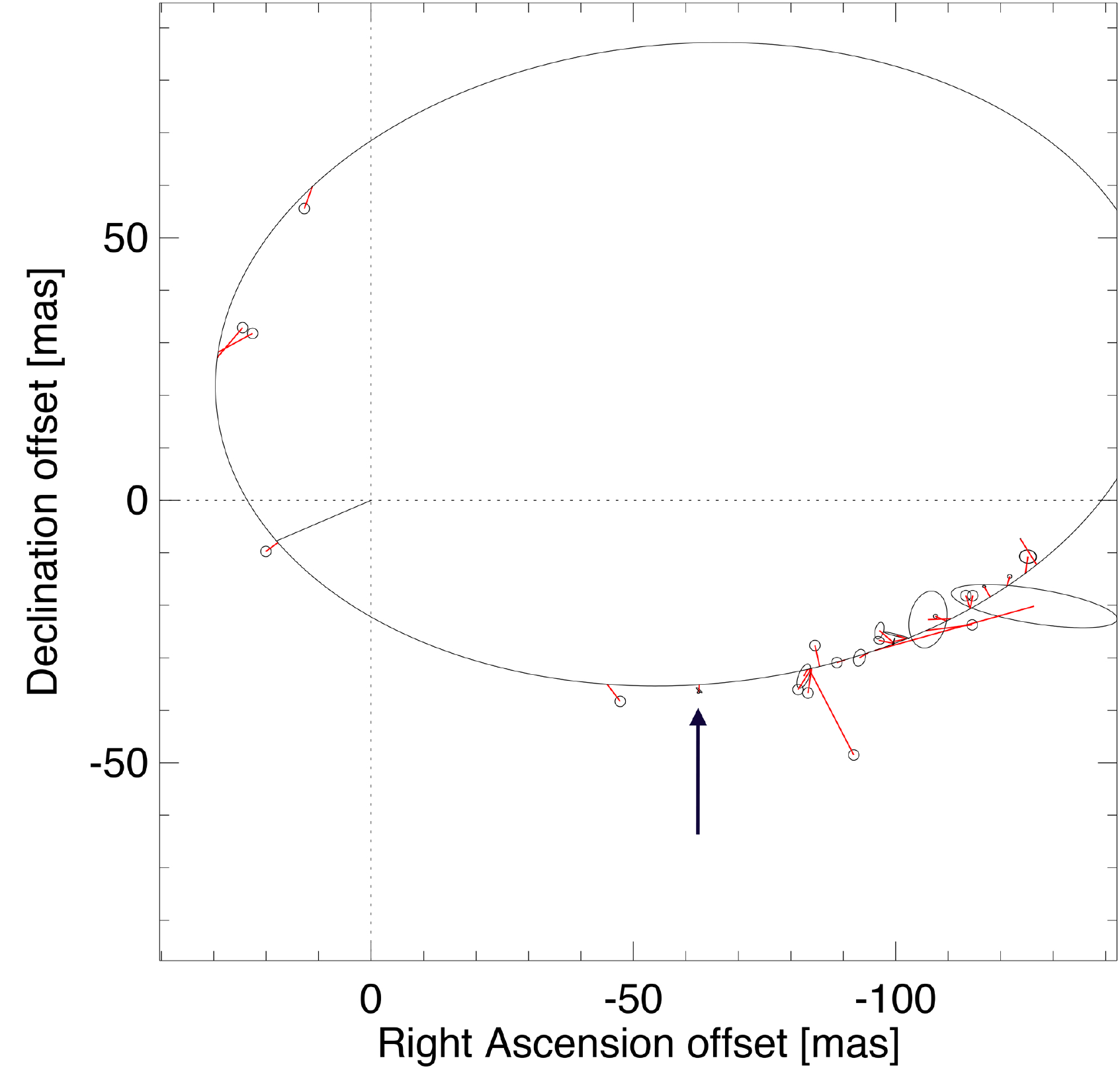}
\epsscale{0.49}
\plotone{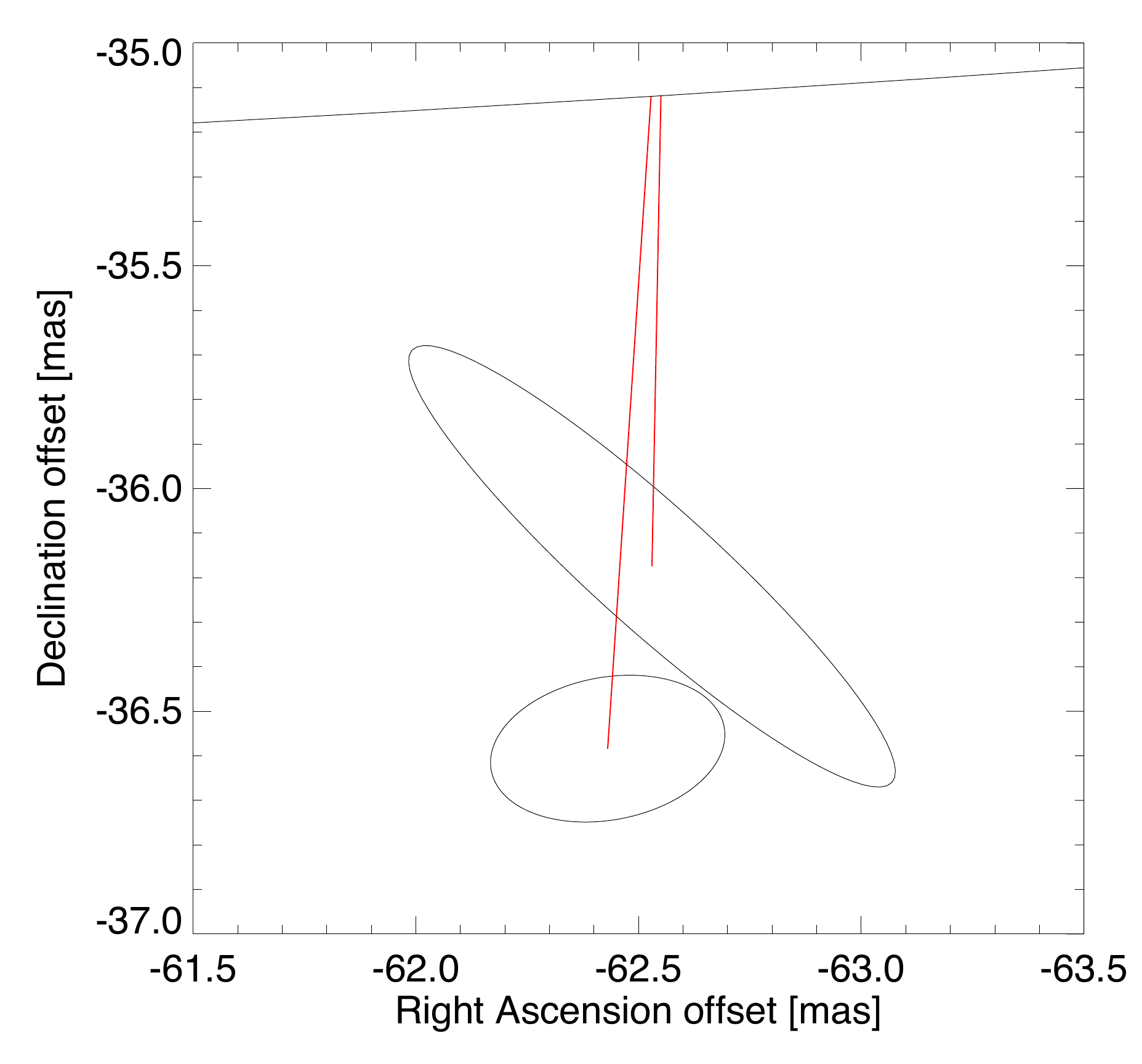}
\caption{Orbit of 15 Mon Aa,Ab: ${\it Left\ Panel}$: Plot of the fitted relative positions (indicated by arrow) at each of the two epochs of NPOI observations (Table~\ref{programfit}) with error ellipses plotted to scale.  Additional observations from the INT4 and \citet{hvd20} are also plotted along with the apparent orbit of \citet{jma19}.  (The position angles of four of the INT4 measurements were adjusted by 180$\arcdeg$ to make a more sensible sequence, and an observation at 2014.2562, with a highly discrepant separation, was not plotted.) The line segment extending from the origin to the orbit marks periastron, while the red line segments indicate the offset of each measurement with respect to the position predicted by the \citet{jma19} orbit at the epoch of observation.  Circles of radius 1 mas are plotted for INT4 measurements lacking error estimates.  ${\it Right\ Panel}$: An expanded version of the lower, center portion of the left panel.  Red line segments indicate the offset of the fitted relative positions for 2002 Dec 13 UT (lower) and 2002 Dec 14 UT (upper) from the \citet{jma19} orbit.}
\label{15-Mon_plot}
\end{figure*}

There are 26 resolved observations of Aa,Ab listed in the INT4, including those made by adaptive optics \citep[AO, combined with aperture masking,][]{sll14}, by short exposure ``lucky imaging,'' \citep{jma10}, or imaging by the High Resolution Camera (HRC) on the HST \citep{jma10}.  These latter observations resolve the quadrant ambiguity of the other observations listed in the INT4.  \citet{hvd20} provide additional observations (2015.1848) made at 692 nm and 880 nm.

The two (consecutive) nights of NPOI observations of 15 Mon listed in Table~\ref{progstarsobs} are rather poor quality, single observations (albeit on multiple baselines) with no usable closure data.  Modeling with the standard GRIDFIT procedure produced a significant $\chi_{\nu}^2$ minimum for each night, but at widely different separations and position angles.  However, a program within OYSTER similar to GRIDFIT was also used to search a grid of 50 mas by 50 mas at 0.5 mas spacings, centered on the position of the secondary star as predicted by the orbit of \citet{jma19}, to produce consistent $\chi_{\nu}^2$ minima.

We then proceeded to detailed modeling of the data from these nights with OYSTER utilizing the orbit of \citet[][a much better fit to the INT4 observations]{jma19}, fixed stellar angular diameters of $\theta_{P}$ = 0.2 mas (from the NPOI planning software) and $\theta_{S}$ = 0.1 mas (assumed), the $\rho$ and $\theta$ values obtained by the OYSTER gridfit procedure, and a $\Delta m_{700}$ = 1.6, typical of INT4 measurements made at similar wavelengths.  The results of the fits are presented in Table~\ref{programfit} and Figure~\ref{15-Mon_plot}.  The O-C values of the final $\rho$, $\theta$ fits with respect to the orbit of \citet{jma19} show good agreement (1.1 -- 1.5 mas), while the fitted $\Delta m_{700}$ = 1.54 $\pm$ 0.03 is consistent with the INT4 measurements made at similar effective wavelengths.

The above listed spectral types of the Aa,Ab components \citep{mab20}, along with other measured stellar properties, can also be used to check the consistency of our $\Delta m_{700}$ value. For Aa, the tables of \citet{sck82} indicate for an O7V spectral type an absolute visual magnitude $M_{V}$(primary) $\approx -5.2$.  For the Be secondary star, the BeSS database lists $\it v$sin$\it i$ = 70 km~s$^{-1}$, which with an inclination angle $\it i$ = 47$\arcdeg$ \citep[][assuming the stellar and orbital axes are parallel]{jma19} yields $\it v$ = 96 km~s$^{-1}$.  The models of \citet{toh04} for rapidly rotating, gravity darkened B stars predict $\it v_{c}$ = 502 km~s$^{-1}$ for spectral type B1, and thus $\it v$/$\it v_{c}$ = 19\%.  Figure 3 of \citet{toh04} then predicts an absolute visual magnitude of $M_{V}$(secondary) $\approx -3.4$ for this combination of spectral type, $\it v$/$\it v_{c}$, and inclination.  Therefore we obtain $\Delta m_{V}$ $\approx$ 1.8, in reasonable agreement with our $\Delta m_{700}$ value of 1.54 $\pm$ 0.03. 

Based on the available evidence, we conclude 15 Mon is likely a hierarchical multiple consisting of four components (Aa, Ab, B, and C).

\subsubsection{53 Boo} \label{53boo}

53 Boo (HIP 76041, HR 5774, HD 138629, WDS 15318+4054AB): This star does not appear in the LIN2, DMSA, SB9, or MSC.  The WDS lists pairs AB,C, and AB,D ($\rho$ $>$ 93$\arcsec$ and $\Delta m$ $\geq$ 8.9) without comment as to their being physical.  Additionally, it lists the pair AB ($\theta$ = 34$\fdg$5, $\rho$ = 48 mas in 2010) noted as having an orbit.  The ORB6 lists the orbital elements of \citet[][$P$ = 9.026~yr, $a$ = 61.5~mas, $e$ = 0.006]{hor12} for this pair.  Discounting as doubtful three observations per the WDS Notes, there are $\approx$ 40 resolved observations of the AB pair listed in the INT4.

\begin{figure*}[ht]
\epsscale{0.45}
\plotone{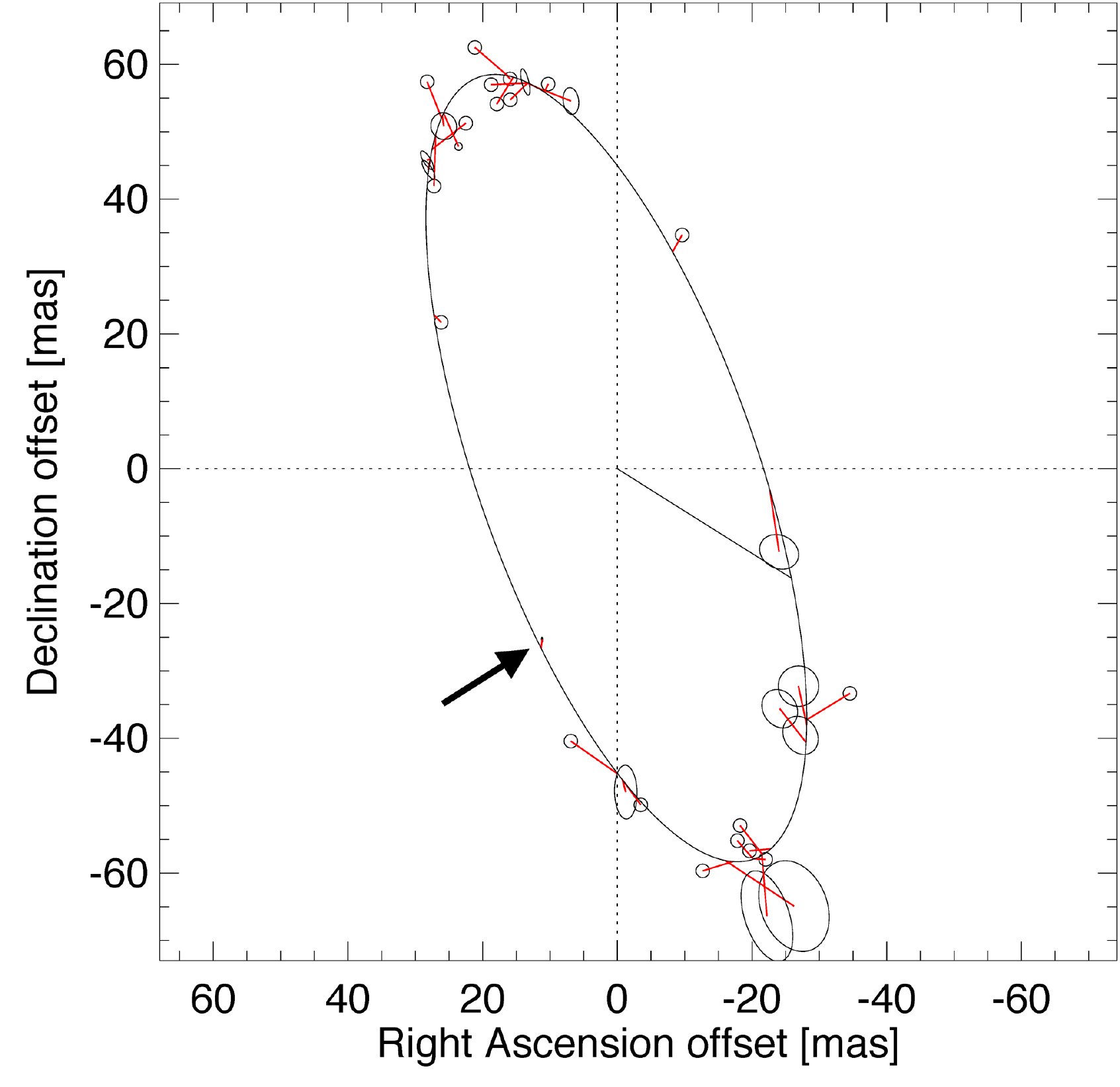}
\epsscale{0.46}
\plotone{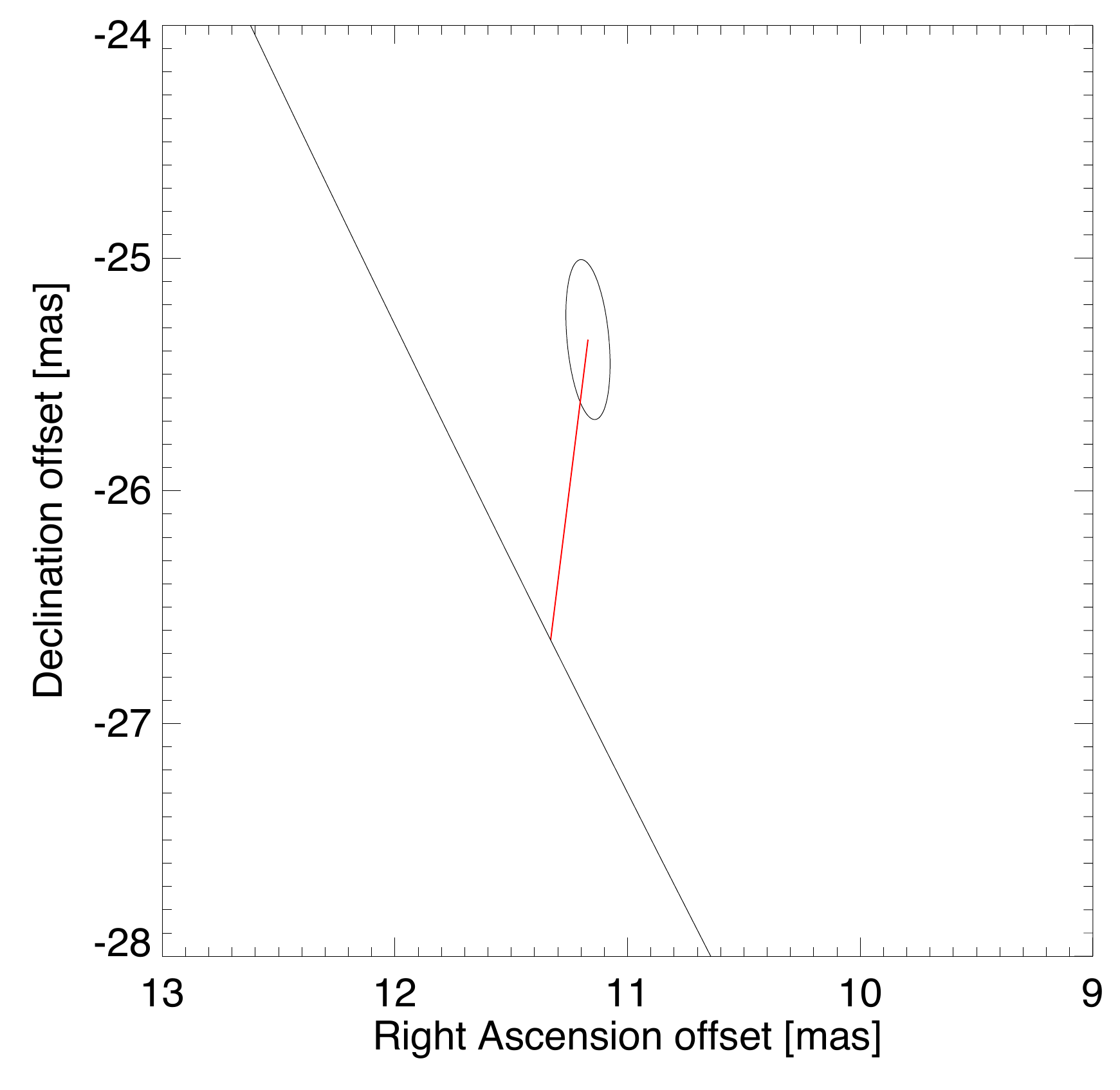}
\caption{Orbit of 53 Boo AB: ${\it Left\ Panel}$: Plot of the fitted relative position from the NPOI observations of 2013 Mar 12 UT (Table~\ref{programfit}) is shown (indicated by arrow), with error ellipse plotted to scale. Additional observations from Table 6 of \citet{hor12}, and from \citet{bbg13}, \citet{hvd20}, and \citet{mut10}, are also plotted along with the apparent orbit of \citet{hor12}.  The line segment extending from the origin to the orbit marks periastron, while the red line segments indicate the offset of each measurement with respect to the position predicted by the \citet{hor12} orbit at the epoch of observation.  Circles of radius 1 mas are plotted for measurements lacking error estimates in the INT4.  ${\it Right\ Panel}$: An expanded version of a section of the lower, left portion of the left panel.  A red line segment indicates the offset of the fitted relative position for 2013 Mar 12 UT from the \citet{hor12} orbit.}
\label{53-Boo_plot}
\end{figure*}

The single night of NPOI observations of 53 Boo AB listed in Table~\ref{progstarsobs} was modeled with GRIDFIT and a significant $\chi_{\nu}^2$ minimum was found~(\S~\ref{modeling}).  We then proceeded to detailed modeling of the data from this night with OYSTER using the orbit of \citet{hor12}, along with fixed stellar angular diameters of $\theta_{P}$ = 0.4 mas (from the NPOI planning software) and $\theta_{S}$ = 0.1 mas (assumed) along with the $\rho$ and $\theta$ values, and the $\Delta m_{700}$ obtained from GRIDFIT as inputs.  The results of the fit are presented in Table~\ref{programfit} and Figure~\ref{53-Boo_plot}.  

%% 53 Boo astrometry fits

The O$-$C value of the final $\rho$, $\theta$ fit with respect to the orbit of \citet{hor12} shows good agreement (O$-$C = 1.3 mas), while the fitted $\Delta m_{700}$ = 1.98 $\pm$ 0.02 is consistent with the range of measurements made at similar effective wavelengths listed in the INT4.  \citet{hor12} suggest for a composite spectral type of A5V for the AB pair (A5Ve, BeSS and Table~\ref{targetlist}), or individual spectral types of A4 and F7.  The tables of \citet{sck82} indicate absolute visual magnitudes of $M_{V}$(primary) $\approx$ 1.73 and $M_{V}$(secondary) $\approx$ 3.83, respectively, for these spectral types and thus $\Delta m_{V}$ $\approx$ 2.1, in reasonable agreement with our $\Delta m_{700}$ value of $1.98\pm 0.02$.

Examination of the GAIA eDR3 results indicates the C and D components are likely optical companions only. The parallaxes and proper motions of C and D are significantly different from the primary A component \citep{gaia20a}. 

From the available evidence, we conclude 53 Boo consists of only two physical stellar components, the AB pair.

\subsubsection{$\delta$ Sco} \label{delsco}

$\delta$ Sco (HIP 78401, HR 5953, FK5 594, HD 143275, WDS 16003-2237AB): This star does not appear in the LIN2 or MSC catalogs.  The WDS lists pair AB ($\theta$ = 352$\fdg$2, $\rho$ = 190 mas in 2016, with $\Delta m = 2.23$) that is noted as having an orbit, and the ORB6 lists the orbital elements of \citet[][$P$ = 3945.4~d, $a$ = 98.94~mas, $i$ = 34$\fdg$12, $e$ = 0.9373]{cmt12} for this pair. The WDS Notes list $\delta$ Sco as a spectroscopic triple and an occultation quadruple, the former perhaps based on earlier reports of an SB1 at short periods, e.g. \citet[][$P$ = 20~d]{vbd63} and \citet[][$P$ = 83~d]{lmm87}, but the INT4 lists no resolved occultation measurements.  The DMSA lists a double system solution ($\theta$ = 344$\fdg$0, $\rho$ = 130 mas) in 1991, but does not list an orbit.  The SB9 lists it as system 1837 and a SB1, including the orbital solution of \citet[][$P$ = 3864.345 d, $e$ = 0.94]{mfb01}, which is based on a combination of spectroscopic and interferometric (speckle) data.

The INT4 lists approximately 250 resolved observations of the AB pair (1973 -- 2015) of which 128 are from the NPOI.  \citet{hvd20} provide additional observations (2015.1830) made at 692 nm and 880 nm.  The first 96 NPOI observations (2000 -- 2010) were used by \citet{taz11}, along with the spectroscopy of \citet{mfb01} to produce a new orbital solution subsequently further improved \citep{cmt12} by the inclusion of an additional 32 nights of NPOI observations and seven CHARA \citep[Center for High Angular Resolution Astronomy Array,][]{tmr05} observations made near the time of periastron passage in 2011.  A search of the NPOI data archive revealed no additional, useful observations (through 2019 Oct 01 UT).  However, we do include the NPOI results of \citet{taz11} and \citet{cmt12} for the 128 earlier NPOI observations in the analyses of the multiplicity statistics (\S~\ref{multstat}) and sensitivity and completeness of our sample (\S~\ref{senscomp}). Inclusion of these results is justified, since as noted in \S~\ref{npoi}, these authors also excluded the H$\alpha$ containing channel from analysis when the binary signature was modeled for this source, therefore requiring no additional modeling on our part.

It is worth noting that the component magnitude difference reported by \citet[][$\Delta m_{V} = 1.5 \pm 0.3$]{trb93}, obtained based on observations from before the 2000 periastron passage when the primary star did not possess strong H$\alpha$ emission, is smaller than those determined by \citet[][$\Delta m_{550} = 1.87 \pm 0.17$, and $\Delta m_{850} = 2.24 \pm 0.26$]{taz11} from observations when the Be primary star's disk was much more prominent, contributing as much as $\approx$ 71\% of the H-band flux \citep{cmt12}.  \citet{cmt12} also note that the integrated $m_{V}$ of $\delta$ Sco was at least 0.3 mag fainter in 2007 than it was in 2011.  \citet[][and references therein]{cmt12} list spectral types of B0.5V and B2V for the primary and secondary star, respectively, and these can be used to check the consistency of the $\Delta m_{V}$ at the earlier epoch.  \citet{gag05} list $\it v$sin$\it i$ = 157 km~s$^{-1}$ for the primary, which with an inclination angle for the primary of $i = 25\arcdeg$ \citep{cmt12} yields $v$ = 371 km~s$^{-1}$.  The models of \citet{toh04} for rapidly rotating, gravity darkened B stars predict $v_{c} = 519$~km~s$^{-1}$ for spectral type B0.5, and thus $\it v$/$\it v_{c}$ = 71\%.  Figure 3 of \citet{toh04} then predicts an absolute visual magnitude of $M_{V}$(primary) $\approx -4.0$ for this combination of spectral type, $\it v$/$\it v_{c}$, and inclination.  For the secondary star with B2V spectral type, the tables of \citet{sck82} predict $M_{V}$(secondary) $\approx -2.45$, and together with the primary component result in $\Delta$$m_{V}$ $\approx$ 1.55, in excellent agreement with the value of $1.5 \pm 0.3$ obtained by \citet{trb93}. 

From the available evidence, we conclude $\delta$ Sco consists of only two physical stellar components, the AB pair.

\subsubsection{66 Oph} \label{66oph}

66 Oph (HIP 88149, HR 6712, HD 164284, WDS 18003+0422AB): This star does not appear in the WDS Notes, LIN2, DMSA, SB9, or MSC.  The WDS lists pair AB ($\theta$ = 197$\fdg$7, $\rho$ = 70~mas in 2019, with $\Delta m$ = 1.5) that is noted as having an orbit, and the ORB6 lists the orbital elements of \citet[][$P$ = 63.9~yr, $a$ = 175~mas, $i$ = 75$\fdg$0, $e$ = 0.37]{hvd20} for this pair.  \citet{chn12} determined this to be a SB2, but provides no additional information.  \citet{shb04} claimed the B component to be a binary of $P \approx 10.8$~d, but this result is not listed in any of the above catalogs.

There are 11 resolved observations (2005 -- 2014) of the AB pair listed in the INT4, including one AO observation \citep{oap10} that lends confidence in the position angles of the other (speckle) observations.  \citet{hvd20} provide additional four observations (2015) made at 692 and 880~nm.  \citet{wgp18}, searching archival $\it IUE$ spectra for hot sdO companions to Be stars, found no indication of such in this system.

\begin{figure*}[ht]
\epsscale{0.51}
\plotone{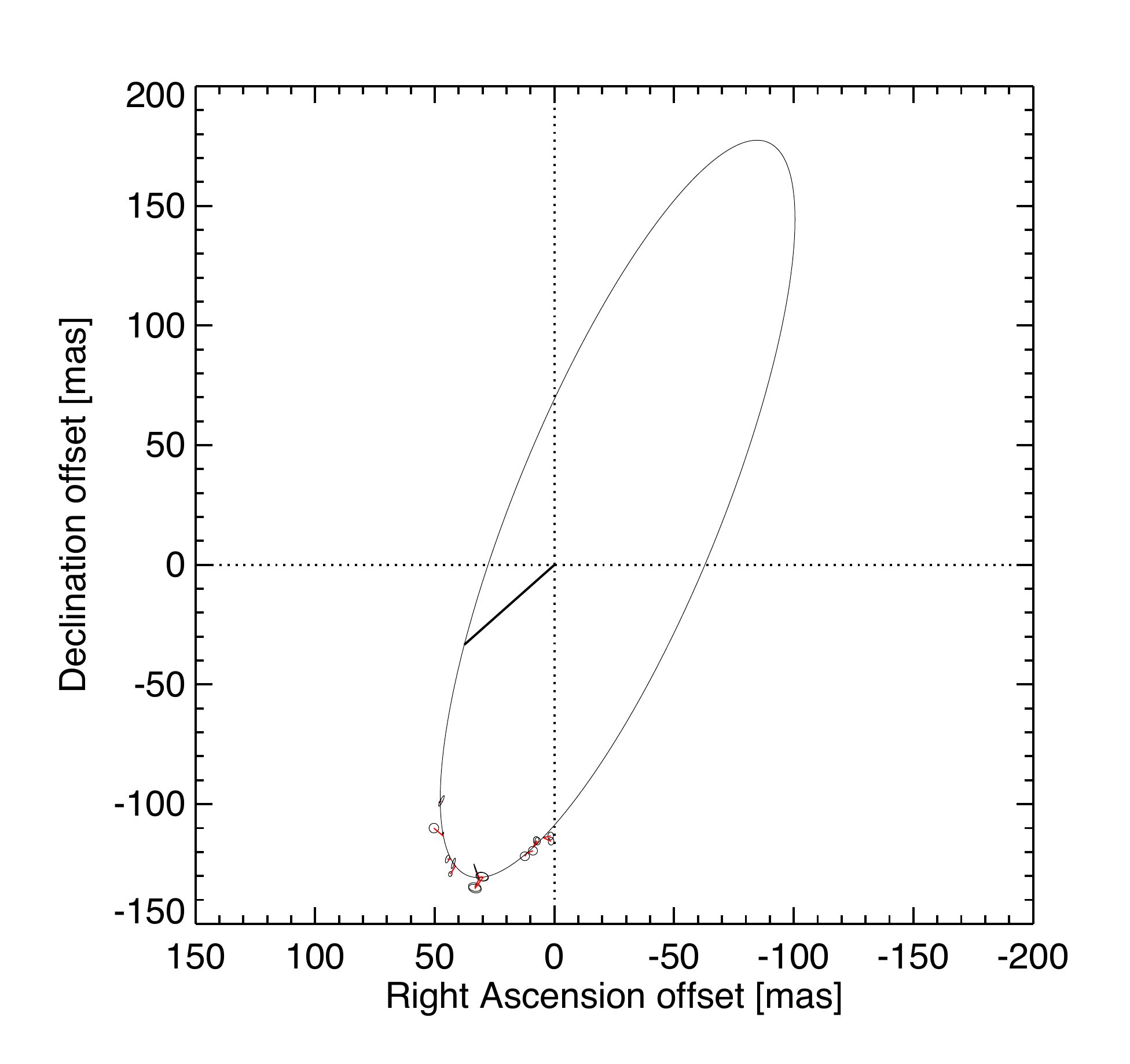}
\epsscale{0.505}
\plotone{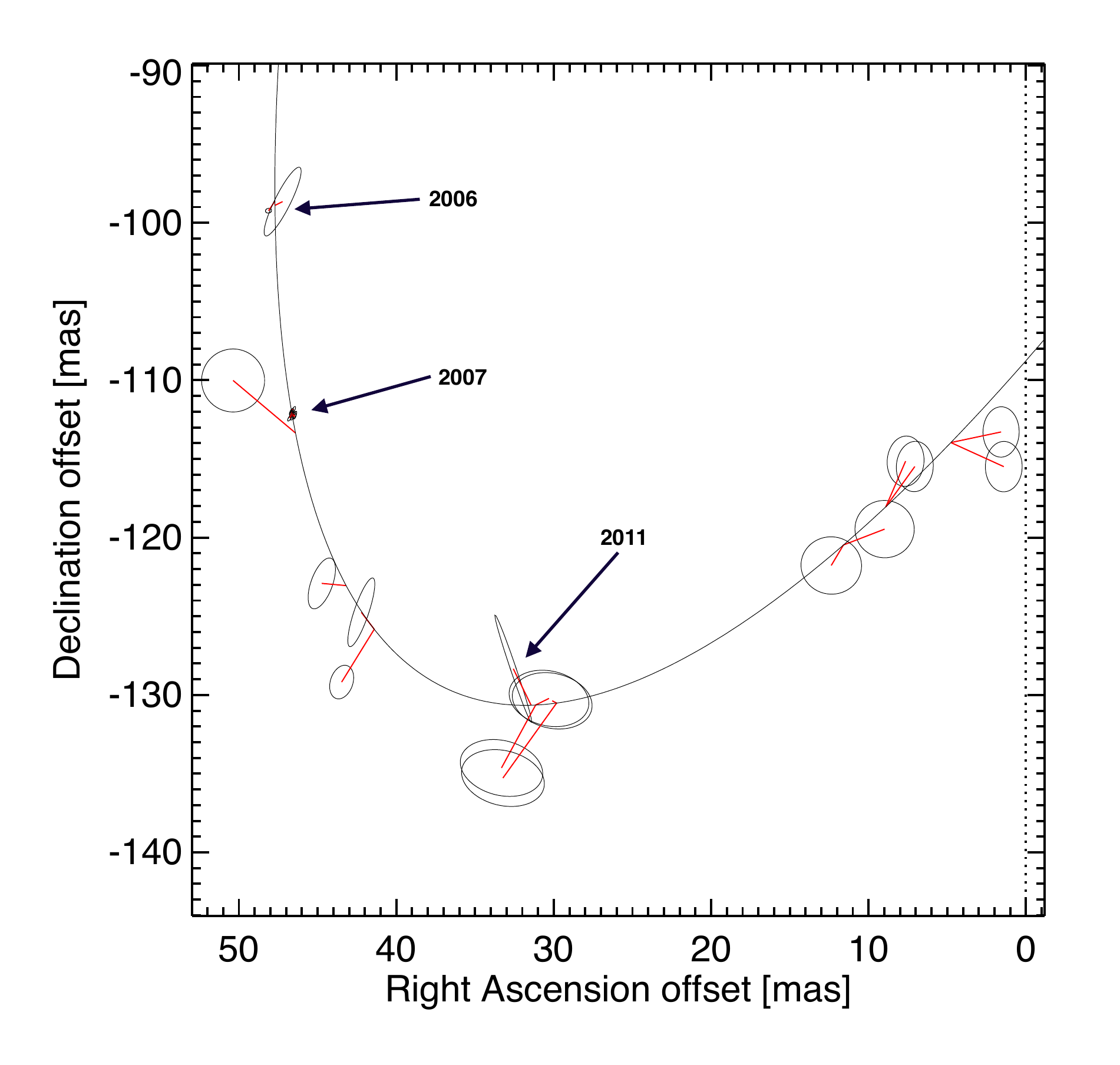}
\caption{Orbit of 66 Oph AB: ${\it Left\ Panel}$: Plot of the fitted relative positions from the 12 nights of NPOI observations (Table~\ref{programfit}) are shown, along with additional observations from the INT4 (and references therein) and \citet{hvd20}.  The apparent orbit corresponding to the best-fit orbital solution to these points (Table~\ref{66-Oph-elements}) is overplotted.  The line segment extending from the origin to the orbit marks periastron.  ${\it Right\ Panel}$: An expanded version of the lower portion of the left panel. Error ellipses are plotted to scale, with circles of radius 2 mas plotted for measurements lacking error estimates.  The NPOI observations are indicated by arrows.  Red line segments indicate the offset of each measurement with respect to the position predicted by the Table~\ref{66-Oph-elements} orbit at the epoch of observation.}
\label{66-Oph_plot}
\end{figure*}

We examined 21 nights of archival NPOI data (2006 -- 2011) using GRIDFIT, obtaining significant detections (\S~\ref{modeling}) on 12 nights during 2006, 2007, and 2011 (Table~\ref{progstarsobs}).  We then proceeded to detailed modeling of the data from each of the 12 nights with OYSTER using fixed stellar angular diameters of $\theta_{P}$ = 0.3 mas (from the NPOI planning software) and $\theta_{S}$ = 0.13 mas, estimated from the component spectral type (``early A'') suggested by \citet{hvd20}, and stellar radii given by \citet{sck82}, along with the nightly $\rho$ and $\theta$ values and the average $\Delta$$m_{700}$ obtained from GRIDFIT as inputs.  The results of the nightly fits are presented in Table~\ref{programfit} and Figure~\ref{66-Oph_plot}. 

%% 66 Oph astrometry fits

%% Table 6 -- 66 Oph orbital elements

\begin{deluxetable*}{lrcl}
%%\tabletypesize{\small}
%%\tablewidth{0pt}
%%\tabletypesize{\scriptsize}
\tablecaption{66 Oph AB -- Orbital Elements \label{66-Oph-elements}}
\tablewidth{0pt}
\tablehead{\\
\colhead{Parameter} & {} & \colhead{Value} \\
}
\startdata
${\it a}$ (mas)			    &     178.36	&	$\pm$	&   1.37	\\
${\it e}$				    &      0.412	&	$\pm$	&   0.006	\\
${\it i}$ (deg)			    &      75.90	&	$\pm$	&   0.69	\\
$\omega$ (deg)		        &     115.24	&	$\pm$	&   0.95	\\
$\Omega$ (deg)		        &     338.87	&	$\pm$	&   0.31	\\
${\it P}$ (days)			&	 23421.1	&	$\pm$	&   4.1 	\\
${\it T}$ (JD - 2440000.0)	&	 12658.5	&	$\pm$	&   50.2	\\
\enddata
\end{deluxetable*}

The weighted mean of the component magnitude difference fits for the 11 of 12 nights for which that parameter could be meaningfully modeled was $\Delta$$m_{700}$ = 2.61 $\pm$ 0.02, which is consistent with the $\Delta m$ values listed in the INT4 for observations made at similar effective wavelengths.  We can use the $\Delta$$m_{700}$ value, along with the spectral type B2Ve \citep[][and BeSS]{hvd20} and $\it v$sin$\it i$ = 262 km~s$^{-1}$ (BeSS) for the primary, and the above cited orbital inclination, to estimate some of the properties of the binary components.  The models of \citet[][and references therein]{toh04} for rapidly rotating, gravity darkened B stars, indicate a mass of 9.6 $M_{\sun}$ and a critical rotation velocity $\it v_{c}$ = 475 km~s$^{-1}$ for a B2 spectral type.  Therefore, for $\it i$ = 75$\arcdeg$, $\it v$ = 271 km~s$^{-1}$ and $\it v$/$\it v_{c}$ = 57\%.  Utilizing Figure 3 of \citet{toh04} for this spectral type and $\it i$ and $\it v$/$\it v_{c}$ values yields an absolute visual magnitude for the Be primary of $M_{V}$(primary) $\approx -2.8$. Our $\Delta m = 2.61$ value then implies $M_{V}$(secondary) $\approx -0.2$, for which the tables of \citet{sck82} indicate, for a main sequence star, a spectral type $\approx$ B8 and $\it M$(secondary) = 3.8 $M_{\sun}$.  The resulting mass sum ($\approx$ 13.4 $M_{\sun}$) is similar to that estimated by \citet[][$\sim$ 14 $M_{\sun}$]{hvd20}.
 
Lastly, we fit a new orbit (Table~\ref{66-Oph-elements} and Figure~\ref{66-Oph_plot}) to the combination of our nightly position fits, those of \citet{hvd20}, and all but one of the resolved INT4 observations, the excluded observation having a very large separation error \citep[$\pm$ 50 mas,][]{oap10}.  The orbital elements of Table~\ref{66-Oph-elements} are quite similar to those of \citet{hvd20}, but with significantly smaller errors, and provide a significantly better fit to all the observations prior to 2011 (Figure~\ref{66-Oph_plot}).

From the available evidence, we conclude 66 Oph consists of only two physical stellar components, the AB pair.

\newpage

\subsubsection{$\beta$ Lyr} \label{betlyr}

$\beta$ Lyr (HIP 92420, HR 7106, FK5 705, HD 174638, WDS 18501+3322Aa,Ab): This star does not appear in the LIN2 or DMSA.  The WDS lists the wide pairs AB through AF, and BE, BF, and EF ($\rho$ $>$ 45$\arcsec$ and $\Delta m$ $\geq$ 0.48) without comment as to their being physical.  Additionally, it lists the close pairs Aa,Ab ($\theta$ = 176$\fdg$3, $\rho$ = 540 mas in 2002, with $\Delta m$ = 4.6) without comment, and Aa1,2 ($\theta$ = 72$\fdg$3, $\rho$ = 1 mas in 2007, with $\Delta m$ = 0.4) noting it as having an orbit.  The Aa1,2 pair constitutes a well-known interacting and eclipsing binary extensively studied since its discovery by \citet{goo85}.  The WDS Notes list Aa1,2 as an eclipsing binary of $P = 12.9$~d and as having been resolved by CHARA \citep{zgm08}, and the ORB6 lists three sets of orbital elements for this pair \citep[][$P = 12.9414$~d, $a = 0.865$ -- 0.993~mas, $e = 0.0$]{zgm08} resulting from CHARA observations.  \citet{zgm08} used the CHARA observations, along with elements obtained from RV and light-curve studies, and adopting $P = 12.9414$~d, to derive $i = 92.25\arcdeg \pm 0.82\arcdeg$, $\Omega = 254.39\arcdeg \pm 0.83\arcdeg$, $a = 0.865 \pm 0.048$~mas, a distance of $312 \pm 17$~pc, and masses for the primary (brighter) and secondary (fainter) stars of $M_{p}$ = 2.83 $\pm$ 0.18 $M_{\sun}$ and $M_{s}$ = 12.76 $\pm$ 0.27 $M_{\sun}$, respectively.  According to \citet[][and references therein]{har02}, the primary or ``donor'' is a B6-B8 II Roche lobe filling, mass-loosing star, while the secondary or ``gainer'' is an early B-type mass-gaining star completely hidden inside an opaque accretion disk.\footnote{While the definition of a classical Be star discussed in \S~\ref{intro} \citep{rcm13} specifically excludes systems with on-going mass transfer between binary components, we nevertheless included $\beta$ Lyr in our sample~(\S~\ref{targsel}) for consistency with the broader definition of a classical Be star used by \citet{nei11}.}  The SB9 lists system 1092 with four orbital solutions with $P \approx 12.91$--$12.94$~d and $e = 0.0$--$0.04$, and notes that the period is known to be increasing by about 18 s/yr.  The MSC lists pairs Aa1,2, Aa,Ab, and AB as all physical (four stars in total), and pair AF as nonphysical, but notes possibly discrepant GAIA DR2 parallaxes \citep{gaia16,gaia18b} for the components of pair AB.  The INT4 lists pair Aa,Ab as unresolved to $<$ 30 mas on eight occasions (1921 -- 1992) by various techniques, but resolved at the above cited PA and separation, and $\Delta m$ = 4.53, on one occasion (2002.6731) at 900 nm by AO \citep{rbr07}.  \citet{rbr07} note that the earlier observations would likely have been insensitive to such a large $\Delta m$ companion.  The INT4 also lists the seven resolved observations of the Aa1,2 pair by \citet{zgm08}.

\begin{figure}
\epsscale{1.0}
\plotone{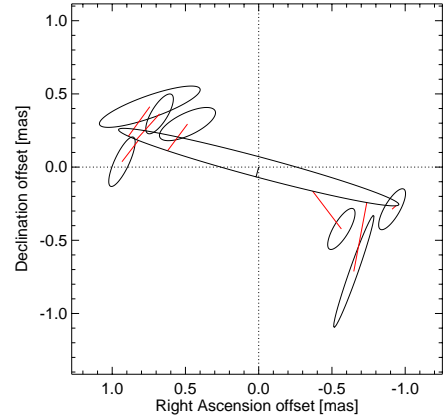}
\caption{Orbit of $\beta$ Lyr Aa1,2: Plot of the fitted relative positions at each of the seven epochs of NPOI observations (Table~\ref{programfit}) with error ellipses plotted to scale, along with the apparent orbit of \citet{spt09}.  The red line segments indicate the offset of each measurement with respect to the position predicted by the orbit of \citet{spt09} at the epoch of observation.}
\label{bet-Lyr_plot}
\end{figure}

The NPOI data archive includes an extensive series of observations spanning the years 1997 -- 2013.  Observations from 2005 and 2013 were used in the modeling of the extended emission in the Aa1,2 system \citep{spt09,mbn18}, as well as determining the parameters of the binary orbit \citep{spt09}.  Here, we report on the analysis of 16 additional nights of observations made in 2002, 2006, and 2007.  The data from all 16 nights were examined using GRIDFIT, with 10 nights having significant binary detections (\S~\ref{modeling}).  We then proceeded to detailed modeling of the data from each of these 10 nights with OYSTER using fixed stellar angular diameters of $\theta_{P}$ = 0.62 mas and $\theta_{S}$ = 1.04 mas \citep{zgm08}, along with the nightly $\rho$ and $\theta$ values from GRIDFIT and $\Delta m_{700}$ = 1.20 \citep{spt09} as input parameters. The results of the nightly fits for the seven nights with reasonable fitting errors ($\leq$ 1~mas) are presented in Table~\ref{programfit} and Figure~\ref{bet-Lyr_plot}.  

%% bet Lyr astrometry fits

The fitted binary positions generally show significantly better agreement with the \citet{spt09} orbit as compared to the \citet{zgm08} orbit, the O-C values with respect to the \citet{spt09} orbit being in the range 0.04 -- 0.50 mas.  As can be seen in Figure~\ref{bet-Lyr_plot}, the binary was detected only at epochs at or near maximum angular separation, not surprising given the longest baselines used for the observations (Table~\ref{progstarsobs}) provided a maximum angular resolution of 0.7 -- 0.9 mas (Table~\ref{baselines}).  The weighted mean of the component magnitude difference fits for the six nights where this parameter could be meaningfully modeled was $\Delta m_{700}$ = 1.10 $\pm$ 0.09, in good agreement with \citet[][$\Delta V = 1.3 \pm 0.1$ and $\Delta R = 1.2 \pm 0.1$]{spt09}.  Simple calculations, based on known properties of the donor star and the accretion disk, can be used to test the reasonableness of these $\Delta m$ results. The average angular diameter of $\approx$ 0.57~mas based on the major and minor axes of the donor star \citep{zgm08} and the distance of $\approx$ 295~pc (Table~\ref{targetlist}) can be used to infer a radius for the donor of $\approx$18.1 $R_{\sun}$ (more consistent with the B6-8II spectral classification of \citet{har02} than the B7Ve value given in the BeSS database).  This radius, along with an effective temperature $T_{\rm eff}$ = 13000~K \citep{har02} and the relations of \citet{sck82}, can be used to derive $M_{\rm bol} = -5.17$, BC $\approx -0.9$, and hence $M_{V}$(donor) = $-4.27$ for this star.  An empirical model for the accretion disk rim by \citet{apl00}, fitted to UBV and OAO2 photometry of $\beta$ Lyr, and requiring a $T_{\rm eff}$ of not less than 9000~K, predicts a bolometric luminosity for the rim of $5.58 \times 10^{36}$ erg $s^{-1}$.  These values imply BC $\approx -0.25$ \citep{sck82} and $M_{\rm bol}$ = $-3.17$, respectively, and hence $M_{V}$(disk) = $-2.92$ for the disk rim.  Therefore, a value of $\Delta V$  $\approx$ 1.35 for the Aa1,2 pair is obtained, similar to the measured values.

From the available evidence, we conclude the $\beta$ Lyr system consists of four physical stellar components: Aa1, Aa2, Ab, and B.

\subsubsection{$\upsilon$ Sgr} \label{upssgr}

$\upsilon$ Sgr (HIP 95176, HR 7342, FK5 727, HD 181615, WDS 19217-1557): This star does not appear in the LIN2, DMSA, or MSC.  While long known as a SB1 \citep{cbl99}, the first positive determination of the secondary orbit was achieved by the application of cross-correlation methods to high-resolution UV spectra obtained with $\it IUE$ \citep{daj90}.  While this star does not appear in the ``main'' WDS catalog\footnote{By the ``main WDS Catalog'' we refer to the file containing the positions (J2000), discoverer designations, epochs, position angles, separations, magnitudes, spectral types, proper motions, Durchmusterung numbers, and links to notes for the components of more than 154000 systems.}, the WDS Notes list it as WDS 19217-1557 and describe the \citet{bnu11} orbit as combining the spectroscopic elements of \citet{khy06} with VLTI interferometric data of \citet{net09} and CHARA array data.  The ORB6 lists the orbital elements of \citet[][including $P$ = 137.9343~d, $a$ = 2.1~mas, $i$ = 50$\arcdeg$, $e$ = 0.0]{bnu11}, while the SB9 lists it as system 1147, including the orbital solution of \citet{khy06}, also with $P$ = 137.9343~d and $e$ = 0.0. \citet[][and references therein]{sgd18} identify the brighter component of the pair as a ``stripped-envelope'' (SES) helium giant, thought to be in a very late phase of helium shell burning following the loss of its hydrogen-rich envelope via Roche-lobe overflow.  \citet{daj90} determined a minimum mass for the primary of $M_{p}$sin$^3 i = 2.52 \pm 0.05 M_{\sun}$ and noted that \citet{dat81} proposed this star could be the progenitor of a core-collapse supernova if $M_{p} > 3 M_{\sun}$.  They further determined a mass ratio of $q = M_{p}$/$M_{s}=0.63 \pm 0.01$, and therefore, a minimum mass for the secondary of $M_{s}$sin$^3$i  = 4.02 $\pm$ 0.10 $M_{\sun}$.  Thus, while it has the higher mass of the two stars, the ``secondary'' was initially detected only through its UV excess.  \citet{bnu11} demonstrated that the H$\alpha$ line forming region in the system is very extended ($\ge 6$~mas), but is off-center from the primary star, following the orbital motion of the secondary.  The circumbinary material was also detected from the strong infrared excess of $\upsilon$ Sgr; LBOI observations in the the mid-IR were modeled by \citet{net09} as dust confined to a geometrically thin disk at an inclination angle $i$ of $50^{+10}_{-20}$ degrees. However, the interferometric observations reported in \citet{net09} and \citet{bnu11}, while useful in characterizing the circumbinary disk and thus helpful in constraining the orbit, did not directly resolve the stellar components. Surprisingly, \citet{wgp18}, searching archival $\it IUE$ spectra for hot sdO companions to Be stars, found no indication of UV excess in the $\upsilon$ Sgr system.  The INT4 lists this star as unresolved by speckle interferometry in the visible on 24 dates (1976 -- 1991) placing a limit of $\rho$ $<$ 30~mas.

%% ups Sgr astrometry fits

\begin{figure}
\epsscale{1.0}
\plotone{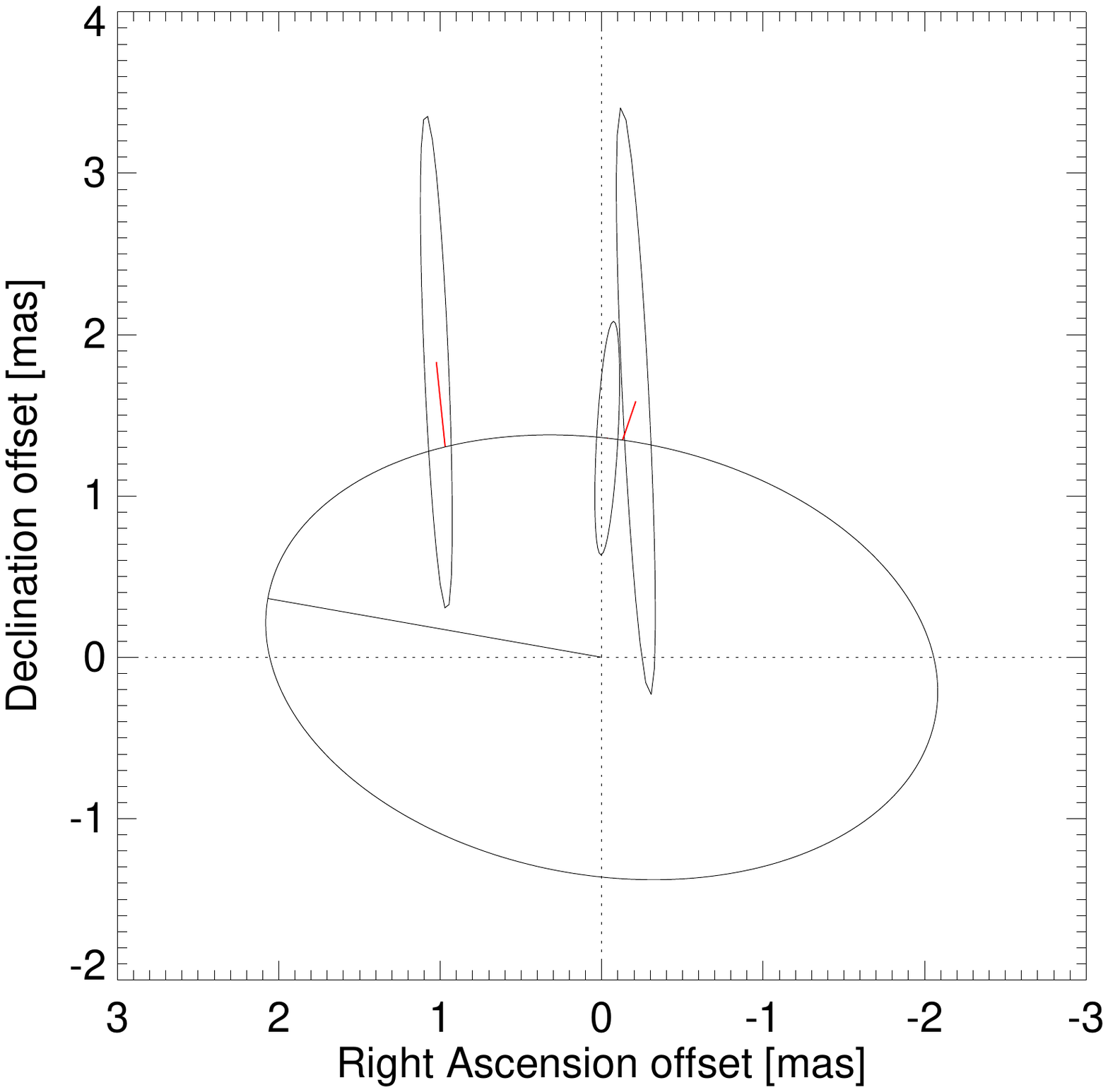}
\caption{Orbit of $\upsilon$ Sgr: Plot of the fitted relative positions at each of the three epochs of NPOI observations (Table~\ref{programfit}) with error ellipses plotted to scale, along with the apparent orbit of \citet{bnu11}.  The line segment extending from the origin to the orbit marks periastron, while the red line segments indicate the offset of each measurement with respect to the position predicted by the orbit of \citet{bnu11} at the epoch of observation.}
\label{ups-Sgr_plot}
\end{figure}

The modeling of the three available nights of archival data (Table~\ref{progstarsobs}) proved difficult as this binary lies near both the limits of angular resolution and $\Delta m$ sensitivity of the NPOI.  The application of the GRIDFIT procedure produced no significant $\chi_{\nu}^2$ minima (\S~\ref{modeling}), while CANDID produced minima at angular separations three times as large as those predicted by the \citet{bnu11} orbit, with position angles differing by $\pm$ 60$\arcdeg$ -- 80$\arcdeg$.  We next used the program analogous to GRIDFIT within OYSTER to search a 10~mas by 10~mas grid, at 0.1~mas spacings, centered on the relative position of the secondary component, as predicted by the orbit of \citet{bnu11}.  This resulted in minima at position angles within $\approx 10\arcdeg$ of those predicted on two of the three nights, and a similar angular separation on one night (2011 Jul 13 UT).  Based on the latter, somewhat encouraging result, we then proceeded to detailed modeling of the data from each of the three nights with OYSTER using the orbit of \citet{bnu11}, along with fixed stellar angular diameters of $\theta_{P}$ = 0.33~mas and $\theta_{S}$ = 0.06~mas \citep{bnu11}, and $\Delta m_{700}$ = 4.5 \citep[roughly the mid-point of the range suggested by][]{bnu11} as inputs.  The results of the fits are presented in Table~\ref{programfit} and Figure~\ref{ups-Sgr_plot}.  The O-C values of the final $\rho$, $\theta$ fits with respect to the orbit of \citet{bnu11} show good agreement ($<$ 0.5~mas), but as indicated by the error ellipses in Figure~\ref{ups-Sgr_plot}, these results are not well constrained in declination, which can be attributed to the array configuration used for the observations with shorter, lower resolution, north-south baseline components (Tables~\ref{baselines} \&~\ref{progstarsobs} and Figure~\ref{Be-star-stations}).  It should be noted, however, the closure phase data in our observations do provide some constraint on the quadrant of the fitted position angles on at least one night.  Thus, it appears our observations likely represent the first direct, visual resolution of the binary.

It was also possible to meaningfully fit the $\Delta m$ of the binary on two of the three nights, the mean value obtained ($\Delta m_{700}$ = 3.49$\pm$ 0.09) being consistent with the conclusion of \citet{bnu11} that the flux ratio of the stellar components is smaller than 0.1 from 500 to 700 nm.  We can also verify that our $\Delta m$ value is reasonable using other known properties of the stars. For example, \citet{kak12} determined an effective temperature $T_{\rm eff}$ = $12300\pm 200$~K and an absolute visual magnitude $M_{V}$ = $-4.73\pm 0.30$ for the SES primary star.  From the $T_{\rm eff}$ value, utilizing \citet{sck82}, we obtain a bolometric correction BC = $-0.8$ and thus a bolometric magnitude $M_{\rm bol} = -5.53$.  Again utilizing \citet{sck82} and these $M_{\rm bol}$ and $T_{\rm eff}$ values, we derive a stellar radius of $\approx$ 24 $R_{\sun}$, consistent with that determined by \citet[][$21 \pm 10 R_{\sun}$]{bnu11}. For the secondary star, our $\Delta m$ value and the $M_{V}$ for the primary of \citet{kak12} imply an absolute magnitude for the secondary star of $M_{V}$(secondary) = $-1.14$.  Assuming the maximum likely orbital inclination of $\it i$ = 60$\arcdeg$ \citep{net09}, $M_{s} \approx 6.2 M_{\sun}$ from \citet{daj90}, and further assuming a main sequence star, \citet{sck82} yields a spectral type $\approx$ B5 with $M_{V} = -1.2$, a good match.  Alternatively, the models of \citet[][and references therein]{toh04} for rapidly rotating, gravity darkened B stars, yield a spectral type closer to B6, with a mass $\approx$ 5 $M_{\sun}$.  However, this would require an even greater orbital inclination angle ($\it i$ $\sim$ 68$\arcdeg$) for consistency with the $M_{s}$sin$^3$i value reported by \citet{daj90}.

Based on the available evidence, we conclude the $\upsilon$ Sgr system consists of two physical stellar components.

\subsubsection{59 Cyg} \label{59cyg}

59 Cyg (HIP 103632, HR 8047, FK5 1551, HD 200120, WDS 20598+4731Aa,Ab): This star does not appear in the LIN2 or SB9.  The WDS lists wide pairs AB through AE ($\rho$ $>$ 21$\arcsec$ and $\Delta m$ $\geq$ 4.69) without comment as to their being physical.  Additionally, it lists the close pair Aa,Ab ($\theta$ = 1$\fdg$8, $\rho$ = 164 mas in 2008), noting it as having an orbit.  The ORB6 lists the orbital elements of \citet[][$P$ = 165.5~yr, $a$ = 208~mas, $e$ = 0.261]{mas11} for Aa,Ab.  The DMSA lists pair AE ($\theta$ = 33$\fdg$0, $\rho$ = 200~mas in 1991.25) with a type-F solution\footnote{Identified as ``fixed double or multiple system (identical proper motions and parallaxes).''}, but this appears to be the Aa,Ab pair as it is listed in the INT4 with the same $\theta$ and $\rho$ values.  The MSC lists pairs AB and AC as questionable CPM pairs and identifies Aa,Ab as a visual binary citing the \citet{mas11} orbital parameters, but additionally lists pair Aa1,Aa2 as a SB1 \citep[][with $P$ = 28.2~d]{hbp02, mrs05}.  \citet{mrs05} detected anti-phase Doppler shifts in the He II $\lambda$4686 absorption line, indicative of a star much hotter than the Be primary star \citep[classified as B1.5Vnne by][]{jrl68}.  Subsequently, \citet{ppg13} detected Aa2 as a hot sdO subdwarf, its spectral signature found in cross-correlation functions of photospheric model spectra with far-UV spectra obtained by $\it IUE$. The RVs from the cross-correlation functions were then used to determine a SB2 orbit ($P$ = 28.1871~d, $e$ = 0.141) and estimates for the stellar masses, radii, and temperatures.  For an estimated inclination range of $60\arcdeg$ -- $80\arcdeg$, they estimated the masses of the Be primary and sdO secondary at $M_{p}$ = 7.85 $\pm$ 1.55 $M_{\sun}$ and $M_{s}$ = 0.77 $\pm$ 0.15 $M_{\sun}$, respectively.  From these masses, along with the orbital period and HIPPARCOS system parallax (Table~\ref{targetlist}) we estimate an angular semimajor axis of $\approx$ 0.9 mas (or 0.6 mas using the GAIA parallax), which is significantly less than the nominal resolution of the longest baselines used in our observations ($\sim$ 1.5 mas; Tables~\ref{baselines} \&~\ref{progstarsobs}).  Likewise, we can use the masses, luminosities, and effective temperatures given in \citet{ppg13} to estimate the magnitude difference of the components.  Using the $M_{p}$ and luminosity estimate of \citet{ppg13} of log($L_{p}/L_{\sun}$) = 3.9 for the Be primary star, Table 1 of \citet{toh04} suggests a spectral type of $\approx$ B2.5.  This spectral type, along with the high inclination and $v\sin i = 379 \pm 27$~km~s$^{-1}$ values of \citet{ppg13}, allow the use of Figure 3 of \citet{toh04} to predict an absolute visual magnitude of $M_{V}$(primary) $\approx -2.3$.  Alternatively, using the B1.5 spectral class of \citet{jrl68} and Figure 3 of \citet{toh04}, a value of $M_{V}$(primary) $\approx -3.0$ is obtained.  For the sdO secondary, \citet{ppg13} estimate log($L_{s}$/$L_{\sun}$) = 3.0, which can be used to estimate \citep[e.g.,][]{sck82} an absolute bolometric magnitude $M_{\rm bol}$(secondary) = $-2.86$, while their value for the effective temperature $\log T_{\rm eff}$(secondary) = 4.7 can be used to predict a bolometric correction BC = $-4.6$ \citep[again using][]{sck82} and thus $M_{V}$(secondary) = +1.74.  Therefore, we estimate a magnitude difference $\Delta m_{V}$ for the Aa1,2 pair in the range 4.0 -- 4.7, again beyond the capabilities of the NPOI at the time of the observations.

Returning to the Aa,Ab pair, we note there are 26 resolved observations of this pair listed in the INT4, including those made by HIPPARCOS, speckle, and the NPOI (Paper I).  \citet{hvd20} provide additional observations (2015.8365) made at 692 and 880~nm.  We examined six nights of archival NPOI data between 2009 Jul 07 UT and 2009 Jul 22 UT using the standard GRIDFIT procedure, but obtained a significant $\chi_{\nu}^2$ minimum (\S~\ref{modeling}) on only one night, with the results for the six nights varying widely in separation and position angle.  As an alternative, we then used the program analogous to GRIDFIT within OYSTER, to search a grid of 50 mas by 50 mas at 0.5 mas spacings, centered on the relative position of the Ab component, as predicted by the orbit of \citet{mas11}.  This resulted in significant $\chi_{\nu}^2$ minima on two of the nights (Table~\ref{progstarsobs}).  We then proceeded to detailed modeling of the data from each of these nights with OYSTER using the orbit of \citet{mas11}, along with fixed stellar angular diameters of $\theta_{P}$ = 0.492 mas and $\theta_{S}$ = 0.2 mas (Paper I), the nightly $\rho$ and $\theta$ values from the OYSTER gridfit procedure, and $\Delta m_{700}$ = 2.83 (Paper I) as inputs.  The weighted mean of the nightly fits for the $\Delta m$ of the binary ($\Delta m_{700}$ = 3.01 $\pm$ 0.12) is consistent with that reported by \citet[][2.85 $\pm$ 0.13]{hvd20}.  We can also use the estimated mass for the Ab component from the MSC (3.29 $M_{\sun}$), and assuming a main sequence star, use the tables of \citet{sck82} to estimate a spectral type of $\sim$ B9 with an absolute magnitude $M_{V}$  = +0.2.  This value, along with our estimates of the absolute magnitudes of the components of the Aa1,2 pair, predict a magnitude difference $\Delta m_{V}$ for the Aa,Ab pair in the range 2.5 -- 3.2, again consistent with our measurements presented here and those in Paper~I.

\begin{figure*}[ht]
\epsscale{0.49}
\plotone{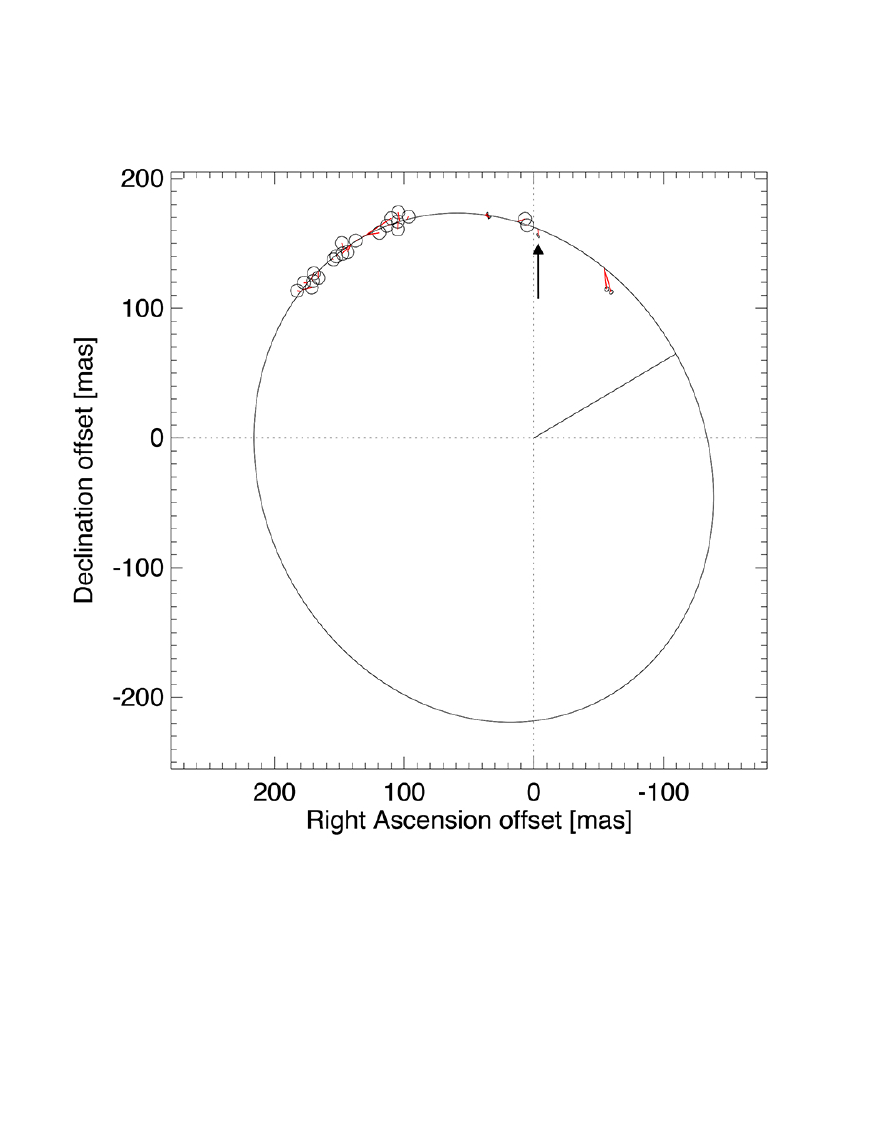}
\epsscale{0.5}
\plotone{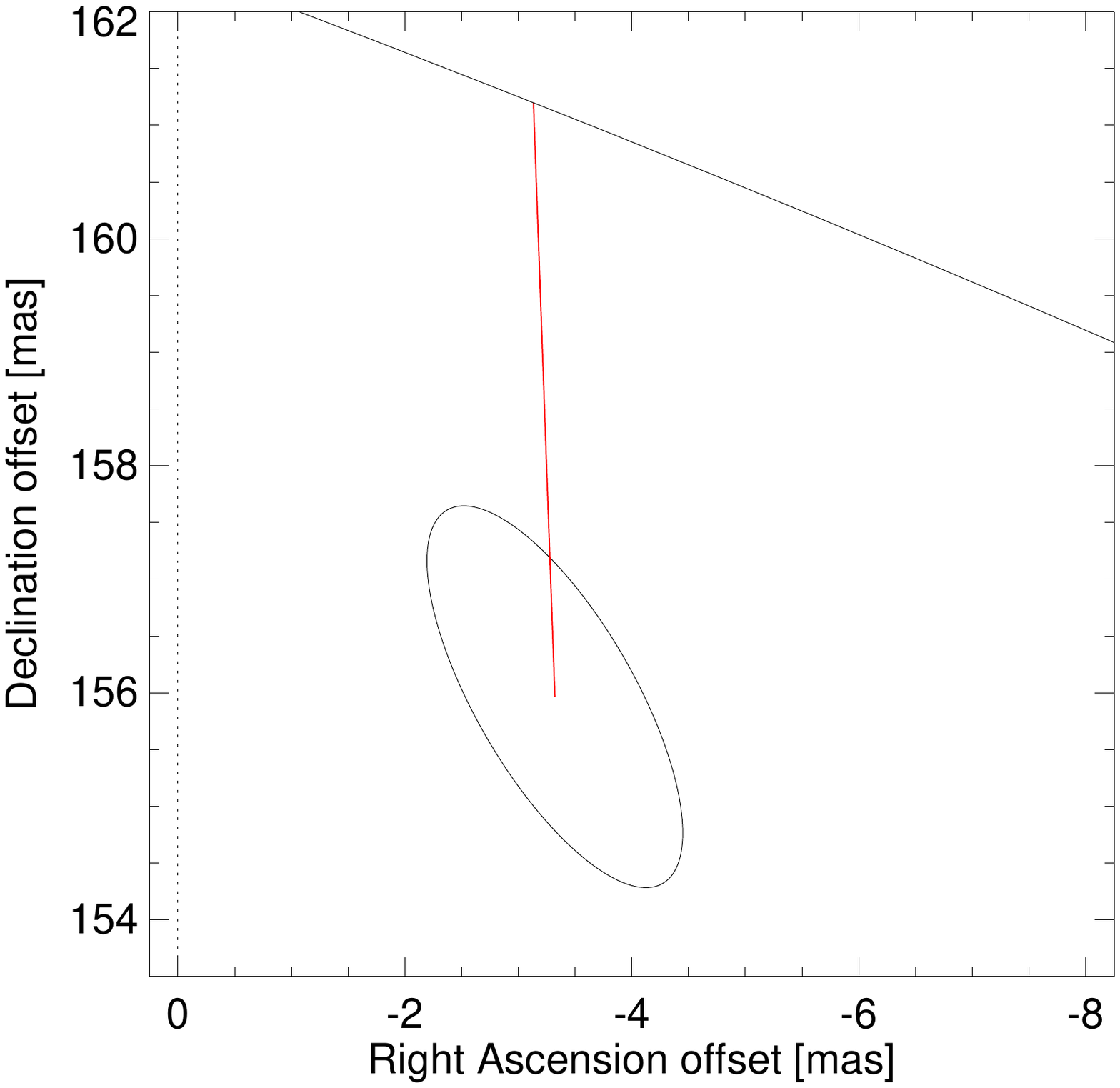}
\caption{Orbit of 59 Cyg Aa,Ab: ${\it Left\ Panel}$: Plot of the fitted relative position (indicated by arrow) from the combined data from the two epochs of NPOI observations (Table~\ref{programfit}) with error ellipse plotted to scale.  Additional observations from the INT4 and \citet{hvd20} are also plotted along with the apparent orbit of \citet{mas11}.  The line segment extending from the origin to the orbit marks periastron, while the red line segments indicate the offset of each measurement with respect to the position predicted by the \citet{mas11} orbit at the epoch of observation.  Circles of radius 5 mas are plotted for INT4 measurements lacking error estimates.  ${\it Right\ Panel}$: An expanded version of the upper, center portion of the left panel.  The red line segment indicates the offset of the fitted relative position for the combined 2009 Jul 07 UT and 2009 Jul 22 UT observations from the \citet{mas11} orbit.}
\label{59-Cyg_plot}
\end{figure*}

However, the independently fitted relative positions of the components on these two nights (2009 Jul 07 and Jul 22) differed by several milliacseconds, unexpected for contemporaneous observations of such a long-period binary.  Examining the $V^2$ data from the individual baselines, we see that their modulations over the spectral channels are generally very small (less than a few percent) on all but the shortest (AC-AE) baseline.  This is not unexpected given the relatively long baselines used for the observations (Table~\ref{progstarsobs}) and the large $\Delta m$ of the Aa,Ab pair, and likely explains the difficulty in modeling the relative astrometry.  In a further attempt to improve our result, the combined data from both nights were again modeled using OYSTER gridfit, this time on a grid of 10 mas by 10 mas at 0.1 mas spacings.  The results of this last fit are presented in Table~\ref{programfit} and Figure~\ref{59-Cyg_plot}, the errors of the fit estimated from the $\chi_{\nu}^2$ surface plot over the grid area. 

%% 59 Cyg astrometry fits

The O-C value of the final $\rho$, $\theta$ fit with respect to the orbit of \citet{mas11} is large ($\approx$ 5.2 mas), but this result, in combination with the even larger offset of the later observations of \citet[][PA $\approx$ 333$\arcdeg$ in left panel of Figure~\ref{59-Cyg_plot}]{hvd20}, may indicate the beginning of a significant divergence from the \citet{mas11} orbit.    

Based on the available evidence, we conclude 59 Cyg is likely a hierarchical multiple consisting of three components (Aa1, Aa2, and Ab, with our observations resolving only Aa,Ab).

\newpage

\subsubsection{$\beta$ Cep} \label{betcep}

$\beta$ Cep (HIP 106032, HR 8238, FK5 809, HD 205021, WDS 21287+7034Aa,Ab): This star does not appear in the LIN2 or DMSA.  The WDS lists pair AB ($\theta$ = 251$\fdg$40, $\rho$ = 13$\farcs$500 in 2016, with $\Delta m$ = 5.46) without comment as to it being a physical pair, and pair Aa,Ab ($\theta$ = 226$\fdg$00, $\rho$ = 172 mas in 2007, with $\Delta m$ = 3.40) as having an orbit, which is listed in the ORB6 \citep[$P$ = 83.0~yr, $a$ = 195~mas, $i$ = 87$\fdg$3, $e$ = 0.732;][]{and06}.  The primary component of this pair (Aa) is identified as the prototype variable star of a class of rapid pulsators \citep{ggs19}, while the secondary (Ab) is a classical Be star \citep{wos09}.  The SB9 lists system 1310 as a SB1, including the orbital solution of  \citet[][$P$ = 10.893~d, $e$ = 0.52]{fit69}, but notes difficulty separating orbital velocities from the variable star pulsations, and that the evidence in this case for a binary nature is ``not completely convincing.''  The MSC likewise lists subsystem (Aa1,Aa2) with these orbital parameters as doubtful, but also lists Aa,Ab with the orbital parameters of \citet{and06} and lists the AB pair as a CPM system.  The INT4 lists resolved speckle observations of the Aa,Ab pair on 59 dates (1971  -- 2007; 38 mas $\leq$ $\rho$ $\leq$ 264 mas).

We examined 13 nights of archival NPOI data (1997 Jul 03 UT -- 2002 Jun 24 UT, Table~\ref{progstarsobs}) using the program analogous to GRIDFIT within OYSTER to search a grid of 25 mas by 25 mas at 0.3 mas spacings, centered on the position of the secondary star, as predicted by the orbit of \citet{and06}. The $\chi^2$ surfaces in the OYSTER gridfit output for the nights in 1997 and later displayed many closely-spaced local minima in the $\rho$, $\theta$ space of similar depth, and thus the position measurement on those nights was ambiguous.

\begin{figure*}[p]
\epsscale{0.4}
\plotone{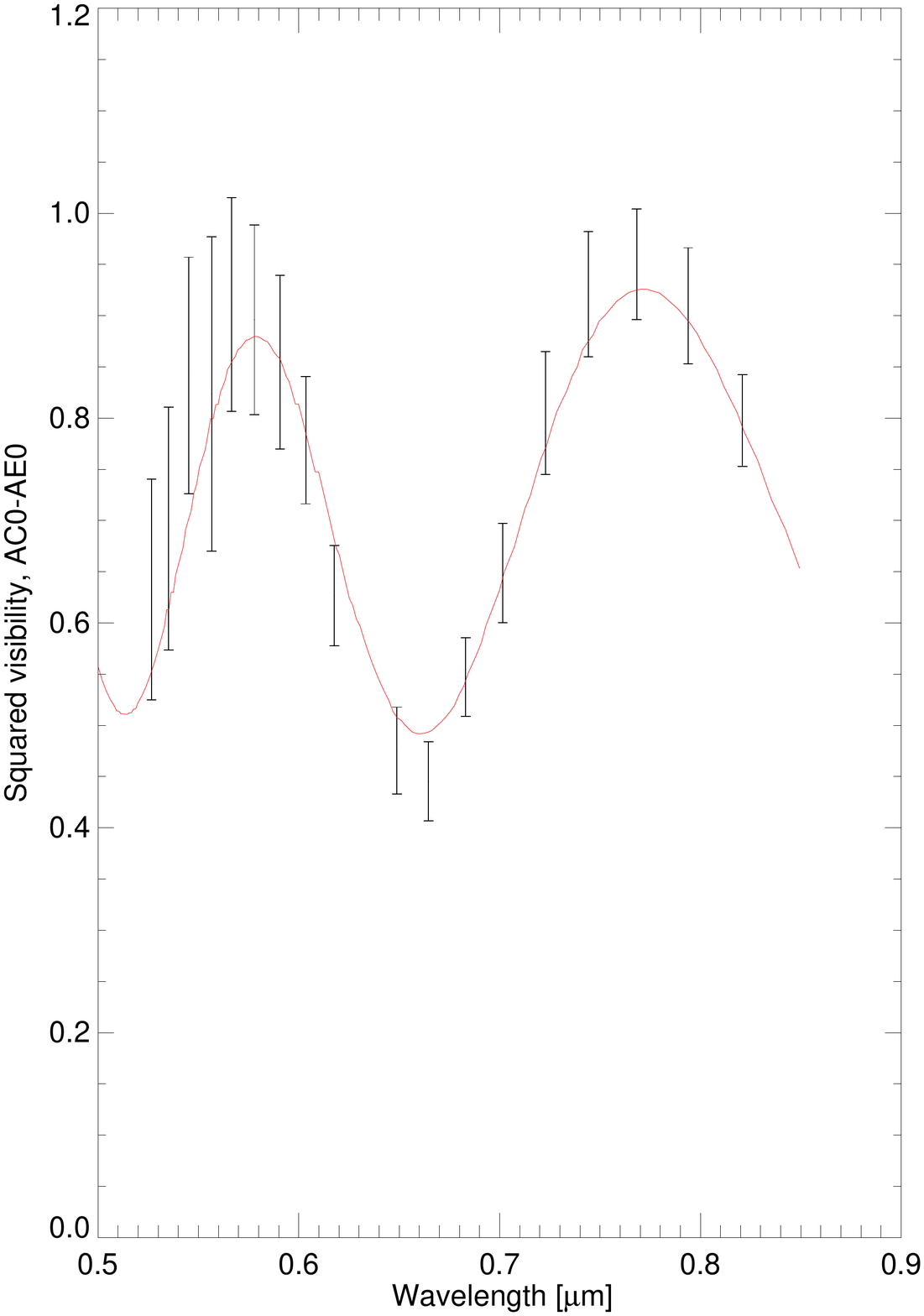}
\epsscale{0.4}
\plotone{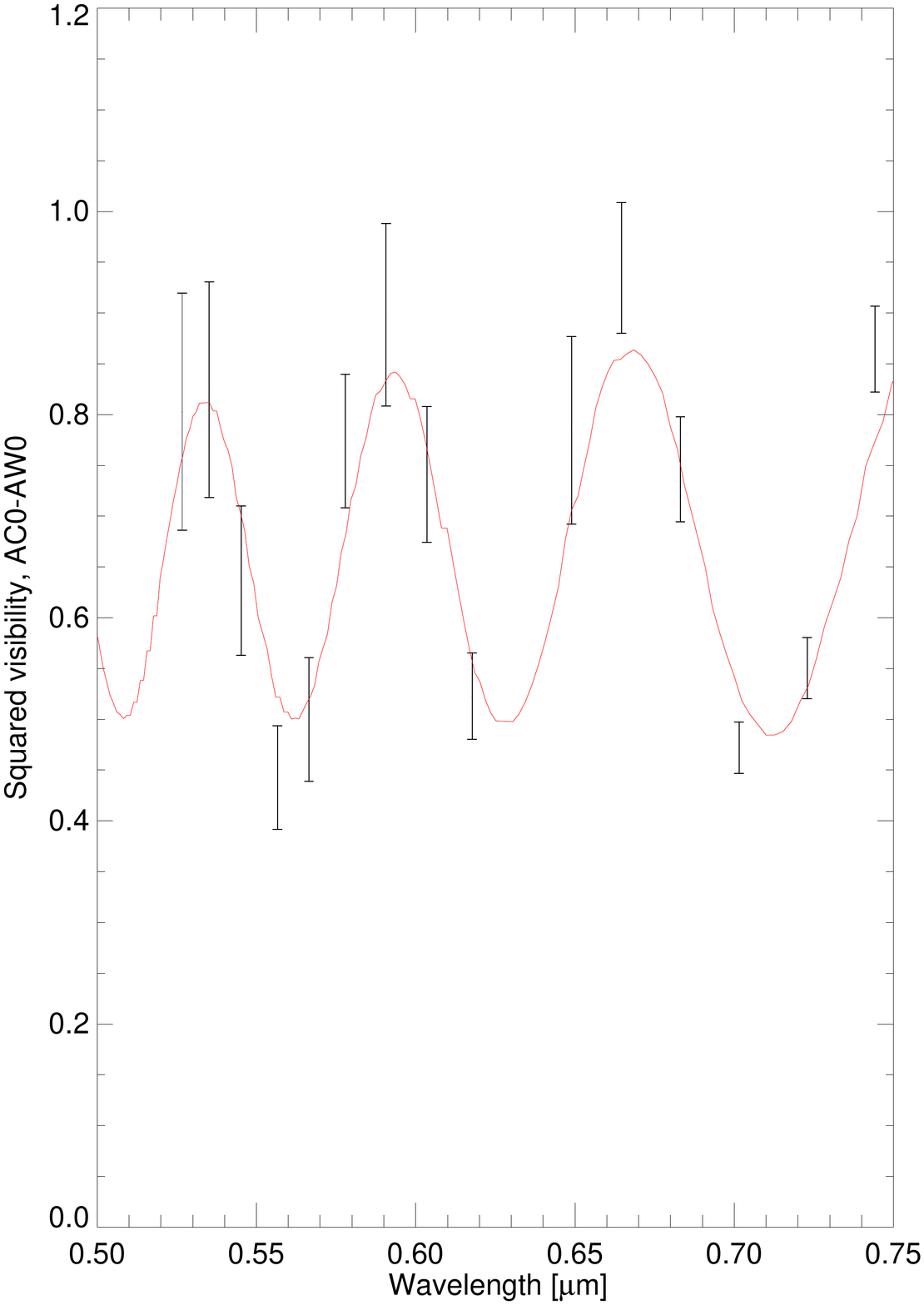}
\epsscale{0.4}
\plotone{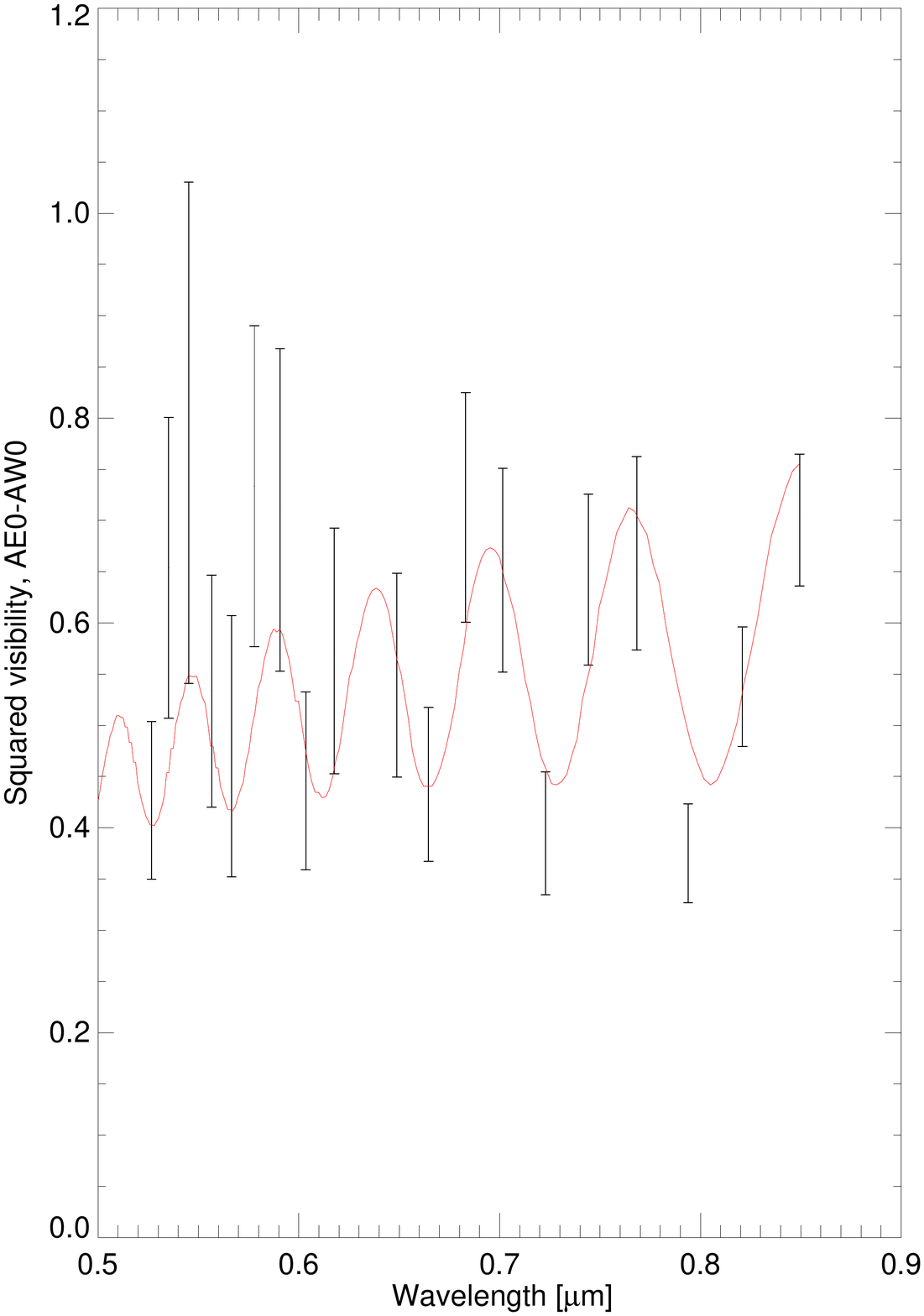}
\epsscale{0.4}
\plotone{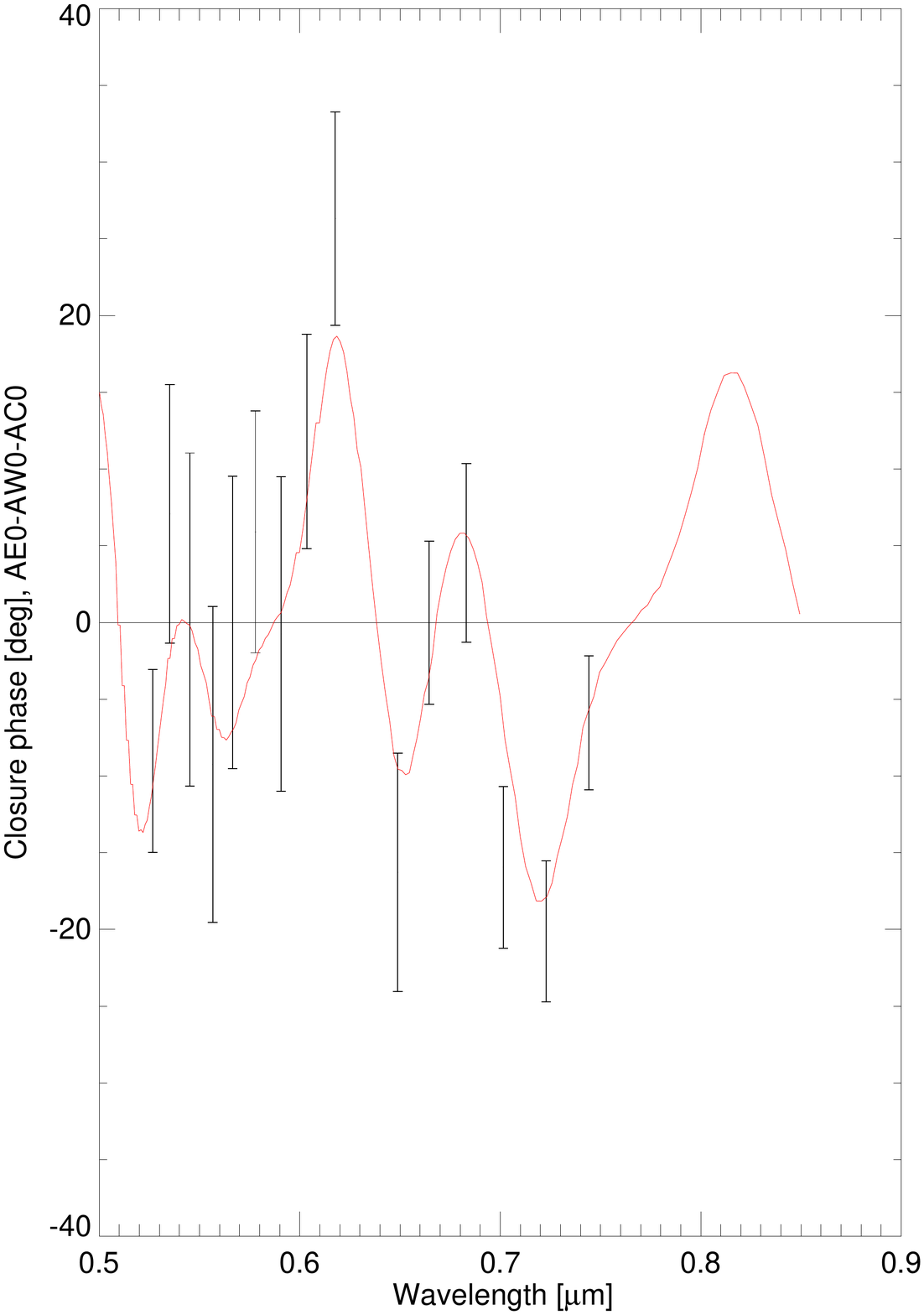}
\caption{Fringe data of $\beta$ Cep Aa,Ab: ${\it Upper\ Left\ Panel}$: Plot of the $V^{2}$ data from a single observation on the AC-AE baseline on 1997 Jul 03 UT.  ${\it Upper\ Right\ Panel}$: Plot of the $V^{2}$ data from the simultaneous observation on the AC-AW baseline.  ${\it Lower\ Left\ Panel}$: Plot of the $V^{2}$ data from the simultaneous observation on the AE-AW baseline.  ${\it Lower\ Right\ Panel}$: Plot of the resulting closure phase data (degrees) on the AC-AE-AW triangle.  The red curve in all four plots represents the equivalent quantities calculated from the best-fit model (Table~\ref{bet-Cep-elements}).} 
\label{bet-Cep-vis-phase_plot}
\end{figure*}

We then devised a method based on the differential correction formulae of \citet{lfr94} and \citet{hwd78} for spectroscopic and visual orbits, respectively.  The equations express the deviation of the observed quantity for each observation (RV, or $\rho$ and $\theta$, respectively) from that predicted by an initial ``input'' orbit as linear combinations of (initially unknown) differential corrections to the orbital elements\footnote{Orbital elements of d$\gamma$ (RV of center of mass), d$K_{1}$ or d$K_{2}$, d$\omega$, d$e$, d$T$, and d$\mu$ (mean motion = 2$\pi/P$) for spectroscopic orbits, or d$\Omega$, d$\omega$, d$i$, d$e$, d$T$, d$\mu$, and d$a$ for visual orbits.}, with the values of the ``constants'' calculated from the elements of the input orbit and the true anomaly at the time of the observation.  The collected expressions for all observations are then used to simultaneously solve for the corrections to all ten elements, each equation of condition first being multiplied by the square root of the weight assigned to that measurement.  After the first such solution, measurements with residuals larger than 3$\sigma$ were assigned zero weight, and the weights of the remaining measurements were adjusted following the work of \citet{iyw96} at each iteration.  In the case of $\beta$ Cep, spectroscopic data taken from the literature \citep{smk53, amg94, hah96, sal03, cal08, hdv13}, along with the resolved observations from the INT4 and all the NPOI observations, plus 10 Mark III observations, (Table~\ref{progstarsobs}) were utilized, with the visual orbit of \citet{and06} and the spectroscopic orbit ($\gamma$ and $K_1$ only) of \citet{pab92} as inputs. The application of these procedures resulted in an orbit solution predicting visibilities and closure phases in excellent agreement with the data in most cases (Figure~\ref{bet-Cep-vis-phase_plot}).

%% bet Cep -- visibilities and closure phase

The orbit solution was further refined by returning it to the standard OYSTER modeling procedures, simultaneously fitting all of the Mark III and NPOI visibility data (Table~\ref{progstarsobs}), along with the RV data cited above. The resulting orbital elements, along with fitted angular diameters of the components and derived properties, are listed in Table~\ref{bet-Cep-elements}.

%% Table 7 -- bet Cep orbital elements

\begin{deluxetable}{lrcl}
%%\tabletypesize{\scriptsize}
\tabletypesize{\small}
\tablecaption{$\beta$ Cep Aa,Ab -- Orbital Elements and \\ Stellar Properties \label{bet-Cep-elements}}
\tablewidth{0pt}
\tablehead{\\
\colhead{Parameter} & {} & \colhead{Value} \\
}
\startdata
${\it a}$ (mas)                 &     206.96    &	$\pm$	&	0.07	\\
${\it e}$				        &   0.747765	&	$\pm$	&	0.00008	\\
${\it i}$ (deg)			        &	   88.80	&	$\pm$	&	0.02	\\
$\omega$ (deg)  		        &     202.43	&	$\pm$	&	0.02	\\
$\Omega$ (deg)	        	    &	  227.83	&	$\pm$	&	0.03	\\
${\it P}$ (days)			    &   29626.21	&	$\pm$	&	0.0004	\\
${\it T}$ (JD - 2440000.0)	    &	50943.38	&	$\pm$	&	0.004	\\
$\gamma$ (km $s^{1}$)	        &	   -6.36	&	$\pm$	&	0.16	\\
$K_{1}$ (km $s^{1}$)	        &	    9.63	&	$\pm$	&	0.26	\\
With RVs:				        &	    		&			&			\\
${\it P}$ (days)			    &	29616.54	&	$\pm$	&	0.0004	\\
${\it T}$ (JD - 2440000.0)      &	50944.45	&	$\pm$	&	0.004	\\
With HIPPARCOS parallax:  &   			&			&			\\
$M_{1}$ $(M_{\sun})$        	&        7.4	&	$\pm$	&	1.5		\\	
$M_{2}$ $(M_{\sun})$	        &        5.0	&	$\pm$	&	1.0		\\
\enddata
\end{deluxetable}

\begin{figure*}[ht]
\epsscale{0.8}
\plotone{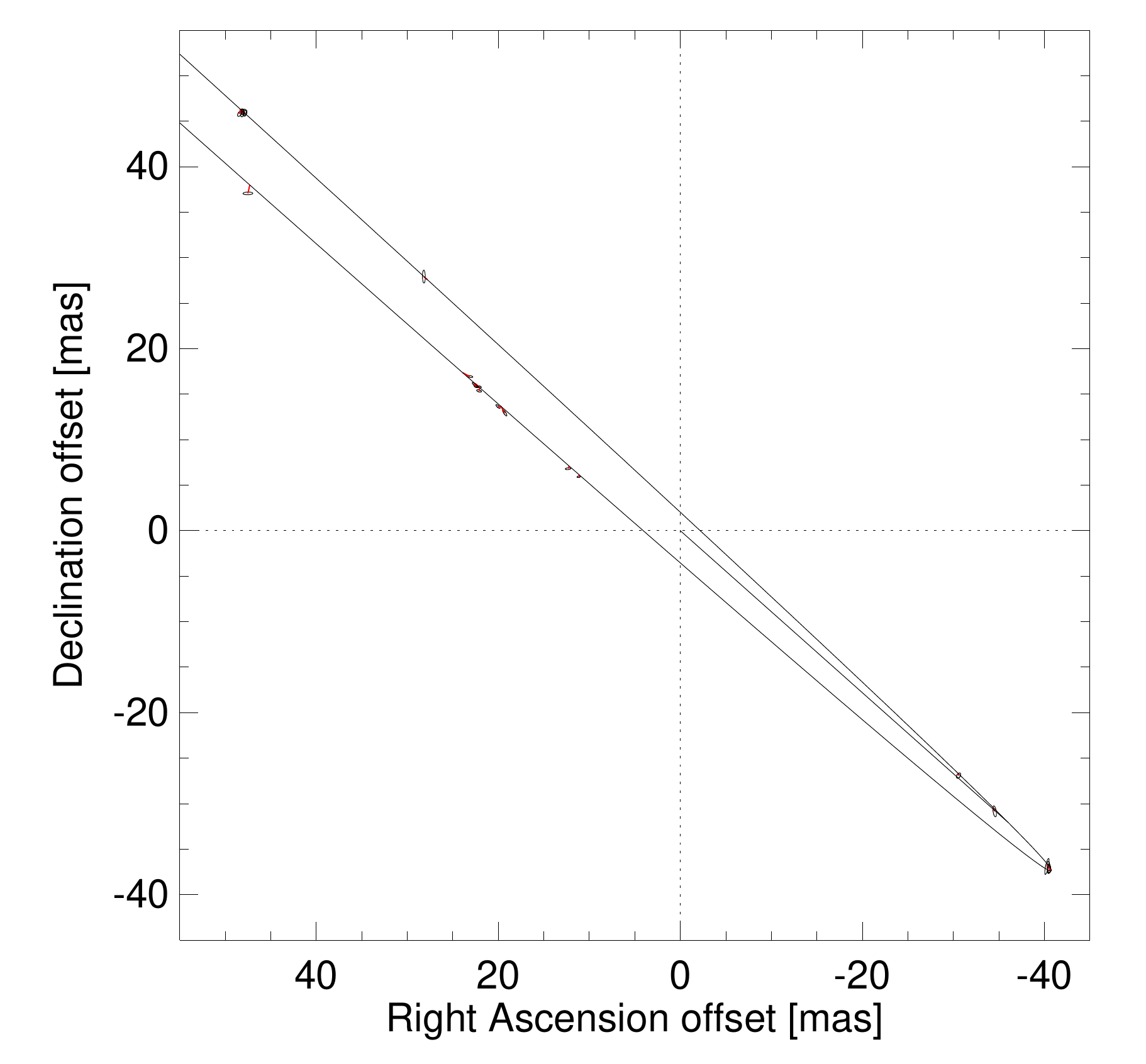}
\epsscale{0.4}
\plotone{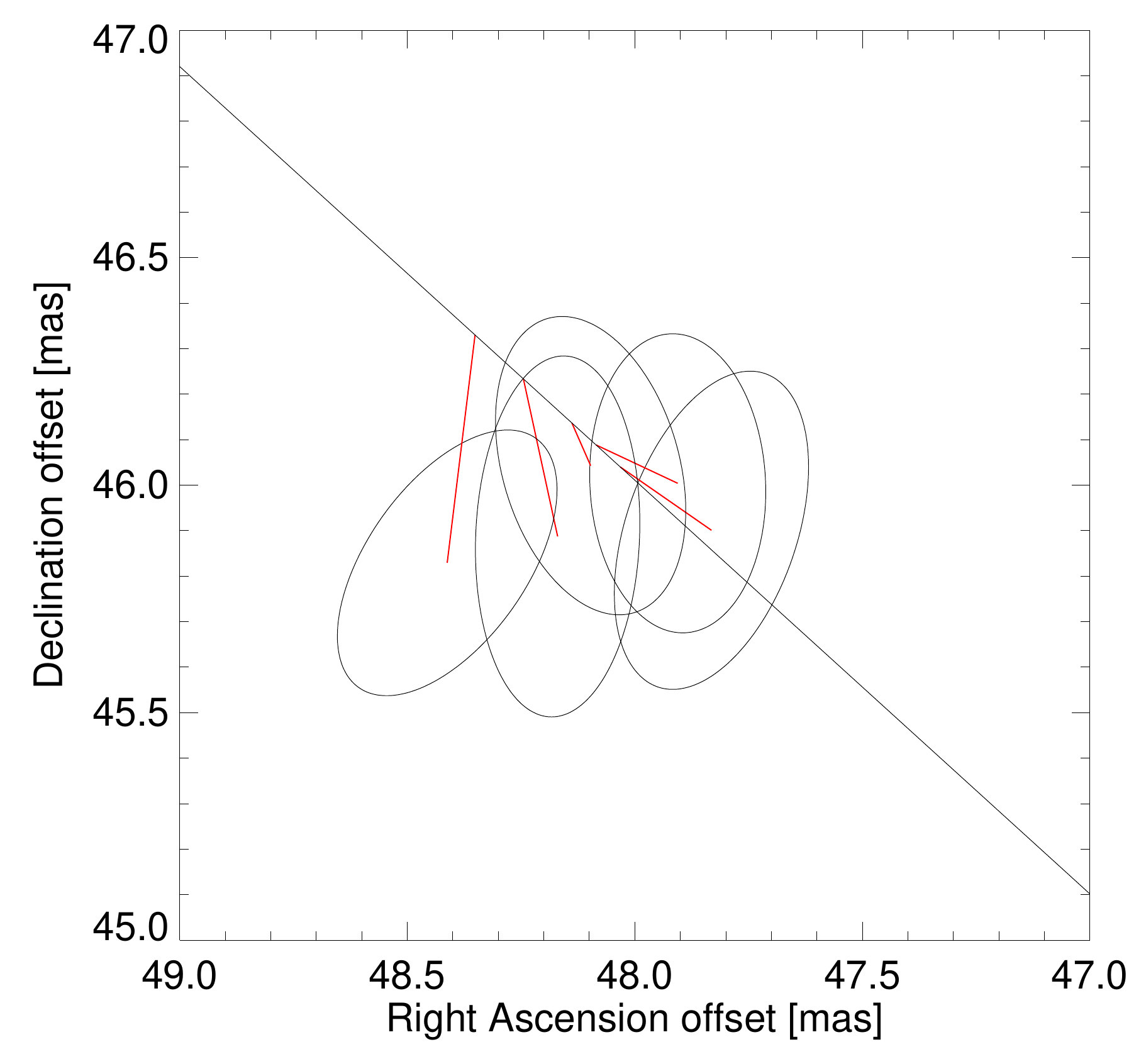}
\epsscale{0.4}
\plotone{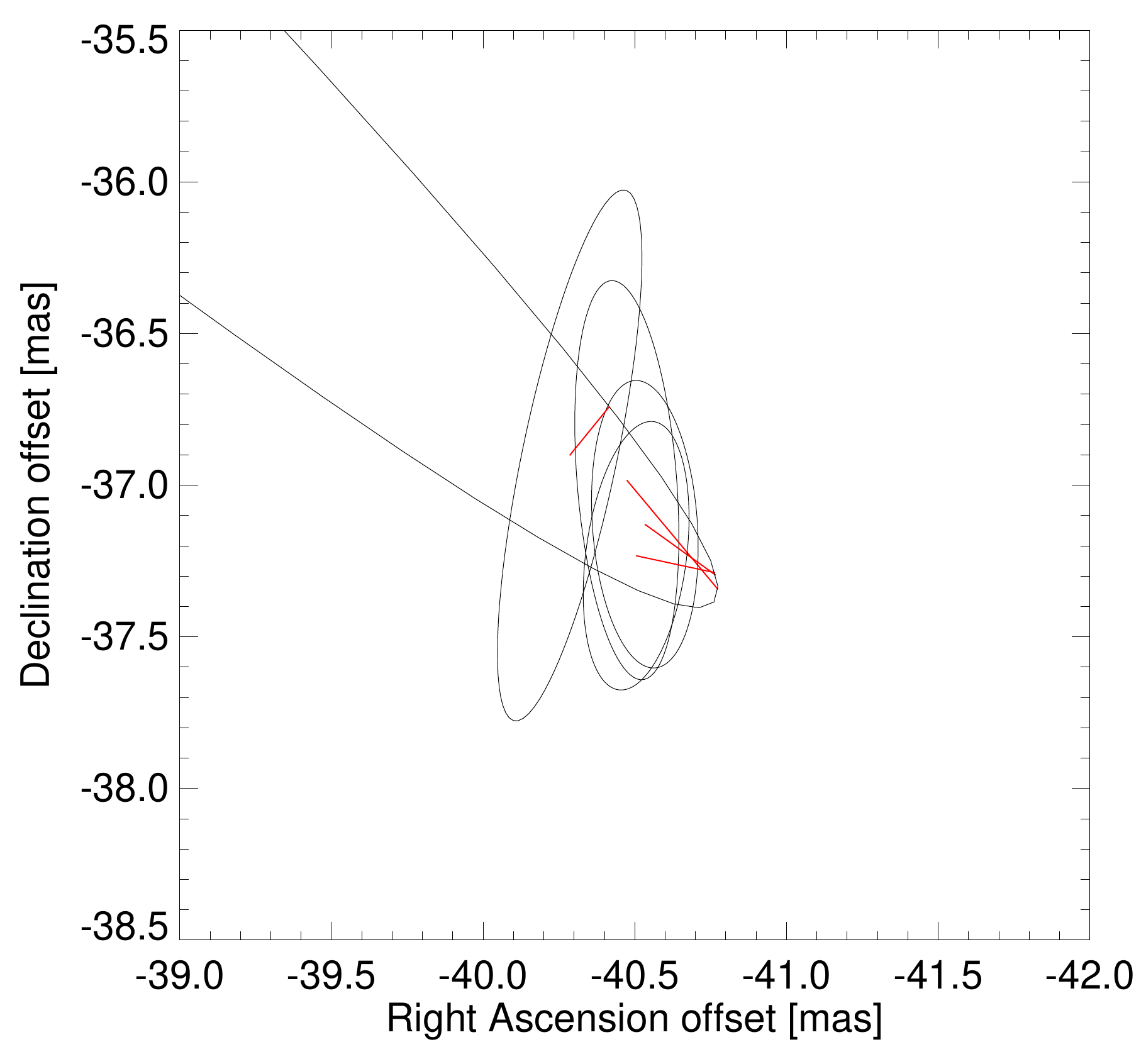}
\caption{Orbit of $\beta$ Cep Aa,Ab: ${\it Upper\ Panel}$: Plot of the fitted relative positions at each of the 23 epochs of Mark III and NPOI observations (Table~\ref{programfit}).  The apparent orbit corresponding to the best-fit orbital solution to these points (Table~\ref{bet-Cep-elements}) is overplotted.  The line segment extending from the origin to the orbit marks periastron.  ${\it Lower\ Left\ Panel}$: An expanded version of the upper left portion of the upper panel.  The red line segments indicate the offset of the fitted relative positions for 2002 June observations from the predicted orbital positions based on the revised orbit~(Table~\ref{bet-Cep-elements}).  ${\it Lower\ Right\ Panel}$: An expanded version of the lower right portion of the upper panel.  The red line segments indicate the offset of the fitted relative positions for 1997 Jul -- Sep UT from the Table~\ref{bet-Cep-elements} orbit.} 
\label{bet-Cep_plot}
\end{figure*}

\begin{figure*}[ht]
\epsscale{0.5}
\plotone{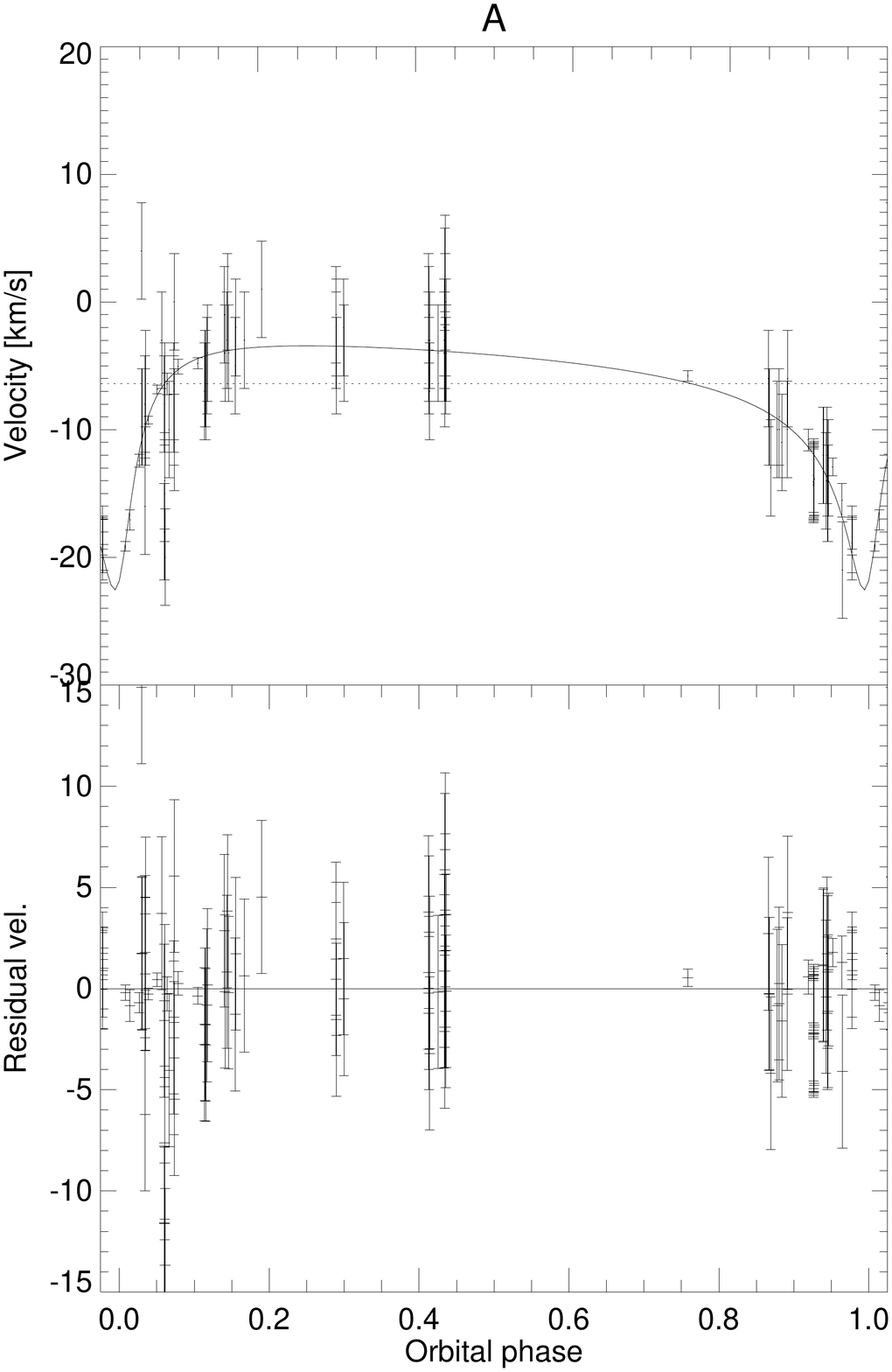}
\caption{Radial velocities of $\beta$ Cep Aa,Ab: ${\it Upper\ Panel}$: Plot of radial velocities taken from the literature (\S~\ref{betcep}), with the RV curve predicted by the best-fit model (Table~\ref{bet-Cep-elements}) overplotted.  ${\it Lower\ Panel}$: Residuals of the observed RVs from the model curve in the upper panel.}
\label{bet-Cep_rv}
\end{figure*}

Since RV data only exist for the primary, the HIPPARCOS parallax (4.76 $\pm$ 0.3 mas; Table~\ref{targetlist}) was used as a constraint to get the total mass, and therefore also the masses of both components.  The final  orbit is plotted in Figure~\ref{bet-Cep_plot}, and the model RV curve is shown in Figure~\ref{bet-Cep_rv}.  

%% bet Cep -- astrometry fits

%% bet Cep -- RVs

The O$-$C values of the nightly $\rho$, $\theta$ fits (Table~\ref{programfit}) with respect to the new orbit show generally good agreement (0.18 -- 0.93 mas).  The fitted component magnitude differences $\Delta m_{500}$ = 2.07 $\pm$ 0.30, $\Delta m_{550}$ = 2.03 $\pm$ 0.20, and $\Delta m_{800}$ = 1.84 $\pm$ 0.10 (from a simultaneous fit to the visibility data from all nights) are smaller than most of the $\Delta m$ values listed in the INT4 ($\Delta m$ $\sim$ 3.0 -- 3.5), but consistent with the $\Delta m$ = 1.8 in the B and R bands determined ``spectroastrometrically'' by \citet{wos09}.  Our mass for the secondary component (the Be star: $M_{2}$  = 5.0 $\pm$ 1.0 $M_{\sun}$; Table~\ref{bet-Cep-elements}) is intermediate between that of \citet[4.4 $\pm$ 0.7 $M_\sun$]{wos09} and \citet[6.2 $\pm$ 0.3 $M_\sun$]{dwb01}.  The models of \citet{toh04} for rapidly rotating, gravity darkened B stars indicate for our secondary star mass a spectral type $\approx$ B6, and with the high inclination of our orbit (assuming the stellar rotation axis is normal to the plane of the orbit), an absolute visual magnitude of $M_{V}$ $\approx -1.1$.  Our $\Delta m_{550}$ = 2.03 ($\approx$ $\Delta m_{V}$) would then predict $M_{V}$ (primary) $\approx -3.1$, for which the tables of \citet{sck82} indicate, assuming a giant, a spectral type B3III.  This is a later spectral type than various previous estimates which range from B0.5IIIs \citep{aac83} to B1IV \citep{hof91}, and B2IIIev \citep[][Table~\ref{targetlist}]{nei11}.  \citet{wos09} determined a spectral type of B2III and a mass of 12.6 $\pm$ 3.2 $M_{\sun}$, while \citet{dwb01} suggest a mass estimate of 12 $\pm$ 1$M_{\sun}$, both significantly larger than our mass for the primary ($M_{1}$  = 7.4 $\pm$ 1.5 $M_{\sun}$; Table~\ref{bet-Cep-elements}).

Based on the available evidence, we conclude $\beta$ Cep is likely a hierarchical triple system consisting of components Aa, Ab, and B.

\newpage

\subsection{Known Binaries - Not Detected by NPOI} \label{knownbin-nondet}

\subsubsection{$\gamma$ Cas} \label{gamcas}

$\gamma$ Cas (HIP 4427, HR 264, FK5 32, HD 5394, WDS 00567+6043AB): This star does not appear in the LIN2, DMSA, or SB9.  However, the WDS lists pairs AB ($\theta$ = 259$\fdg$0, $\rho$ = 2$\farcs$07 in 2002, with $\Delta m$ = 8.7) and AD ($\theta$ = 178$\fdg$6, $\rho$ = 1274$\farcs$52 in 2010, with $\Delta m$ = 3.41) as physical, and pair AC ($\theta$ = 345$\fdg$3, $\rho$ = 53$\farcs$26 in 2015, with $\Delta m$ = 10.7) without comment (we assume it to be nonphysical).  The WDS Notes mention the very small relative motion of B despite the high proper motion of A reveals AB to be a CPM pair.  The INT4 lists a single AO observation of AB \citep[$\theta$ = 259$\fdg$0, $\rho$ = 2$\farcs$07 and $\Delta m$ = 6.85 at 0.9 $\mu$m on 2002.6845]{rbr07}.    

Additionally, the A component consists of three stars: First, the WDS Notes mention the ``photocentric motion of (probably) the Aa pair'' detected by \citet{gon01}, and the ORB6 \citep{gon00} lists $P = 60$~yr, $a$ = 150.0~mas for this astrometric binary.  The MSC identifies this as the Aab,Ac pair, but also lists Aa,Ab ($P$ = 203.59 d) as an SB1 \citep{hhs00}.  \citet{nhk12} confirmed the period of Aa,Ab at $P$ = 203.52~d ($e$ = 0) for the latter pair through 16.84 years of optical spectral monitoring.  Adopting the inclination value $\it i$ = 45$\arcdeg$ and the primary star mass $M_{1}$ = 13 $M_{\sun}$ suggested by \citet{hhs00}, they estimated the secondary star mass $M_{2}$ = 0.98 $M_{\sun}$ and suggested it might be a ``hot helium star'' that could be directly detectable in the UV.  However, \citet{wgp17}, searching archival $\it IUE$ spectra for hot sdO companions to Be stars, found no indication of such in the $\gamma$ Cas system ($<$ 0.6\% contribution to total UV flux).  It is noteworthy that the mass sum for pair Aa,Ab (13.98 $M_{\sun}$) along with the orbital period of \citet{nhk12}, and the HIPPARCOS parallax listed in Table~\ref{targetlist} predict an angular semimajor axis of $\it a$ = 9.7 mas, certainly within the range of angular separation resolvable by NPOI observations.  Likewise, we can predict a range for the component magnitude difference ($\Delta m_{V}$) from plausible properties of the stars.  For a primary mass $M_{1}$ = 13 $M_{\sun}$, Table 1 of \citet{toh04} predicts a spectral type of B1, similar to that listed in Table~\ref{targetlist} (B0IVpe, taken from BeSS).  This table also predicts a critical rotation velocity $\it v_{c}$ = 502 km~s$^{-1}$, so the value of $\it v$sin$\it i$ = 432 km~s$^{-1}$ listed in the BeSS database for $\gamma$ Cas indicates close to $\it v_{c}$ rotation.  Adopting these values for the spectral type, inclination, and $\it v$/$\it v_{c}$, Figure 3 of \citet{toh04} then predicts an absolute visual magnitude $M_{V}$(primary) $\approx$ $-3.7$.  If the secondary is a main sequence dwarf, the tables of \citet{sck82} predict, for $M_{2}$ = 0.98 $M_{\sun}$, a spectral type $\sim$ G2 and an absolute magnitude $M_{V}$(secondary) $\approx$ +4.7, and thus $\Delta m_{V}$ $\approx$ 8 for $\gamma$ Cas Aa,Ab, far beyond the reach of the current NPOI.  Alternatively, if the secondary star were a sdO subdwarf, its mass would be intermediate between those of the sdO stars $\phi$ Per (\S~\ref{phiper}) and 59 Cyg (\S~\ref{59cyg}).  Were the luminosity and $T_{\rm eff}$ also of intermediate values, a bolometric magnitude $M_{\rm bol}$(secondary) = $-3.86$, bolometric correction BC = $-4.7$, and thus $M_{V}$(secondary) = +0.84 and a system $\Delta m_{V}$ $\sim$ 4.5 would be predicted, also beyond the capabilities of the current NPOI.

The INT4 lists component A as unresolved on 10 dates (1975 -- 1994) implying $\rho < 24$~mas in the visible, and \citet{hvd20} also lists it as unresolved (2015.8338) at a 692 nm with detection limit of $\Delta m$ of 3.66 at 200~mas.  The five nights of NPOI observations of $\gamma$ Cas listed in Table~\ref{progstarsobs} were modeled with GRIDFIT with null result; while the GRIDFIT minima met our significance criteria (\S~\ref{modeling}), the indicated angular separations were well below the angular resolution of the baselines used in the observations (Table~\ref{baselines}), hence indicating a single source.  Therefore, the NPOI observations failed to resolve either the astrometric Aab,Ac pair or the SB1 Aa,Ab pair.

The WDS Notes further mention that the D component of WDS 00567+6043 is the A component of WDS 00568+6022 (HIP 4440, HR 266, HD 5408).  The HIPPARCOS re-reduction \citep{vL07} provides very similar (within 1$\sigma$) parallaxes and proper motions for these components indicating that the two stars have common space motions.  The DMSA lists the AB pair of WDS 00568+6022 as a ``linear double or multiple system'' (solution type ``L''), while the MSC lists it as a visual binary with a computed orbit ($P$=83.1~yr, $\rho$=245~mas).  As in WDS 00567+6043, the A component of WDS 00568+6022 consists of three stars detected as astrometric and spectroscopic binaries.  The astrometric binary (Aa,Ab in the WDS Notes) has orbital elements listed in the ORB6 \citep{doc06} with $P = 4.85$~yr, $a$ = 32~mas, and $\Delta m$ = 0.71, while the MSC identifies it as Aab,Ac.  \citet{cfh92} lists this pair as Bab,c, and it also appears as system 1726 in the SB9.  The Aa component (per WDS Notes) is in turn a SB2 ($P = 4.24$~d) discovered by Young \citep{phy21} and is system 49 in the SB9 \citep[who list it as Bab]{cfh92}, whereas MSC lists the SB2 as pair Aa,Ab. The INT4 lists speckle observations in the visible of the astrometric-visible binary on five dates between 1994 and 2000 (16~mas $\leq$ $\rho$ $\leq$ 30~mas), as well as $\sim$ 75 observations of AB (1923 -- 2010) at 117 mas $\leq$ $\rho$ $\leq$ 423 mas covering the full range of position angles.  WDS 00568+6022 does not appear in the LIN2. 

Based on the available evidence, we conclude the combined WDS 00567+6043 plus WDS 00568+6022 system is a hierarchical multiple consisting of eight components.

\subsubsection{$\zeta$ Tau} \label{zettau}

$\zeta$ Tau (HIP 26451, HR 1910, FK5 211, HD 37202): This star does not appear in the WDS, WDS Notes, LIN2, ORB6, DMSA, or MSC.  The SB9 lists system 344 \citep{und52} as a SB1 with $P$ = 132.91~d, but with orbital elements that ``must be considered uncertain because of the superposed effects of the shell on the spectrum.''  Subsequently, \citet{rbh09} improved the parameters of the orbit ($P$ = 132.987~d, $e$ = 0) using H$\alpha$ spectroscopic and UBV photometric observations spanning about a century.  Adopting $M_{1}$ = 11 $M_{\sun}$ \citep{har84} for the primary star, and several plausible values of the orbital inclination, they estimated the secondary mass ($M_{2}$) to be in the range of 0.87 -- 1.02 $M_{\sun}$.  Assuming the mid-range value for $M_{2}$ (0.95 $M_{\sun}$ for $\it i$ = 75$\arcdeg$), the resulting mass sum, along with the orbital period of \citet{rbh09}, and the HIPPARCOS parallax listed in Table~\ref{targetlist} predict an angular semimajor axis of $\it a$ = 8.5 mas, certainly within the range of the NPOI observations.  Likewise, we can predict a range for the component magnitude difference ($\Delta m_{V}$) from plausible properties of the stars.  For a primary mass $M_{1}$ = 11 $M_{\sun}$, Table 1 of \citet{toh04} predicts a spectral type of B1.5, similar to that listed in Table~\ref{targetlist} (taken from BeSS) and consistent with the range B1 -- B4 given by various sources \citep[][and references therein]{har84}.  This table also predicts a critical rotation velocity $\it v_{c}$ = 491 km~s$^{-1}$.  Assuming $\it i$ = 75$\arcdeg$ (and the rotational axis of the star to be normal to the orbital plane) and $\it v$sin$\it i$ = 245 km~s$^{-1}$ (BeSS), a rotational velocity of the primary $\approx$ 254 km~s$^{-1}$ (52\% of critical) is predicted.  Adopting these values for the spectral type, inclination, and $\it v$/$\it v_{c}$, Figure 3 of \citet{toh04} then predicts an absolute visual magnitude $M_{V}$(primary) $\approx -3.1$.  The IR spectroscopic data of \citet{fhm89} indicated the companion star in this system is likely not a luminous giant, so if a main sequence dwarf, the tables of \citet{sck82} predict, for $M_{2}$ = 0.95 $M_{\sun}$, a spectral type $\sim$ G5 and an absolute magnitude $M_{V}$(secondary) $\approx$ +5.0, and thus $\Delta m_{V}$ $\approx$ 8 for $\zeta$ Tau, far beyond the reach of the current NPOI.  Alternatively, if the secondary star were a sdO subdwarf, its mass would be intermediate between those of the sdO stars $\phi$ Per (\S~\ref{phiper}) and 59 Cyg (\S~\ref{59cyg}).  Were the luminosity and $T_{\rm eff}$ also of intermediate values, a bolometric magnitude $M_{\rm bol}$(secondary) = $-3.86$, bolometric correction BC = $-4.7$, and thus $M_{V}$(secondary) = +0.84 and a system $\Delta m_{V}$ $\sim$ 3.9 for $\zeta$ Tau would be predicted, the latter only marginally beyond the capabilities of the current NPOI.

\citet{wgp17}, searching archival $\it IUE$ spectra for hot sdO companions to Be stars, found no indication of such in the $\zeta$ Tau system ($\leq$ 1\% contribution to total UV flux).  The INT4 lists this star as unresolved by visual and speckle interferometry on five dates (1921-1997) implying $\rho < 30$~mas, and \citet{hvd20} also lists it as unresolved on two dates in 2015 at 692~nm with detection limit of $\Delta m$ of 4.17 at 200~mas.  Interferometric observations at the VLTI \citep[Very Large Telescope Interferometer,][]{gaa03, src09} and CHARA \citep{sgm10} also failed to detect a stellar companion.
  
The five nights of NPOI observations of $\zeta$ Tau listed in Table~\ref{progstarsobs} were modeled with both GRIDFIT and CANDID, but with significantly different results: GRIDFIT produced a significant ``detection'' (\S~\ref{modeling}) on only one night (2007 Feb 06) as an unresolved source, while CANDID produced results on the first four (consecutive) nights at angular separations $\sim$ 6 mas, with position angles over a range of 190$\arcdeg$ $\lesssim$ $\theta$ $\lesssim$ 215$\arcdeg$, and at $\sim$ 3 mas and 81$\arcdeg$ on 2007 Feb 06. As a third alternative, we used the program within OYSTER, similar to GRIDFIT, to search a grid of 50 mas by 50 mas at 0.5 mas spacings, centered on the primary star; this produced weak, ill-defined $\chi_{\nu}^2$ minima on each night, at positions differing from both the GRIDFIT and CANDID results, with the exception of 2007 Feb 06 where a solution corresponding to an unresolved source was again preferred.  Therefore, we conclude that our observations have failed to unambiguously detect a companion.

Based on the available evidence, we conclude $\zeta$ Tau consists of two physical stellar components.

\subsubsection{$\kappa$ Dra} \label{kapdra}

$\kappa$ Dra (HIP 61281, HR 4787, FK5 472, HD 109387): This star does not appear in the WDS, LIN2, ORB6, DMSA, SB9, or MSC catalogs.  However, \citet{skk04} found $T_{\rm eff}$ = 14000~K and $\it v$sin$\it i$ = 170 km~s$^{-1}$ for the primary star, and then used these values along with a HIPPARCOS parallax to derive a mass of $M_{p}$ = 4.8 $\pm$ 0.8 $\it M_{\sun}$ and radius $R_{p}$ = 6.4 $\pm$ 0.5 $R_{\sun}$.  Subsequently, \citet{skh05} present a SB1 orbital solution ($P$ = 61.555~d, $e$ = 0), and calculated several possible values of the secondary mass $M_{s}$ for different inclination angles.  Adopting the $R_{p}$ and $v\sin i$ values of \citet{skk04}, and assuming the rotation axis of the primary is normal to the orbital plane and the primary rotates near its break-up velocity, they derive an inclination angle $\it i$ $\sim$ 30$\arcdeg$ and a mass for the secondary star of $M_{s}$ $\approx$ 0.8 $\it M_{\sun}$.  

Using the mass sum (5.6 $M_{\sun}$), the orbital period of \citet{skh05}, and the HIPPARCOS parallax listed in Table~\ref{targetlist}, we estimate an angular semimajor axis of the binary orbit of $\it a$ $\approx$ 3.6 mas (or 3.8 mas using the GAIA parallax).  Given all of our observations of $\kappa$ Dra (Table~\ref{progstarsobs}) included at least one baseline with nominal angular resolution of 1.7 mas (Table~\ref{baselines}), it is likely at least some of our observations should have resolved this system provided the component magnitude difference is $\Delta m$ $\lesssim$ 3.5.  

We can estimate a likely range of $\Delta m$ values from the information provided above.  For a primary star of $M_{p}$ = 4.8 $\it M_{\sun}$, Table 1 of \citet{toh04} indicates a spectral type of B6 (consistent with that listed in Table~\ref{targetlist}) and a critical rotation velocity $\it v_{c}$ = 418 km~s$^{-1}$.  At $\it i$ = 30$\arcdeg$, the rotational velocity of the primary would then be $\approx$ 340 km~s$^{-1}$ (81\% of critical), and we can use this along with the spectral type and inclination angle to estimate an absolute magnitude of $M_{V}$(primary) $\approx -1.4$ using Figure 3 of \citet{toh04}.  Noting the estimated mass of the secondary star is similar to the secondary star in 59 Cyg Aa1,2 (\S~\ref{59cyg}), and assuming it to be an identical, hot sdO subdwarf of the same absolute magnitude $M_{V}$(secondary) = +1.74, we then obtain $\Delta m_{V}$ $\approx$ 3.1 for the $\kappa$ Dra binary, which should be detectable with the NPOI.  However, if the secondary is a main sequence star, then at $M_{s}$ = 0.8 $\it M_{\sun}$ the tables of \citet{sck82} indicate a spectral type of $\approx$~K0 with $M_{V}$(secondary) = +5.9, and thus $\Delta m_{V}$ $\approx$ 7.3 for $\kappa$ Dra, far beyond the reach of the current NPOI.

\citet{wgp17}, searching archival $\it IUE$ spectra for hot sdO companions to Be stars, found no indication of such in the $\kappa$ Dra system ($\leq$ 1\% contribution to total UV flux).  The INT4 lists $\kappa$ Dra as unresolved in one observation in 2008 implying a limit of $\rho$ $<$ 36 mas in the visible \citep{mas09}, and \citet{hvd20} also lists it as unresolved (2015.1828) at 692~nm with detection limit $\Delta m$ of 4.25 at 200~mas.  The five nights of NPOI observations of $\kappa$ Dra listed in Table~\ref{progstarsobs} were modeled with GRIDFIT with null result. Therefore, our observations have not detected the SB1 companion.  

Based on the available evidence, we conclude $\kappa$ Dra consists of two physical stellar components.

\newpage

\subsection{Stars Not Previously Confirmed as Binaries} \label{nondet}

\subsubsection{$\epsilon$ Cas} \label{epscas}

$\epsilon$ Cas (HIP 8886, HR 542, FK5 63, HD 11415): This star does not appear in any of the catalogs searched (WDS, WDS Notes, LIN2, ORB6, INT4, DMSA, SB9, or MSC).  \citet{wgp18}, searching archival $\it IUE$ spectra for hot sdO companions to Be stars, found no indication of such in this system.

The seven nights of NPOI observations of $\epsilon$ Cas listed in Table~\ref{progstarsobs} were modeled with both GRIDFIT and CANDID with null result.  While GRIDFIT produced significant $\chi_{\nu}^2$ minima (\S~\ref{modeling}) on four nights, these minima occurred at widely differing separations and position angles.  The CANDID results also varied widely in position angle and separation and displayed no correlation with the GRIDFIT results.  Based on the available evidence, we conclude $\epsilon$ Cas is a single star.

\subsubsection{BK Cam} \label{bkcam}

BK Cam (HIP 15520, HR 985, HD 20336, WDS 03200+6539Aa,Ab): The WDS lists pairs Aa,Ab ($\theta$ = 43$\fdg$70, $\rho$ = 132 mas in 2007, with $\Delta m$ = 1.2) and AB ($\theta$ = 58$\fdg$30, $\rho$ = 120$\farcs$740 in 2003, with $\Delta m$ = 8.04) without comment as to their being physical pairs.  However, the WDS further indicates essentially no change in relative position of the AB pair over the period 1911 -- 2003. On the other hand, the GAIA DR2 and HIPPARCOS proper motions of BK Cam A are consistent and do not agree with the DR2 proper motion for the B component.  The INT4 lists only a single speckle interferometry observation of Aa,Ab \citep{mas09}.

Furthermore, the WDS Notes list BK Cam as a possible long-period spectroscopic binary, but neither pair appears in the ORB6, SB9, DMSA, LIN2, or MSC catalogs.  BK Cam was reported as a possible spectroscopic variable star by Espin \citep{esp1896} as a result of his spectroscopic sweeps at the Wolsingham Observatory in the U.K. \citep{esp1895}.  The Henry Draper catalog \citep{hd-bkCam} lists variability in the H{$\beta$} and the H{$\gamma$} lines as a remark in the Annals of the Harvard College Observatory by Maury \citep{mp1897}.  However, this remark appears to refer to BD{$+23\arcdeg\ 516$} (20 Tau) and not BK Cam.  \citet{cur12} studied the emission line variability of BK Cam, and suggested that a Doppler variation should be applied cautiously as the emission line variations did not adhere to changes attributable only to Doppler effects.  This caution was borne out when the spectroscopic binary hypothesis was put to rest by \citet{McL63} using spectra spanning 1903 to 1961.  The variations in the lines failed to show a regular variation.  \citet{McL63} suggested a ring or shell of circumstellar material as the best scenario to explain the irregular spectral line variations. 

\begin{figure}
\epsscale{1.0}
\plotone{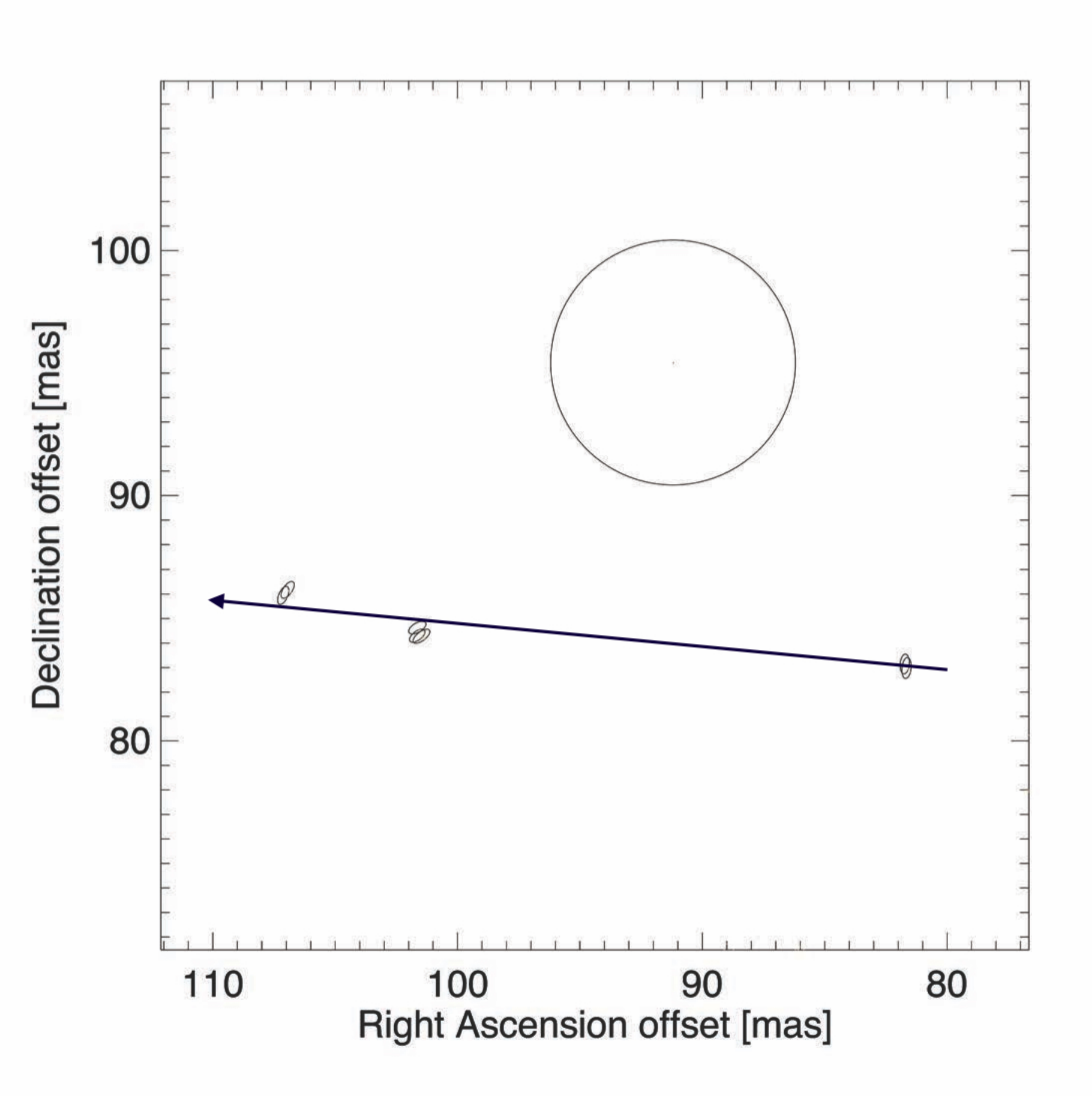}
\caption{Relative astrometry of the Aa,Ab components of BK Cam.  The fitted relative positions at each of the seven epochs of NPOI observations (Table~\ref{programfit}) are shown with error ellipses plotted to scale.  The arrow represents a linear regression fit to the NPOI observations.  The speckle interferometry observation of \citet{mas09} is also plotted, with an assumed 5 mas radius error circle, a value typical of speckle interferometry observations \citep{mhr11}.}
\label{BK-Cam_plot}
\end{figure}

While BK Cam is an unlikely spectroscopic binary, and the lack of relative motion of the wide AB is not supportive of a wide binary, a binary resolvable by optical interferometry is still a possibility to consider.  To investigate the possibility of a small separation binary system in BK Cam, we examined 23 nights of archival NPOI data (2005 -- 2012) using GRIDFIT, obtaining consistent and significant detections (\S~\ref{modeling}) on five nights during 2006, 2009, and 2010.  An additional two nights produced results at similar positions, although at slightly below our significance criterion.  We then proceeded to detailed modeling of the data from each of these seven nights (Table~\ref{progstarsobs}) with OYSTER using fixed stellar angular diameters of $\theta_{P}$ = 0.2 mas (from the NPOI planning software) and $\theta_{S}$ = 0.1 mas (assumed) along with the nightly $\rho$ and $\theta$ values and the average $\Delta m_{700}$ obtained from GRIDFIT as inputs.  The results of the nightly fits are presented in Table~\ref{programfit} and Figure~\ref{BK-Cam_plot}.  

%% BK Cam astrometry fits

As none of our observations included closure phase data, there may be a 180$\arcdeg$ ambiguity in the position angle of our results.  However, the observation of \citet{mas09} was calibrated in position angle using a double-slit mask and observations of wide binaries of well established orientation, which lends confidence in our reported position angles.   Our results cover too small a range of position angle to suggest anything more than linear relative motion (6.42 mas/yr in RA and 0.65 mas/yr in Dec, per the regression line plotted in Figure~\ref{BK-Cam_plot}).  The weighted mean magnitude difference from our models is $\Delta m_{700}$ = 2.45 $\pm$ 0.03, significantly larger than the $\Delta m_{550}$ = 1.2 $\pm$ 0.5 value obtained by \citet{mas09}.  However, \citet{mas09} note this value as only an estimate, not a ``catalog value.''

From the evidence discussed above, the close binary BK Cam Aa,Ab resolved by the NPOI, if physical, would have a period of many 10's of years long.  Using the distance listed in Table~\ref{targetlist} the angular separations in Table~\ref{programfit} convert to distances of 25--31~AU projected on the sky. Longer term monitoring will be required to complete a relative astrometric orbit.  Furthermore, we conclude that the wide BK Cam AB pair is a likely a chance alignment. Therefore, based on the available evidence, we conclude BK Cam consists of at most two physical stellar components, the Aa,Ab pair.

\subsubsection{$\psi$ Per} \label{psiper}

$\psi$ Per (HIP 16826, HR 1087, HD 22192): This star does not appear in the WDS, WDS Notes, LIN2, ORB6, DMSA, SB9, or MSC catalogs.  The INT4 lists it as unresolved on six dates (1980 -- 1996), placing a constraint of $\rho$ $<$ 36 mas in the visible \citep{mas97}, and \citet{hvd20} also lists it as unresolved on two dates in 2015 at 692~nm with detection limit of $\Delta m$ of 4.23 at 200~mas.  \citet{wgp18}, searching archival $\it IUE$ spectra for hot sdO companions to Be stars, found no indication of such in this system.

The five nights of NPOI observations of $\psi$ Per listed in Table~\ref{progstarsobs} were modeled with GRIDFIT with null result.  Based on the available evidence, we conclude $\psi$ Per is a single star.

\subsubsection{17 Tau} \label{17tau}
  
17 Tau (HIP 17499, HR 1142, FK5 136, HD 23302, WDS 03449+2407A): This star does not appear in the LIN2, ORB6, or DMSA.  The WDS lists pairs AB through AE ($\rho$ $>$ 99$\arcsec$ and $\Delta m$ $\geq$ 7.3) along with close pairs Aa,Ab ($\rho$ = 1.450 mas in 1987 -- see below, with $\Delta m$ = 3.10) and Aa, Ac ($\rho$ = 196 mas in 1988, with $\Delta m$ = 3.60) without comment as to any of them being physical (we assume AB--AE to be nonphysical).

The INT4 lists three occultation measurements of Aa,Ab (1972 -- 1987) at 1.5~mas $\leq \rho \leq$ 10~mas, with 0.2 $\leq$ $\Delta m$ $\leq$ 3.4, and a single occultation measure for Aa,Ac \citep{rcl96} at the above cited separation.  Neither pair is resolved in $\sim$ 30 visible speckle and HIPPARCOS observations, placing a limit of $\rho$ $<$ 30~mas.  \citet{hvd20} also lists it as unresolved (2015.8342) at 692~nm with detection limit of $\Delta m$ of 4.24 at 200~mas.

The SB9 lists system 188 \citep{abb65}, as a SB1 with $P$ = 100.46~d.  The MSC identifies this pair as Aab,Ac and an occultation binary and questionable SB1 at an estimated $\rho$ $\approx$ 8.0 mas.  The MSC also lists pair Aa,Ab \citep[$P$ = 4.292~d;][]{jhs89} as another questionable SB1 at an estimated separation of 0.9 mas.  This pair does not seem related to either the Aa,Ab, or the wider Aa,Ac pair listed in the WDS/INT4.  \citet{chn12} lists 17 Tau as a constant RV source, and \citet{wgp18}, searching archival $\it IUE$ spectra for hot sdO companions to Be stars, found no indication of such in the 17 Tau system.

The four nights of NPOI observations of 17 Tau listed in Table~\ref{progstarsobs} were modeled with GRIDFIT with null result.  Given the lack of information on the AB -- AE pairs, the lack of confirming observations of either the purported occultation or spectroscopic binaries, and the failure of either speckle or our observations to resolve the A component, we conclude 17 Tau is likely a single star.

\subsubsection{23 Tau} \label{23tau}
  
23 Tau (HIP 17608, HR 1156, HD 23480, WDS 03463+2357Aa,Ab): This star does not appear in the LIN2, ORB6, DMSA, SB9, or MSC.  The WDS lists pairs AB and AC ($\rho$ $>$ 110$\arcsec$ and $\Delta m$ $\geq$ 8.78) along with a close pair Aa,Ab ($\rho$ = 250 mas and $\Delta m$ = 3.96 in 2002) without comment as to any of them being physical.  The INT4 lists a single detection of Aa,Ab by AO \citep[][2002.096, at 0.8 $\mu$m]{dbp11} at the above cited position and $\Delta m$.  Aa,Ab is also listed as a HIPPARCOS ``suspected non-single,'' but is also unresolved by speckle on four occasions (1987 -- 2005) placing a limit of  $\rho <$ 35~mas in the visible.  \citet{hvd20} also lists it as unresolved (2015.8343) at 692~nm with detection limit of $\Delta m$ of 4.33 at 200~mas.  \citet{chn12} lists 23 Tau as a constant RV source, and \citet{wgp18}, searching archival $\it IUE$ spectra for hot sdO companions to Be stars, found no indication of such in the 23 Tau system.

The five nights of NPOI observations of 23 Tau listed in Table~\ref{progstarsobs} were modeled with GRIDFIT with null result.  While the companion Ab reported by \citet{dbp11} is well within the range of angular separations detectable by the NPOI observations (\S~\ref{npoi}), the $\Delta m$ is likely beyond the range of the NPOI (Paper I).  Given the lack of information on the AB and AC pairs, and the lack of any further observations of the Aa,Ab pair to indicate any orbital motion, we conclude 23 Tau is likely a single star.

\newpage

\subsubsection{$\eta$ Tau} \label{etatau}

$\eta$ Tau (HIP 17702, HR 1165, FK5 139, HD 23630, WDS 03475+2406A): This star does not appear in the LIN2, ORB6, DMSA, or SB9.  The WDS lists pairs AB -- AH and BC, BD, and CD ($\rho$ $>$ 54$\arcsec$ and $\Delta m$ $\geq$ 0.5) along with close pairs Aa,Ab ($\theta$ and $\rho$ unlisted, with $\Delta m$ = 1.6) and Da,Db ($\theta$ = 292$\fdg$5, $\rho$ = 330 mas in 2010, with $\Delta m$ = 0.6) without comment as to any of them being physical.  The WDS Notes list A as an occultation and spectroscopic binary and B as an occultation binary (without reference), and pairs BC, BD, and CD as possible CPM pairs \citep[][ although the author estimates the probability of them being physical pairs at only 60$\%$]{hal86}.  The MSC lists AB as a CPM pair (without reference), pair Aa,Ab as a possible occultation binary at $\rho$ = 31 mas, and Aa1,Aa2 as a possible SB1 with $P = 4.13$~d and $\rho$ $\sim$ 0.9 mas \citep[][who list the 4.13 d period as the most likely of three possibilities]{jhs89}.  

The INT4 lists Aa,Ab as resolved by occultation at 1.0 mas, 1.7 mas, and 31 mas  ($\Delta m$ = 1.2 -- 1.6) on three dates between 1971 and 1972 \citep{dev76, bth75}, but as unresolved on 20 dates (1975 -- 2008) with limit of $\rho$ $<$ 30 mas in the visible by speckle and HIPPARCOS, as well as by AO at 0.9~$\mu$m with a limit of $\rho$ $<$ 100~mas~\citep{rbr07}.  \citet{hvd20} lists it as unresolved (2015.8343) at 692~nm with detection limit $\Delta m$ of 4.31 at 200~mas.  The INT4 lists Da,Db as resolved ($\theta$ = 267$\fdg$4, $\rho$ = 197~mas) on two occasions in 2005 by speckle \citep[][listed as HD 23608]{mas09}, and at $\rho$ = 330 mas in 2010, with $\Delta m$ = 1.1 by occultation \citep{loa12}.  \citet{chn12} lists $\eta$ Tau as a constant RV source, and \citet{wgp18}, searching archival $\it IUE$ spectra for hot sdO companions to Be stars, found no indication of such in the $\eta$ Tau system.

The five nights of NPOI observations of $\eta$ Tau listed in Table~\ref{progstarsobs} were modeled with GRIDFIT with null result.  Given the wide pairs AB -- AH do not appear in any of the above cited references as possible physical pairings (except AB, in the MSC, without citation), and the evidence for the Aa1,Aa2 SB appearing weak, we judge $\eta$ Tau to contain at most the Aa,Ab pair.  The cited $\Delta m$ in the range of 1.2 -- 1.6 would certainly have made the pair detectable in our observations if the pair were as widely separated as per \citet{bth75}, but could have remained unresolved if as closely separated as reported by \citet{dev76}, the nominal resolution of our baselines being $\sim$ 1.5 mas (Tables~\ref{baselines} \&~\ref{progstarsobs}).  We conclude at present no convincing case can be made for $\eta$ Tau being more than a single star.

\subsubsection{48 Per} \label{48per}

48 Per (HIP 19343, HR 1273, FK5 152, HD 25940): This star does not appear in the WDS, WDS Notes, LIN2, ORB6, INT4, DMSA, SB9, or MSC catalogs.  \citet{jhs89} suggested a possible SB1 at a period of 16.6~d, but stated confirmation of a binary would ``require substantially more data.''  \citet{hvd20} lists it as unresolved (2015.8343) at 692~nm with detection limit $\Delta m$ of 4.72 at 200~mas, and \citet{wgp18}, searching archival $\it IUE$ spectra for hot sdO companions to Be stars, found no indication of such in this system.  \citet{jsg17} spatially resolved the H$\alpha$ emitting region using NPOI and determined that disk emission originates in a moderately large disk with a radius of 25 stellar radii, thus excluding the possibility of very close binary companions.

The five nights of NPOI observations of 48 Per listed in Table~\ref{progstarsobs} were modeled with GRIDFIT with null result.  Based on the available evidence, we conclude 48 Per is likely a single star.

\subsubsection{$\omega$ Ori} \label{oori}

$\omega$ Ori (HIP 26594, HR 1934, FK5 2423, HD 37490): This star does not appear in the WDS, WDS Notes, LIN2, ORB6, DMSA, SB9, or MSC catalogs.  The INT4 lists it as unresolved on seven dates (1989 -- 1997) placing a limit of $\rho$ $<$ 35 mas in the visible \citep{mas97}.  

The five nights of NPOI observations of $\omega$ Ori listed in Table~\ref{progstarsobs} were modeled with GRIDFIT with null result.  Based on the available evidence, we conclude $\omega$ Ori is a single star.

\subsubsection{$\beta$ CMi} \label{betcmi}

$\beta$ CMi (HIP 36188, HR 2845, FK5 285, HD 58715, WDS 07272+0817A): This star does not appear in the LIN2, ORB6, DMSA, SB9, or MSC catalogs.  The WDS lists pairs AB through AG, and FG and FH ($\rho$ $>$ 4$\arcsec$ and $\Delta m$ $\geq$ 0.5) without comment, and since they do not appear in any of the above listed catalogs we assume them all to be nonphysical.  The additional pairs AI and AJ ($\rho$ $\geq$ 7.5$\arcsec$ and $\Delta m$ $\geq$ 14.7) listed in the WDS are listed as nonphysical.  The INT4 lists them as having been observed with AO on two occasions in 2008 and 2010 \citep{jan11}. 

The WDS Notes list component A as a spectroscopic binary, as does \citet[][as a SB1]{chn12}, but it does not appear in the SB9.  $\beta$ CMi has been identified as a nonradial pulsator on the basis of photometry with the $\it Microvariability\ and\ Oscillations\ of\ Stars\ (MOST)$ satellite \citep{sck07}, and it was historically reported as RV variable and classified as a likely SB of unknown orbit \citep{stv25, fbs26}.  More recently, \citet{jhs89} and \citet{drg17} report SB1 orbits, but they differ widely in period, eccentricity, and RV amplitude, and \citet{hsk19}, using a large set of spectra, concluded there was no convincing evidence of significant RV changes.  However, \citet{kcr19} defend the results of \citet{drg17}, claiming many of the observations used in \citet{hsk19} lacked sufficient spectral resolution.  \citet{wgp18} searched for hot sdO companions to Be stars in archival $\it IUE$ spectra, but were unable to detect such a companion to $\beta$ CMi.  The INT4 lists component A as unresolved in the visible on two occasions (1988, 1991), placing a limit of $\rho <$ 38~mas, and \citet{hvd20} also lists it as unresolved (2015.8347) at 692~nm with detection limit $\Delta m$ of 4.38 at 200~mas.   

The five nights of NPOI observations of $\beta$ CMi listed in Table~\ref{progstarsobs} were modeled with GRIDFIT with null result.  Based on the available evidence, we conclude there is not yet convincing evidence that $\beta$~CMi is more than a single star.

\subsubsection{$\phi$ Leo} \label{phileo}

$\phi$ Leo (HIP 55084, HR 4368, FK5 1292, HD 98058, WDS 11167-0339A): The WDS lists only one nonphysical companion ($\theta$ = 290$\fdg$5, $\rho$ = 86$\farcs$78 in 2015, with $\Delta m$ = 5.27) with linear elements (LIN2), and there are no entries in the ORB6, DMSA, SB9, or MSC catalogs.  The INT4 lists observations by HIPPARCOS (unresolved with limit of $\rho <$ 0$\farcs$1) and by $\it Tycho$ ($\theta$ = 290$\fdg$1, $\rho$ = 89$\farcs$29) in 1991.  

The five nights of NPOI observations of $\phi$ Leo listed in Table~\ref{progstarsobs} were modeled with GRIDFIT with null result.  Based on the available evidence, we conclude $\phi$ Leo is a single star.

\subsubsection{48 Lib} \label{48lib}

48 Lib (HIP 78207, HR 5941, FK5 1417, HD 142983): This star does not appear in the WDS, LIN2, ORB6, DMSA, SB9, or MSC catalogs.  The INT4 lists this star as unresolved on five dates in 1982 and 2008, placing a limit of $\rho <$ 34~mas in the visible, and $\rho < 100$~mas with $\Delta m$ limit of 8.5 at 2.1~$\mu$m.  Additionally, \citet{slc12}, based on spectroscopic and VLTI interferometric observations of the circumstellar disk, concluded a binary star model was ``unattractive'' for 48 Lib.  
Similarly, \citet{sjc16} successful modeled the quasi-cyclic variations ($P \sim 12$~yr) of the asymmetric double-peaked H$\alpha$ profiles of 48~Lib with one-armed density wave precessing in the disk that extended at least $\sim 40$ stellar radii, excluding the possibility of very close companions that would result in a disk truncation mechanism. \citet{wgp18}, searching archival $\it IUE$ spectra for hot sdO companions to Be stars, found no indication of such in this system.

The five nights of NPOI observations of 48 Lib listed in Table~\ref{progstarsobs} were modeled with GRIDFIT with null result.  Based on the available evidence, we conclude 48 Lib is a single star.

\subsubsection{$\chi$ Oph} \label{chioph}

$\chi$ Oph (HIP 80569, HR 6118, FK5 3298, HD 148184): This star does not appear in the WDS, WDS Notes, LIN2, ORB6, DMSA, or MSC.  The SB9 lists system 903 \citep{aal78} as a SB1 with $P$ = 138.8~d, but with orbital elements that are ``described as `marginal' by Abt and Levy themselves.''  \citet{har87} and \citet{lmm87} also classified $\chi$ Oph as a SB1, but there is no agreement on the period, \citet{har87} reporting a period of 34.12 d and \citet{lmm87} adopting the elements of \citet{aal78}.  The INT4 lists this star as unresolved by speckle on two dates in 2008 placing a limit of $\rho <$ 34~mas at 0.55~$\mu$m, as well as by LBOI with limit of $\rho <$ 0.2~mas and with $\Delta m$ limit of 2.5 at 2.2~$\mu$m \citep[2010.352,][]{grk15}.  \citet{tjs08} used NPOI H$\alpha$ observations to spatially resolve the inner region of the circumstellar disk for the first time, but could not confirm a binary companion.

The four nights of NPOI observations of $\chi$ Oph listed in Table~\ref{progstarsobs} were modeled with GRIDFIT with null result, confirming the result of \citet{tjs08} that $\chi$ Oph is a single star.  It should be noted these are the same data used by \citet{tjs08}, but these were independently calibrated and modeled.  Based on the available evidence, we conclude $\chi$ Oph is likely a single star.

\subsubsection{$\zeta$ Oph} \label{zetoph}

$\zeta$ Oph (HIP 81377, HR 6175, FK5 622, HD 149757, WDS 16372-1034): $\zeta$ Oph does not appear in the main WDS, LIN2, ORB6, DMSA, SB9, or MSC catalogs, but the WDS Notes cite the discussion of \citet{vvd96} regarding the runaway trajectory of this star originating in a binary scenario, where it was ejected after the SN event of its primary.  \citet{chn12} lists $\zeta$ Oph as an SB2 without further information.  The INT4 lists this star as unresolved on 11 dates (1988 -- 2012) placing a limit of $\rho <$ 30~mas in the visible and $\rho <$ 1~mas with detection limit $\Delta m$ of 5.0 at 1.7~$\mu$m, the latter result by LBOI \citep{sll14}.  \citet{wgp18}, searching archival $\it IUE$ spectra for hot sdO companions to Be stars, found no indication of such in this system.

The five nights of NPOI observations of $\zeta$ Oph listed in Table~\ref{progstarsobs} were examined with both GRIDFIT and CANDID with inconsistent results.  GRIDFIT obtained a significant binary detection (\S~\ref{modeling}) on only one night (2011 April 17), while CANDID produced results that varied widely in position angle and separation, and displayed no correlation with the GRIDFIT result.  Based on the available evidence, we conclude $\zeta$ Oph is a single star.

\subsubsection{o Her} \label{oher}
 
o Her (HIP 88794, HR 6779, FK5 681, HD 166014, WDS 18075+2846AB): This star does not appear in the LIN2, ORB6, DMSA, SB9, or MSC catalogs.  The WDS lists components AB ($\theta$ = 86$\fdg$6, $\rho$ = 61 mas in 1983), while the WDS Notes list it as a spectroscopic and interferometric binary (without reference).  The INT4 lists observations of this star on seven dates (1921 -- 2008), with binary detection at $\rho$ $\approx$ 60 mas on three dates in 1981 and 1983 using a phase grating interferometer, but subsequent observations by HIPPARCOS and \citet{mas09} failed to resolve this star implying a limit of $\rho$ $<$ 36~mas in the visible.  \citet{edg07} found no evidence of RV variability \citep{tgs13}.  \citet{wgp18}, searching archival $\it IUE$ spectra for hot sdO companions to Be stars, found no indication of such in the o Her system.

The four nights of NPOI observations of o Her listed in Table~\ref{progstarsobs} were examined with GRIDFIT with null result.  Therefore, based on the available evidence, we conclude o Her is a single star.

\subsubsection{$\upsilon$ Cyg} \label{upscyg}

$\upsilon$ Cyg (HIP 105138, HR 8146, FK5 1559, HD 202904, WDS 21179+3454A): This star does not appear in the LIN2, ORB6, SB9, or MSC catalogs.  The WDS lists pairs AB through AE ($\rho$ $>$ 15$\arcsec$ and $\Delta m$ $\geq$ 5.59) without comment and we assume them to be nonphysical.  The WDS also lists components BC ($\theta$ = 141$\fdg$3, $\rho$ = 13$\farcs$15 and $\Delta m$ = 0.85 in 2015), with the WDS Notes pointing to an acceleration double solution (multiplicity flag ``G'') in the main HIPPARCOS catalog.  However, this system does not appear in the DMSA, and \citet{aac84} list component B as a foreground star (G8.5V), and C as not having a common proper motion.

The INT4 lists component A as unresolved on three dates (1980 -- 2008), placing a limit of $\rho <$ 30~mas in the visible, as well as an observation of components AC ($\theta$ = 183$\fdg$3, $\rho$ = 21$\farcs$77 and $\Delta m$ = 5.591 in 1991.45) by $\it Tycho$.  \citet{aal78} found the RV of $\upsilon$ Cyg is constant, and \citet{wgp18}, searching archival $\it IUE$ spectra for hot sdO companions to Be stars, found no indication of such in this system.  \citet{nfh05} found that a longer timescale RV variation might be present, with a period of $\approx$ 11 years, which could be associated with a binary companion.

The five nights of NPOI observations of $\upsilon$ Cyg listed in Table~\ref{progstarsobs} were examined with GRIDFIT with null result.  Therefore, based on the available evidence, we conclude $\upsilon$ Cyg is likely a single star.

\subsubsection{o Aqr} \label{oaqr}

o Aqr (HIP 108874, HR 8402, FK5 3765, HD 209409): This star does not appear in the WDS, LIN2, ORB6, DMSA, SB9, or MSC catalogs.  \citet{chn12} list this as a constant RV source.  The INT4 lists this star as unresolved on four dates (1980 -- 2005) placing a limit of $\rho <$ 55~mas in the visible and $\rho <$ 100~mas with detection limit $\Delta m$ of 10.0 at 2.2~$\mu$m \citep{oap10}.  \citet{hvd20} also lists it as unresolved (2015.8335) at 692~nm with detection limit $\Delta m$ of 4.29 at 200~mas.  \citet{wgp18}, searching archival $\it IUE$ spectra for hot sdO companions to Be stars, found no indication of such in this system. Also, based on LBOI observations that resolved the H$\alpha$ emitting disk of o~Aqr, \citet{stj15} determined that 90\% of the H$\alpha$ emission arises from within 9.5 stellar radii, thus excluding the possibility of very close companions that would result in disk truncation.

The five nights of NPOI observations of o Aqr listed in Table~\ref{progstarsobs} were examined with both GRIDFIT and CANDID with null result.  GRIDFIT produced no statistically significant $\chi_{\nu}^2$ minima, while the CANDID results varied widely in position angle and separation.  Based on the available evidence, we conclude o Aqr is a single star.

\subsubsection{31 Peg} \label{31peg}

31 Peg (HIP 110386, HR 8520, FK5 843, HD 212076): This star does not appear in the WDS, LIN2, ORB6, DMSA, SB9, or MSC catalogs.  \citet{chn12} list this as a constant RV source.  INT4 lists this star as unresolved on two dates in 1985 and 2008, placing a limit of $\rho < 34$~mas in the visible, and \citet{hvd20} also lists it as unresolved (2015.8336) at 692~nm with detection limit $\Delta m$ of 3.98 at 200~mas.

The five nights of NPOI observations of 31 Peg listed in Table~\ref{progstarsobs} were examined with GRIDFIT with null result.  Based on the available evidence, we conclude 31 Peg is a single star.

\subsubsection{$\beta$ Psc} \label{betpsc}

$\beta$ Psc (HIP 113889, HR 8773, FK5 1602, HD 217891): This star does not appear in the WDS, LIN2, ORB6, DMSA, SB9, or MSC catalogs.  \citet{chn12} list this as a constant RV source.  The INT4 lists this star as unresolved on four dates (1980 -- 2008) placing a limit of $\rho < 34$~mas in the visible and $\rho < 80$~mas at 0.9 $\mu$m.  \citet{wgp18}, searching archival $\it IUE$ spectra for hot sdO companions to Be stars, found no indication of such in this system.

The five nights of NPOI observations of $\beta$ Psc listed in Table~\ref{progstarsobs} were examined with both GRIDFIT and CANDID with null result.  GRIDFIT produced no statistaically significant $\chi_{\nu}^2$ minima, while the CANDID results varied widely in position angle and separation.  Based on the available evidence, we conclude $\beta$ Psc is a single star.

\section{Discussion} \label{resdisc}

\subsection{Multiplicity Statistics} \label{multstat}

Our detailed results from \S~\ref{bindet} can be summarized in terms of the overall multiplicity statistics of our sample and examined in terms of survey sensitivities and completeness (\S~\ref{senscomp}).  At face value, our evaluation of the number of likely physical components for each of the systems implies that 14 of the 31 systems in our sample are multiple (Table~\ref{targetlist}, Column~12), and that the primaries in the multiple systems are accompanied by a total of 28 companions.  Therefore, these statistics imply a Multiplicity Frequency (MF), the fraction of multiple systems in the sample \citep{dak13} of MF = 14/31 = 45\%, roughly consistent with the conclusions of \citet{dak13} that MF $\ge$ 50\% for intermediate-mass stars ($\it M_{*}$ $\approx$ 1.5 -- 5$\it M_\sun$; B5 -- F2) and MF $\sim$ 45 $\pm$ 5\% for visual companions of high-mass stars ($\it M_{*}$ $\gtrsim$ 8$\it M_\sun$; B2 and earlier).

The corresponding Companion Frequency (CF), the average number of companions per target in the sample, which can exceed 100\% when significant numbers of higher-order multiple systems are present \citep{dak13}, is CF = 28/31 = 90\%. This result is consistent with the assessment of \citet{dak13} that for systems with primaries of 8 -- 16 $M_\sun$ and mass ratio $q = M_{2}$/$M_{1}$ $\gtrsim$ 0.1, CF $\approx$ 100 $\pm$ 20\%, and that CF = 100 $\pm$ 10\% for $\it M_{*}$ = 1.5 -- 5$\it M_\sun$ primaries \citep{kbp07}.

\citet{dak13} also note that there is evidence that the frequency of visual companions decreases with the mass of the primary star \citep[e.g.,][]{pbh99}.  If we divide our sample into two roughly equal subgroups by spectral type ($\leq$ B3; 16 sources, and $>$ B3; 15 sources, including $\upsilon$ Sgr; \S~\ref{upssgr}), the corresponding ratios are MF = 9/16 = 56\% and CF = 19/16 = 119\% for the early type, higher mass primaries, and MF = 5/15 = 33\% and CF = 9/15 = 60\% for the later type, lower mass primaries, confirming the suggested trend \citep[although the CF value for the lower mass stars is below that of][cited above] {kbp07}.  However, \citet{dak13} note that the completeness of multiplicity surveys is very uncertain for systems with $q < 0.1$.  We will discuss this issue with regard to our sample and its relevance to theories of the origin of the Be star phenomenum in \S~\ref{future}.

\subsection{Survey Sensitivities and Completeness} \label{senscomp}

\begin{figure*}[ht]
\epsscale{1.2}
\plotone{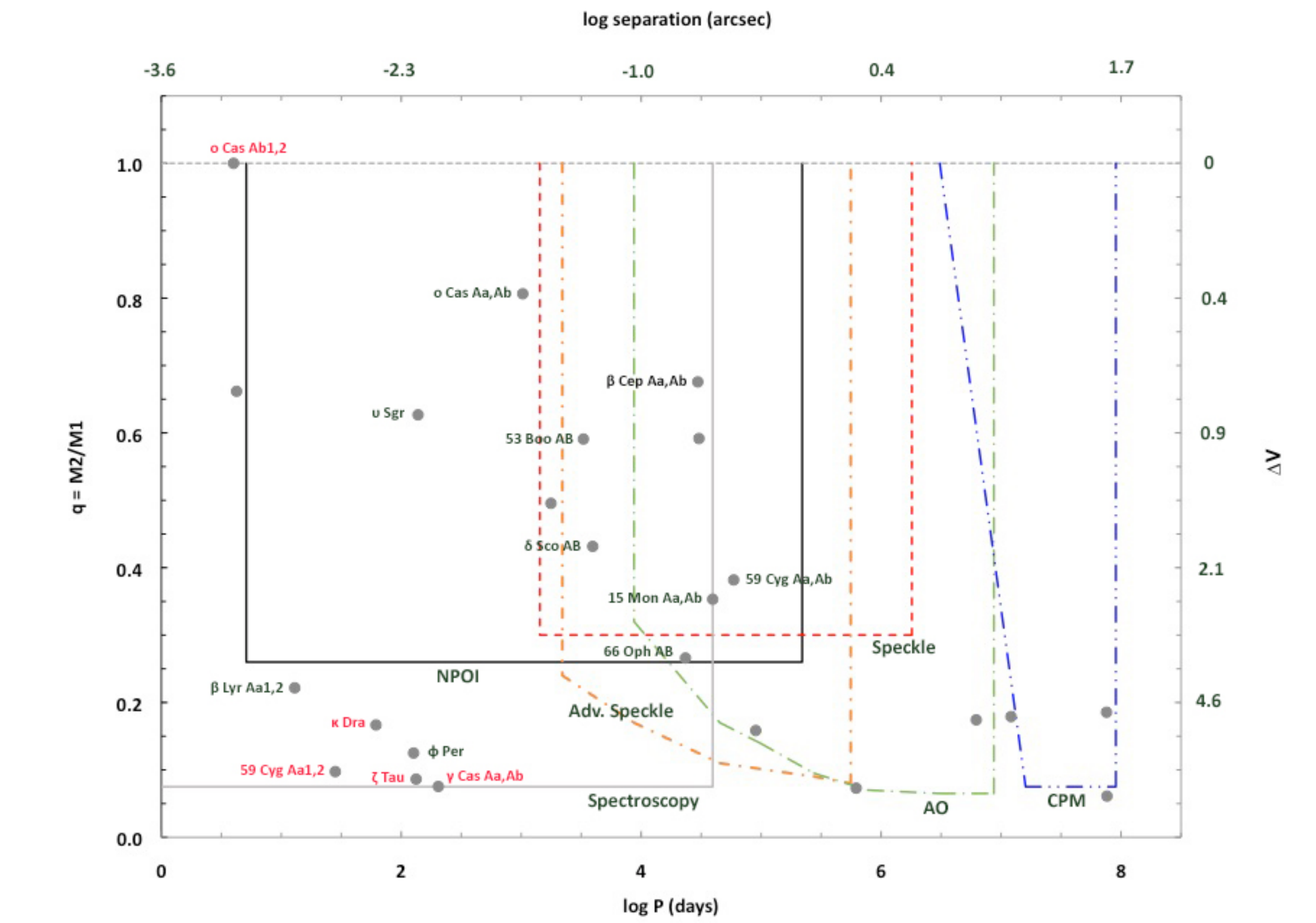}
\caption{Detection limits of the various companion search methods by period (log P (days)) and mass ratio ($q = M_2/M_1$).  The periods are converted to equivalent angular separations (log separation (arcsec); top axis) using the average mass sum (16.3~$M_\sun$) of the identified physical pairs with measured or estimated masses (Table~\ref{binmassper}) and their median source distance of 210.08 pc (Table~\ref{targetlist}).  The mass ratios are converted to approximate $\Delta V$ values (right axis) using an empirical fit to the relation between the $q$ values calculated using nominal masses for a B3V primary and main sequence secondary stars ranging from B5V through M0V and the corresponding $\Delta V$ values calculated from their absolute $V$ magnitudes \citep[masses and absolute magnitudes taken from][]{sck82}.  The horizontal, solid (gray) line at $q = 0.075$, turning vertical at $\log P = 4.596$ represents the empirical limit of the region (toward upper left) where previous studies have detected SB1 and SB2 systems (\S~\ref{bindet}).  The solid black line encloses the detection region of the NPOI (this study, and Papers I and II), the dashed (red) line marks the region of most speckle interferometry surveys, the (orange) dash-dot line the region for newer ``advanced'' speckle cameras, the heavy (green) dash-dot line the region of AO detection, and the dash double-dot (blue) line encloses the region of CPM searches \citep{rag10}.  The derivation of the detection limit boundaries is discussed in \S~\ref{senscomp}.  The overplotted points show the binary components in our Be star sample with measured or estimated masses  (Table~\ref{binmassper}, \S~\ref{bindet}), with those measured by the NPOI labeled in black, while those not observed or resolved in the NPOI observations, but discussed in \S~\ref{future}, are labeled in red.} 
\label{q-vs-logP}
\end{figure*}

To identify the detection limits of the various survey methods mentioned in \S~\ref{bindet}, and to better discuss their likely completeness, we created a plot (Figure~\ref{q-vs-logP}) of system period (log $P$, where period is in days) versus mass ratio ($q = M_2/M_1$), similar to those of \citet{rag10} and \citet{tok14}.  To better illustrate the limits of the various detection methods, the periods are converted to equivalent angular separations (top axis) using the average mass sum (16.3~$M_\sun$) of the identified physical pairs with measured or estimated masses (Table~\ref{binmassper}) and their median HIPPARCOS source distance of 210.08 pc (Table~\ref{targetlist}). Likewise, the mass ratios are converted to approximate $\Delta V$ values (right axis) using an empirical fit to the relation between $q$ values calculated using nominal masses for a B3V primary (the median spectral type in Table~\ref{targetlist}) and main sequence secondary stars ranging from B5V through M0V, and the corresponding $\Delta V$ values calculated from their absolute $V$ magnitudes \citep[masses and absolute magnitudes taken from][]{sck82}.  The overplotted points show the binary components in our Be star sample with measured or estimated masses  (Table~\ref{binmassper}, \S~\ref{bindet}).  The labeled points refer to binary pairs discussed further in \S~\ref{future}.

%% q vs logP plot

The detection limits and observational coverage for the stars in our sample for each method were defined as follows. The horizontal, solid (gray) line (Figure~\ref{q-vs-logP}) at $q = 0.075$, turning vertical at log~$P$ = 4.596 is the empirical limit of the region (toward the upper left) where previous studies have detected SB1 or SB2 systems among the stars in our sample (\S~\ref{bindet}).  Our literature search revealed that 24 of the 31 systems in our sample (77\%) were subjected to RV monitoring over timespans sufficient to confirm periodic variations.  The RV measurement errors in the most recent studies of each system, where indicated, range widely, from $<$ 1 km~s$^{-1}$ to 10 km~s$^{-1}$, with a median value $\sim 4$~km~s$^{-1}$.  While spectroscopic studies have detected RV variations in systems in our sample with orbital periods of order one century (15 Mon Aa,Ab and $\beta$ Cep Aa,Ab; Table~\ref{binmassper}, \S~\ref{bindet}), simulations by \citet{sge09} and \citet{evn10} have shown that for $\sigma_{\mathrm RV}$ = 5 km~s$^{-1}$, the detection probability for binary companions drops steeply for orbital periods greater than a couple hundred days, but is still significant at $\sim$ 10 years.  Thus, higher-precision ($\sigma_{\mathrm RV}$ $\lesssim$ 1 km~s$^{-1}$) RV surveys of the remaining stars of our sample, conducted over sufficient timespans, might hold the prospect of detecting lower-mass companions of longer periods.  Analogously, while spectroscopy is biased towards finding close companions that are a significant fraction of the primary's mass \citep[$q > 0.1$,][]{chn12} the detection of $\gamma$ Cas Aa,Ab at $q = 0.075$ \citep[Table~\ref{binmassper},][]{nhk12} and $\zeta$ Tau at $q = 0.086$ \citep[Table~\ref{binmassper},][]{rbh09} by means of observations over timescales of decades points towards potential for further advances in the spectroscopic detection of lower mass companions.   

The solid black line (Figure~\ref{q-vs-logP}) encloses the detection region of the NPOI; $\Delta V$ $\leq$ 3.5 for separations 3 -- 860 mas (Papers I and II), here extended down to separations $\lesssim$ 1 mas by the inclusion of the 64--79~m baselines (baselines AE-W07, AN-W07, and E06-W07; Tables~\ref{baselines} \&~\ref{progstarsobs}).  Per \S~\ref{bindet}, 100\% of our primary Be star sources were observed, and all the known binaries falling within this region (excluding two binary pairs in the extended $\gamma$ Cas - WDS 00568+6022 system; \S~\ref{gamcas}) were successfully detected, including the first direct, visual resolution of the binary in $\upsilon$ Sgr (\S~\ref{upssgr}).  No new companions to these or any other program stars were found, but two binaries lying outside this region ($\phi$ Per and $\beta$ Lyr Aa1,2; labeled in black in Figure~\ref{q-vs-logP}) were also resolved in our observations (\S~\ref{phiper}, \S~\ref{betlyr}).  The likely reasons why these binaries were detected, while other low mass ratio pairs in a similar range of periods (labeled in red) were not, are discussed later in \S~\ref{future}.

The dashed red line in Figure~\ref{q-vs-logP} encloses the detection region for most speckle interferometry surveys, covering the range $\Delta V$ $\leq$ 3.0 for separations $\sim$ 30 mas to 3.5 arcsec \citep[][and \S~\ref{bindet}]{mas17}.  Per \S~\ref{bindet}, the primary sources in 28 of the 31 systems (90\%) in our sample were observed.  Additionally, two of the outer components in the extended $\gamma$ Cas - WDS 00568+6022 system were also observed.  The orange dash-dot line encloses the detection region for higher sensitivity ``advanced'' speckle cameras \citep{tok10, hvb09} that were used for observations of 16 of our 31 targets (52\%).  The plotted detection limits as a function of angular separation are an average of those quoted in \citet{tok10} and \citet{hvd20}.  Together ``speckle'' and ``advanced speckle'' observations have been reported on 29 of our 31 sample targets (94\% of the sample) and represent the great majority of the visual observations of the systems within their respective regions of Figure~\ref{q-vs-logP}.

The green dash-dot line (Figure~\ref{q-vs-logP}) encloses the detection region of AO observations (generally made at I and K bands).  The boundary of this region is derived from data plotted in Figure~1 of \citet{ljv14}.  Twelve of our 31 sources (39\% of the sample) were observed by AO, including the first, and so far only, observations of the pairs $\gamma$ Cas AB, 23 Tau, and $\beta$ Lyr Aa,Ab.  Additionally, one of the outer components in the extended $\gamma$ Cas - WDS 00568+6022 system was also observed. Lastly, the dash double-dot blue line (Figure~\ref{q-vs-logP}) encloses the region of CPM searches \citep{rag10}.  Based on references in the DMSA, main HIPPARCOS catalog, LIN2, MSC, SB9, and WDS (\S~\ref{bindet}), we estimate that at a minimum 12 of our 31 sources (39\% of the sample) have been the subjects of some type of CPM searches.

While not appearing in Figure~\ref{q-vs-logP}, we should note a significant number (11 of 31 = 36\%) of the stars in our sample were previously observed with LBOI techniques listed in the INT4, including observations made with CHARA, NPOI, the Mount Wilson 100-inch telescope rotating interferometer \citep{mer22}, PTI \citep[Palomar Testbed Interferometer,][]{cwh99}, SUSI \citep[Sydney University Stellar Interferometer,][]{dtb99}, and VLTI.  While some of these instruments may meet or exceed the $\Delta V$ sensitivity of the NPOI, many do not have the capability to detect binaries at the wider angular separations accessible with the NPOI (\S~\ref{npoi} and Paper I).  Also, as discussed in detail in \S~\ref{bindet}, a majority (21 of 31 = 67\%) of the stars in our sample were also previously studied using $\it IUE$ data, resulting in the detection of the sdO components in the $\phi$ Per \citep[\S~\ref{phiper},][]{tbg95} and 59 Cyg \citep[\S~\ref{59cyg},][]{ppg13} systems.  Additionally, small numbers of observations made with various other techniques, including visual interferometers on smaller-aperture telescopes, lunar occultations, ``lucky'' imaging and spectroscopy, and the HST were also reported in the INT4.

The above discussion serves to demonstrate the need for additional observations, particularly by AO, high-contrast speckle, and higher precision spectroscopy over longer timespans.  The prospects for future LBOI observations to detect additional stellar companions are discussed in the following section.

%% Table 8 -- Binary delV, Masses & Periods

\begin{longrotatetable}
\begin{deluxetable*}{lclclclclc}
%%\tiny
%%\center
\tablecaption{Binary Component Magnitude Differences, Masses, and Orbital Periods \label{binmassper}}
\tablewidth{700pt}
%%tabletypesize{scriptsize}
\tablehead{
\colhead{Name} & \colhead{Comp.} & \colhead{$\Delta m$} & \colhead{Ref.$^a$ / Sec.} & \colhead{M1} & \colhead{Ref.$^a$ / Note / Sec.} & \colhead{M2} & \colhead{Ref.$^a$ / Sec.} & \colhead{log P} & 
\colhead{Ref.$^a$ / Sec.} \\
\colhead{} & \colhead{} & \colhead{} & \colhead{} & \colhead{($\it M_\sun$)} & \colhead{} & \colhead{($\it M_\sun$)} & \colhead{} & \colhead{(days)} & \colhead{} \\
\colhead{(1)} & \colhead{(2)} & \colhead{(3)} & \colhead{(4)} & \colhead{(5)} & \colhead{(6)} & \colhead{(7)} & \colhead{(8)} & \colhead{(9)} & \colhead{(10)} \\
}
\startdata
%% Name           Comps       Del V               ref./sec.            M1                    ref./note/sec.        M2                    ref./sec.                    logP(d)    ref..sec.
o Cas           & Ab1,2     & 0                 & ref 1             & 2.5$^b$               & ref 1             & 2.5$^b$               & ref 1                     & 0.602     & ref 2             \\
                & Aa,Ab     & 2.89 $\pm$ 0.02	& \S~\ref{omicas}   & 6.2$^c$               & ref 1             & 5                     & ref 1                     & 3.013     & ref 1             \\
$\gamma$ Cas    & Aa,Ab     & 4.5 -- 8			& \S~\ref{gamcas}	& 13			        & ref 3,4		    & 0.98		            & ref 4		                & 2.309     & ref 4		        \\
                & AB		& 6.85              & ref 5				& 13.98			        & sum Aa,Ab 	    & 1.02                  & ref 6                     & 5.792     & ref 6             \\
00568+6022      & Aa,Ab     & 1.30				& ref 6				& 3.4 $\pm$ 0.8         & ref 7			    & 2.25 $\pm$ 0.16       & ref 7                     & 0.628     & ref 7             \\
                & Aab,Ac	& 0.71              & ORB6              & 5.65			        & sum Aa,Ab 	    & 2.8				    & ref 7                     & 3.248     & ref 7             \\
                & AB		& 0.45  			& ref 6				& 8.45			        & sum Aab,Ac        & 5					    & ref 7		                & 4.482     & ref 7             \\
$\phi$ Per      & binary	& 3.00 $\pm$ 0.11	& \S~\ref{phiper}   & 9.6 $\pm$ 0.3         & ref 8			    & 1.2 $\pm$ 0.2         & ref 8                     & 2.103     & ref 8		        \\
$\zeta$ Tau     & binary	& 3.9 -- 8		    & \S~\ref{zettau}   & 11                    & ref 9			    & 0.95$^d$			    & ref 9		                & 2.124     & ref 9             \\
15 Mon          & Aa,Ab 	& 1.54 $\pm$ 0.04	& \S~\ref{15mon}	& 31			        & ref 6			    & 10.95				    & ref 6		                & 4.596 	& ref 10            \\
                & AB		& 3.20  			& ref 6				& 41.95			        & sum Aa,Ab         & 7.31                  & ref 6                     & 6.793     & ref 6             \\
                & AB,C      & 5.20  			& ref 6				& 49.26			        & sum AB		    & 3.01				    & ref 6		                & 7.883 	& ref 6		        \\
$\kappa$ Dra    & binary    & 3.1 -- 7.3        & \S~\ref{kapdra}   & 4.8 $\pm$ 0.8         & ref 11		    & 0.8                   & ref 12                    & 1.789     & ref 12            \\
53 Boo          & AB		& 1.98 $\pm$ 0.02	& \S~\ref{53boo}	& 2.2			        & ref 13		    & 1.3				    & ref 13                    & 3.518     & ref 13            \\
$\delta$ Sco    & AB		& 1.50 $\pm$ 0.3	& ref 14			& 13.9                  & ref 15            & 6					    & ref 15	                & 3.596 	& ref 15            \\
66 Oph          & AB        & 2.61 $\pm$ 0.02   & \S~\ref{66oph}    & 9.6                   & \S~\ref{66oph}    & 3.8                   & \S~\ref{66oph}            & 4.369     & \S~\ref{66oph}    \\
$\beta$ Lyr     & Aa1,2     & 1.10 $\pm$ 0.08	& \S~\ref{betlyr}	& 2.83 $\pm$ 0.18       & ref 16            & 12.76 $\pm$ 0.27	    & ref 16	                & 1.112 	& ref 17            \\
                & Aa,Ab     & 4.60				& ref 6				& 15.59				    & sum Aa1,2         & 2.47				    & ref 6		                & 4.955	    & ref 6		        \\
                & AB        & 3.77				& ref 6				& 18.06			        & sum Aa,Ab	        & 3.35				    & ref 6		                & 7.879     & ref 6             \\
$\upsilon$ Sgr  & binary	& 3.59 $\pm$ 0.19	& \S~\ref{upssgr}	& 2.52 $\pm$ 0.05$^e$   & ref 18            & 4.02 $\pm$ 0.10$^e$   & ref 18                    & 2.140     & ref 19            \\
59 Cyg          & Aa1,2     & 4.0 -- 4.7		& \S~\ref{59cyg}    & 7.85 $\pm$ 1.55$^f$	& ref 20	        & 0.77 $\pm$ 0.15$^f$	& ref 20	                & 1.450	    & ref 20            \\
                & Aa,Ab     & 3.01 $\pm$ 0.17	& \S~\ref{59cyg}	& 8.62				    & sum Aa1,2         & 3.29                  & ref 6                     & 4.771     & ref 21            \\
$\beta$ Cep     & Aa,Ab     & 1.94$^g$          & \S~\ref{betcep}   & 7.4 $\pm$ 1.5         & \S~\ref{betcep}   & 5 $\pm$ 1.0           & \S~\ref{betcep}           & 4.472     & \S~\ref{betcep}	\\
                & AB		& 4.61				& ref 6				& 12.4				    & sum Aa,Ab	        & 2.22				    & ref 6		                & 7.083     & ref 6		        \\
\enddata
\tablecomments{Col. (1): name (Bayer or Flamsteed designation) from the $\mathit{SIMBAD}$ database \citep{wen00}, and references therein.  Col. (2): component designations (\S~\ref{bindet}).  
Col. (3): component magnitude difference.  Col. (4): section or literature reference.  Col. (5): mass of primary component in $\it M_\sun$.  Col. (6): note, section, or literature reference.  
Col. (7): mass of secondary component in $\it M_\sun$.  Col. (8): section or literture reference.  Col. (9): logarithm of orbital period (days).  Col. (10): section or literature reference.}  
\tablenotetext{a}{Literature references: 1 = \citet{khh10}; 2 = \citet{tgs13}; 3 = \citet{hhs00}; 4 = \citet{nhk12}; 5 = \citet{rbr07}; 6 = \citet{tok18a}; 7 = \citet{cfh92}; 8 = \citet{mmm15}; 
9 = \citet{rbh09}; 10 = \citet{jma19}; 11 = \citet{skk04}; 12 = \citet{skh05}; 13 = \citet{hor12}; 14 = \citet{trb93}; 15 = \citet{cmt12}; 16 = \citet{zgm08}; 17 = \citet{spt09}; 18 = \citet{daj90}; 19 = \citet{khy06}; 20 = \citet{ppg13}; 21 = \citet{mas11}.}
\tablenotetext{b}{M2 (next line)/2.}
\tablenotetext{c}{Mass derived from orbit and parallax.}
\tablenotetext{d}{Middle of range of possible mass values.}
\tablenotetext{e}{Minimum mass.}
\tablenotetext{f}{Middle of range of quoted masses.}
\tablenotetext{g}{Average of $\Delta$$m_{550}$ and $\Delta$$m_{800}$ values.}
\end{deluxetable*}
\end{longrotatetable}

\subsection{Future Directions} \label{future}

The binary pairs in our sample that were successfully detected by the NPOI (\S~\ref{knownbin-det}, labelled in black in Figure~\ref{q-vs-logP}) have measured $\Delta m$ values (Table~\ref{binmassper}, Column 3) that in some cases disagree with the right ($\Delta V$) scale by upwards of 3 magnitudes.  As noted in \S~\ref{senscomp}, this scale was derived assuming the properties of only main sequence stars, pairing a B3V primary with a range of secondary stars of equal or later spectral types.  However, the detailed discussions of each observed pair in \S~\ref{knownbin-det} illustrate that in many cases binary components include a range of increasingly ``exotic'' objects producing increasingly larger disagreement with the $\Delta V$ scale of Figure~\ref{q-vs-logP}.  For pairs where both components are, or are close to, main sequence B stars (66 Oph AB, 15 Mon Aa,Ab), and have relatively weak disk emission at the time of observation ($\delta$ Sco AB) the $\Delta m$ deviations are only 0.2 to 1~mag.  When the primary star in the binary differs significantly in either spectral type (53 Boo AB) or luminosity class ($\beta$ Cep Aa,Ab) from B-main sequence, but are likely the products of normal, single-star evolution, the deviations are 1.1 -- 1.3 mag.  For systems where a third stellar component is present at an angular separation below the resolution limit of the NPOI (59 Cyg, o Cas) the $\Delta m$ differences are 0.7--2.5~mag.  The most extreme discrepancies are seen for $\beta$ Lyr Aa1,2 ($\sim$ 3.1 mag, where the secondary star, the Be, is essentially invisible, the ``companion'' flux originating from its surrounding, opaque disk), or when the primary or secondary is likely the product of extensive interaction and mass transfer ($\phi$ Per, $\upsilon$ Sgr; $\Delta m$ = 2.9--3.6~mag).  Indeed, in the cases of $\beta$ Lyr Aa1,2 and $\phi$ Per it is only the high luminousity of one of the constituent components (disk or sdO subdwarf, respectively), resulting in such anomolously small $\Delta m$ values relative to those expected for pairs of main sequence stars of such extreme mass ratios ($q \lesssim 0.25$) that they were at all detectable as binaries by the NPOI (i.e., they lie outside the nominal range of NPOI sensitivity in Figure~\ref{q-vs-logP}).

The above discussion naturally leads to consideration of the four known spectroscopic binary pairs in our sample $\it not$ detected by our NPOI observations (Figure~\ref{q-vs-logP}, labelled in red). As discussed in \S~\ref{bindet}, $\gamma$ Cas Aa,Ab (\S~\ref{gamcas}), $\zeta$ Tau (\S~\ref{zettau}), and $\kappa$ Dra (\S~\ref{kapdra}), were most likely not detected due to a very large $\Delta m$ differences between their respective components (Table~\ref{binmassper}, Column 3), rather than lack of angular resolution in the NPOI observations (the estimated angular separations for these three binaries are in the range of 4--10~mas, whereas the baselines used provided nominal angular resolutions of 0.7 -- 1.6 mas).  In the case of the fourth pair, 59 Cyg Aa1,2 (\S~\ref{59cyg}) the NPOI observations likely lacked both the $\Delta m$ sensitivity and the angular resolution needed to detect the pair, although the angular resolution might have been sufficient had the longest baseline available at the time of observations (E06-W07) been utilized.  Therefore, it would appear that future observations made at the equivalent of the highest current NPOI angular resolutions, but with a $\Delta m$ sensitivity $\gtrsim$ 5 magnitudes, such as those achieved at CHARA  \citep{gal19} might visually resolve at least some of these pairs.  In this regard, the radio observations of \citet{kcr19} searching for signs of SED turndown in a sample of 57 classical Be stars may provide further encouragement.  \citet{kcr19} argue that such turndowns indicate the presence of close binary companions at orbital periods of tens to hundreds of days that disrupt or truncate the outer part of the Be star disk.  While their source sample only partially overlaps ours, they detected SED turndowns in 13 of our sources.  Not unexpectedly per their paradigm, SED turndowns were seen for the three systems just discussed ($\gamma$ Cas Aa,Ab, $\zeta$ Tau, and $\kappa$ Dra), along with $\phi$ Per (\S~\ref{phiper}), all of which contain Be star -- low mass companion pairs of periods 62 -- 204 days.  Somewhat unexpectedly, an SED turndown was also detected in 66 Oph (\S~\ref{66oph}), perhaps indicating an as yet unresolved component at a period much shorter than the $\approx$ 64 year period of the pair observed by the NPOI.  The remaining eight SED turndown detections of \citet{kcr19} are all among the targets of \S~\ref{nondet} that do not contain previously confirmed spectroscopic or visual binaries.  Therefore, the results of \citet{kcr19} would seem to offer real hope of success in detecting new Be star companions by optical interferometry at the highest currently available angular resolutions and $\Delta m$ sensitivities.  However, on a cautionary note, we remind the reader that in estimating the likely range of $\Delta m$ values for $\gamma$ Cas Aa,Ab, $\zeta$ Tau, and $\kappa$ Dra, we assumed an inherently luminous sdO subdwarf in calculating the minimum likely $\Delta m$ value (for 59 Cyg Aa1,2, where the companion is a known sdO, the range of likely $\Delta m$ reflects uncertainty in the absolute magnitude of the Be component).  In this regard the evolutionary models of \citet{sgd18} indicate that low-mass, stripped envelope stars exist in an overluminous, shell-burning sdO phase for only 2 -- 3\% of their total lifetime as subdwarfs, otherwise they spend most of their time in a core-burning phase at an order of magnitude lower luminosity.  On the assumption that low-mass, stripped envelope stars are ubiqitous, close companions to Be stars (as a source of earlier mass transfer responsible for ``spin up'' of the Be star), then our most optimistic $\Delta m$ estimates would have to be increased by $\sim$ 2.5~mag in nearly all cases.  Referencing Table~\ref{binmassper}, our most optimistic $\Delta m$ estimates for $\gamma$ Cas Aa,Ab, $\zeta$ Tau, and $\kappa$ Dra would then be in the range of 5.6--7~mag, significantly dimming the probability of interferometric detection of such companions even with the most advanced current systems, unless far larger source samples are examined.

\section{Conclusions} \label{conclus}

We have presented the results of observations of 31 stars from a magnitude-limited sample of classical Be stars using the NPOI and Mark III interferometers.  With the exception of $\delta$ Sco, reported elsewhere, 1857 coherent, multi-baseline observations were obtained over 150 nights spanning 1989 -- 2018.  To place our results in the larger context of previous multiplicity studies we also performed an extensive search of the relevant literature.
 
The systematic examination and modeling of the calibrated NPOI and Mark III visibility data for all 31 program stars indicates these sources can be placed into one of three groups.  The 10 sources of the first group each contain a previously confirmed spectroscopic or visual binary at separations $\lesssim$ 0$\farcs$8 that were also detected and modeled in our NPOI and Mark III data.  For two of the sources (66 Oph and $\beta$ Cep) the number and timespan of our new relative position fits allowed new orbital solutions, while for a third source ($\upsilon$ Sgr) our observations provide the first direct, visual detection of the hot sd0 secondary star of the binary.  Four of the 10 sources (o Cas, 15 Mon, $\beta$ Lyr, and $\beta$ Cep) also contain wider binary components that our literature search indicated are physical companions to the narrow binaries, thus forming hierarchical multiple systems.  The three sources of the second group each contain a previously confirmed spectroscopic binary, but were not detected as binaries in our NPOI data, most likely due to a very large magnitude difference ($\Delta m$) between the components rather than lack of angular resolution in the NPOI observations.  The 18 sources of the third group do not contain previously confirmed spectroscopic or visual binaries at separations $\lesssim$ 0$\farcs$8.  Of these, 17 were not detected as binary in our NPOI data, nor were they indicated as having wider, physical binary companions based on our literature search.  The remaining source, BK Cam was resolved on a number of nights with the NPOI and modeled in detail, but the resulting position fits indicate only linear relative motion between the stars over the relatively short timespan of our observations.  However, the close physical proximity of the stars suggests a physical binary.

In evaluating these results, we also attempted to assess the likely number of physical components in each system.  The results of this analysis indicate a multiplicity frequency of 45\% and a companion frequency of 90\%, both trending downward with decreasing primary mass, as suggested by other studies.  We also compared the sensitivity and completeness of our survey with other surveys made at complementary ranges of component angular separation and magnitude difference, demonstrating the need for additional observations, particularly by AO, high-contrast speckle, and higher precision spectroscopy over longer timespans. Lastly, the future role of LBOI in multiplicity studies of Be stars was discussed, concluding that significant progress in our understanding of the frequency of low mass ($q \leq 0.2$) companions will require observations capable of detecting component magnitude differences of $\Delta m$ in the range of 5 to 7~mag, or greater.

\vspace{5mm}
%%\acknowledgments

The authors would like to thank an anonymous referee for comments, suggestions, and questions that helped to significantly improve this paper.  We gratefully acknowledge NPOI observers D. Allen, L. Bright, B. Burress, C. Denison, L. Foley, J. Gannon, A. Guth, W. Johnson, C. Kyte, I. Nisley, B. O'Neill, T. Pugh, C. Sachs, M. Sakosky, J. Sanborn, S. Strosahl, D. Theiling, W. Wack, R. Winner, and S. Zawicki for their careful and efficient operation of the NPOI over the many epochs of data collection required for this paper.  We also gratefully acknowledge B. Mason, U.S. Naval Observatory, for his assistance in the acquisition and analysis of observations of $\beta$ Cep.  D.J.H. and C.T. also acknowledge, with thanks, the support of Central Michigan University.  C.T. acknowledges support from the National Science Foundation through grant AST-1614983. The NPOI project is funded by the Oceanographer of the Navy and the Office of Naval Research.  This research has made use of the SIMBAD astronomical database \citep{wen00} and the VizieR catalog access tool \citep{och00}, both operated at CDS, Strasbourg, France; the Washington Double Star Catalog and associated catalogs, maintained at the U.S. Naval Observatory, Washington, DC (with mirror site at CHARA), The Multiple Star Catalog maintained by Andrei Tokovinin, and the NASA Astrophysics Data System Abstract Service.    Access to Ralph H. Curtis' 1912 paper was made possible courtesy of Hathi Trust.  R.T.Z. gratefully acknowledges the research assistance provided by Morgan Aronson, Librarian, U.S. Naval Observatory Melvin Gilliss Library.  This work has made use of data from the European Space Agency (ESA) mission
{\it Gaia} (\url{https://www.cosmos.esa.int/gaia}), processed by the {\it Gaia} Data Processing and Analysis Consortium (DPAC, \url{https://www.cosmos.esa.int/web/gaia/dpac/consortium}).  Funding for the DPAC has been provided by national institutions, in particular the institutions participating in the {\it Gaia} Multilateral Agreement.

\vspace{5mm}
\facilities{GAIA, HIPPARCOS, Mark III, NPOI}
\software{CANDID, GRIDFIT, OYSTER}

%%                                 REFERENCE LIST
%%\clearpage

\end{document}